\documentclass[11pt,a4paper]{article}

\usepackage{jheppub}
\usepackage{amsmath}
\usepackage{mathtools}
\usepackage{nccmath}
\usepackage{amsthm}
\usepackage{amssymb}
\usepackage{amsfonts}
\usepackage{anyfontsize}
\usepackage{mathrsfs}
\usepackage{bbm}
\usepackage{graphicx}
\usepackage{caption}
\usepackage{bookmark,hyperref}
\usepackage{12many}
\usepackage{parskip}
\usepackage{setspace}
\usepackage{multirow}
\usepackage{arydshln}
\usepackage{anyfontsize}
\usepackage{changepage}
\usepackage{todonotes}

\newcommand{\diff}{\text{d}}
\newcommand{\lagr}{\mathscr{L}}
\newcommand{\viel}{\mathcal{V}}
\newcommand{\graviph}{\mathcal{A}}
\newcommand{\massmatr}{\mathbbm{M}}
\newcommand{\tr}{\text{Tr}}
\newcommand{\sofivefive}{\text{Spin$(5,5)$}}
\newcommand{\spinfourfour}{\text{Spin$(4,4)$}}
\newcommand{\sofourfour}{\text{SO$(4,4)$}}
\newcommand{\sofive}{\text{SO$(5)$}}
\newcommand{\sofour}{\text{SO$(4)$}}
\newcommand{\spinfive}{\text{Spin$(5)$}}
\newcommand{\spinfour}{\text{Spin$(4)$}}

\newcommand{\sutwo}{\text{SU$(2)$}}
\newcommand{\uspfour}{\text{USp$(4)$}}
\newcommand{\usptwo}{\text{USp$(2)$}}
\newcommand{\glfive}{\text{GL$(5)$}}
\newcommand{\uone}{\text{U$(1)$}}

\DeclareMathOperator{\Tr}{Tr}
\DeclareMathOperator{\diag}{diag}
\numberwithin{equation}{section}

\begin{document}

\begin{titlepage}
 \thispagestyle{empty}
 \begin{flushright}
 \hfill{Imperial-TP-2020-CH-01 }\\
 \end{flushright}

 \vspace{80pt}

 \begin{center}
 
     {\fontsize{20}{24} \bf {Black holes in string theory with duality twists}}

     \vspace{50pt}

{\fontsize{13}{16}\selectfont {Chris Hull$^1$, Eric Marcus$^2$, Koen Stemerdink$^2$ and Stefan Vandoren$^2$}} \\[10mm]

{\small\it
${}^1$ 
{{The Blackett Laboratory}},
{{Imperial College London}}\\
{{Prince Consort Road}}, 
{{London SW7 2AZ, U.K.}}\\[3mm]

${}^2$ Institute for Theoretical Physics {and} Center for Extreme Matter and Emergent Phenomena \\
Utrecht University, 3508 TD Utrecht, The Netherlands \\[3mm]}

\vspace{1.7cm}

{\bf Abstract}

\vspace{0.3cm}
   
\end{center}

\begin{adjustwidth}{12pt}{12pt}
We consider 5D supersymmetric black holes in string theory compactifications that partially break supersymmetry.
We  compactify type IIB  on $T^4$ and then further  compactify on a circle with a duality twist  to give Minkowski vacua preserving partial supersymmetry ($\mathcal{N}=6,4,2,0$) in five dimensions. The effective supergravity theory is given by a Scherk-Schwarz reduction with  a Scherk-Schwarz supergravity potential on the moduli space, and the lift of this to string theory imposes a quantization condition on the mass parameters. In this theory, we study black holes with three charges that descend from various ten-dimensional brane configurations. For each black hole we choose the duality twist to be a transformation that preserves the solution, so that it remains a supersymmetric solution of the twisted  theory with partially broken supersymmetry. We discuss the quantum corrections arising from the twist to the pure gauge and mixed gauge-gravitational Chern-Simons terms in the action and the resulting corrections to the black hole entropy.
\end{adjustwidth}

\vspace{20pt}

\newcommand\blfootnote[1]{%
  \begingroup
  \renewcommand\thefootnote{}\footnote{#1}%
  \addtocounter{footnote}{-1}%
  \endgroup
}

\blfootnote{c.hull@imperial.ac.uk \quad e.j.marcus@uu.nl \quad k.c.stemerdink@uu.nl \quad s.j.g.vandoren@uu.nl}

\noindent

\end{titlepage}

\begin{spacing}{1.15}
\tableofcontents
\end{spacing}

\newpage

\section{Introduction}

The D1-D5-P system in type IIB string theory on $T^4\times S^1$ (or $K3\times S^1$) provides a set-up for the study of BPS black holes in five spacetime dimensions, both microscopically and macroscopically. It can be described as an asymptotically flat three-charge 1/8 BPS black hole solution of 5D $\mathcal{N}=8$ supergravity (or 1/4 BPS in $\mathcal{N}=4$ supergravity). The entropy can be computed microscopically from a 2D $(4,4)$ CFT dual to the near horizon geometry of the black hole \cite{Strominger:1996sh}. This system has U-dual formulations in terms of F1 and NS5-branes, and in terms of  intersecting  D3-branes.

It is  interesting to consider extensions of this to  black holes in theories with less supersymmetry.   
Black holes in compactifications  preserving eight supersymmetries in five dimensions can be constructed in M-theory on $CY_3$ \cite{Maldacena:1997de} or in F-theory on  $CY_3\times S^1$ \cite{Vafa:1997gr,Haghighat:2015ega}. In these cases, the microscopic field theory dual to the black hole horizon geometry is a 2D $(0,4)$ CFT. These CFTs are considerably more complicated  than  the $(4,4)$ CFTs on the symmetric product of $T^4$ (or $K3$) as they have less  supersymmetry \cite{Strominger:1996sh}.

In this paper, we consider a different way to reduce supersymmetry, namely string compactifications with a duality twist \cite{Dabholkar:2002sy}, which are the lifts to string theory of Scherk-Schwarz reductions in supergravity \cite{Scherk:1978ta,Scherk:1979zr}. Such compactifications allow for partial supersymmetry breaking and include string vacua preserving no supersymmetry at all (though this won't be the focus in this work). This gives rise to 5D Minkowski vacua preserving $\mathcal{N}=6,4,2,0$ supersymmetry. We investigate 5D supersymmetric black holes in these theories that lift to 10D systems of branes in the string theory picture. An important point is that we choose the twist inducing the supersymmetry breaking 
to be a duality transformation that leaves the original system of branes invariant, and so the 
5D black hole solution of the untwisted theory remains a solution of the twisted theory.
The fields sourcing the system of branes are invariant under the twist, so that the fields appearing in the solution remain massless and the same solution remains as a solution of the twisted theory. This makes it possible to consider the effect of the twist on the corresponding CFT and so to investigate the microscopic aspects of these black holes.
 
This work is a follow-up to the ideas proposed earlier in \cite{Gaddam:2014mna} in an M-theory setting in which supersymmetry is completely broken. Completely broken supersymmetry is not a well controlled situation, and for that  reason we
will focus on the twists preserving some supersymmetry.
 The current work studies string vacua with partial supersymmetry breaking and the macroscopic supergravity description of black holes in such vacua. The microscopic description of the dual CFTs is left for future study.

Scherk-Schwarz reduction of supergravity theories has been extensively studied in the literature; see e.g. \cite{Scherk:1978ta,Scherk:1979zr,Cremmer:1979uq,Kaloper:1999yr,Dabholkar:2002sy,Hull:2003kr,Andrianopoli:2004xu,Hull:2005hk,Hull:2007jy,Hull:2009sg,Hull:2017llx,Gautier:2019qiq} and references therein. IIB supergravity compactified on $T^4$ gives maximal $\mathcal{N}=(2,2)$ supergravity in six dimensions which has a $\sofivefive$ duality symmetry. The maximal compact subgroup of this global symmetry is $\spinfive\times \spinfive$; we shall refer to this as the R-symmetry group. (Note that this global R-symmetry should not be confused with the $\spinfive\times \spinfive$ local symmetry that is introduced in some formulations of the theory.) The 6D scalar fields take values in the coset $\sofivefive/\spinfive\times \spinfive$. We will be interested in the Scherk-Schwarz reduction of this theory to five dimensions, which has been considered previously in  \cite{Hull:2003kr,Andrianopoli:2004xu}.
This uses ans\"atze of the type $\hat{\psi}(x^{\mu},z) = g(z)\,\psi(x^{\mu})$ where $z$ is the $S^1$ coordinate and $g(z)$ is a local element of $\sofivefive$. On going round the circle $z\to z+ 2\pi R$, the fields pick up a monodromy $\mathcal{M}=g(2\pi R)\in\sofivefive$. Such a reduction gives a consistent truncation to a 5D gauged supergravity theory for the fields $\psi(x^{\mu})$, in which there is a Scherk-Schwarz potential for the scalar fields and mass terms are generated for all fields charged under the monodromy.

If the twist $g(z)$ is compact, i.e. it is an element of the R-symmetry group, then the potential is non-negative and has stable five-dimensional Minkowski vacua \cite{Dabholkar:2002sy}. Such a twist can be specified by four parameters $m_1,m_2,m_3,m_4$ which become mass parameters in the reduced theory. The amount of supersymmetry that is preserved in the vacuum depends on the number of parameters $m_i$ that are equal to zero: if $r$ of the parameters $m_i$ are zero, then $\mathcal{N}=2r$ supersymmetry is preserved. This yields 5D supergravities with $\mathcal{N}=8,6,4,2,0$ Minkowski vacua \cite{Andrianopoli:2004xu} (where the case $r=4$ is the untwisted reduction to 5D $\mathcal{N}=8$ supergravity, and the case $r=0$ is the twisted reduction that breaks all supersymmetry). This reduction is a straightforward generalization of the Scherk-Schwarz reduction of 5D $\mathcal{N}=8$ supergravity to 4D with four mass parameters and $\mathcal{N}=8,6,4,2,0$ vacua \cite{Cremmer:1979uq}.

The lift of these supergravity reductions to full compactifications of string theory involves a number of subtle features  \cite{Dabholkar:2002sy}. These  have been worked out in detail for compactifications of IIA string theory on K3 or the heterotic string on $T^4$ followed by a reduction on a circle with a duality twist in \cite{Hull:2017llx,Gautier:2019qiq}. Here we draw on these and the results of \cite{Dabholkar:2002sy} for our construction, which is IIB string theory compactified on $T^4\times S^1$ with a U-duality twist around the circle. Type IIB on $T^4$ has a $\sofivefive$ supergravity duality symmetry that, on the level of the full string theory, is broken to the discrete U-duality subgroup Spin$(5,5;\mathbb{Z})$ by quantum corrections \cite{Hull:1994ys}. The moduli space is the scalar coset space $\sofivefive/[(\spinfive\times \spinfive)/\mathbb{Z}_2]$ identified under the action of Spin$(5,5;\mathbb{Z})$. A key requirement  for there to be a lift to string theory is that   the Scherk-Schwarz monodromy lies in the U-duality group Spin$(5,5;\mathbb{Z})$, imposing a `quantization' condition on the twist parameters $m_i$ \cite{Hull:1998vy,Dabholkar:2002sy}.

There is still an action of the continuous group $\sofivefive$ on the theory, but only the subgroup Spin$(5,5;\mathbb{Z})$ is a symmetry. Reduction of the theory on a circle with a duality twist introduces a monodromy ${\cal{M}}$ which is required to be an element of Spin$(5,5;\mathbb{Z})$. If the monodromy acts as a diffeomorphism of $T^4$, which requires that it is in a GL$(4;\mathbb{Z})$ subgroup of Spin$(5,5;\mathbb{Z})$, then this corresponds to compactification of the IIB string on a $T^4$ bundle over $S^1$. If the monodromy acts as a T-duality of $T^4$, which requires that it is in an $\text{SO}(4,4;\mathbb{Z})$ subgroup of $\text{Spin}(5,5;\mathbb{Z})$, then this constructs a T-fold background, while for general U-duality monodromies this is a U-fold \cite{Hull:2004in}.

A point in the scalar coset will be a minimum of the scalar potential giving a stable Minkowski vacuum if and only if it is a fixed point under the action of the  monodromy ${\mathcal{M}}\in \text{Spin} (5,5;\mathbb{Z})$  \cite{Dabholkar:2002sy}. The monodromy will then generate a $\mathbb{Z}_p$ subgroup of $\text{Spin}(5,5;\mathbb{Z})$ for some integer $p$. Furthermore, at this critical point the construction becomes a $\mathbb{Z}_p$ generalized orbifold of IIB string theory on $T^5$, where the theory is quotiented by the  
$\mathbb{Z}_p$ generated by $\mathcal{M}$ acting on the IIB string on $T^4$ combined with a  shift by $2\pi R/p$ on the circle. When the monodromy is a T-duality, this is a $\mathbb{Z}_p$ asymmetric orbifold.

Regarded as an element of  $\sofivefive$, the monodromy is conjugate to an R-symmetry transformation: ${\cal M}= k R k^{-1}$ for some $R\in \spinfive\times \spinfive$ and $k\in \sofivefive$ \cite{Dabholkar:2002sy}. The rotation $R$ conjugate to a given monodromy is specified by four angles, which are given by the four parameters $m_i$. For $\mathcal{N}=2$ supersymmetry to be preserved, one of the parameters must be zero so that $R$ in fact lies in an $\sutwo\times \spinfive$ subgroup. For $\mathcal{N}=4$ supersymmetry to be preserved, two of the parameters must be zero so that $R$ lies either in a $\spinfive$ subgroup or a $\sutwo\times\sutwo$ subgroup (with one $\sutwo$ factor in each $\spinfive$).
 These two options lead to theories that have the same massless sector, but differ in their massive sectors. We use the notation $(0,2)$ and $(1,1)$ to distinguish these two $\mathcal{N}=4$ theories, 
 reflecting whether the massive states are in $(0,2)$ or $(1,1)$ BPS supermultiplets, using the terminology of  \cite{Hull:2000cf}.
 Lastly, for $\mathcal{N}=6$ supersymmetry to be preserved, three of the parameters must be zero so that $R$ lies in an $\sutwo$ subgroup of the R-symmetry.

As mentioned above, if $\cal M$ is a perturbative symmetry (i.e. a T-duality) in $ \text{Spin} (4,4;\mathbb{Z})$, the theory in the Minkowski vacuum is an asymmetric orbifold. In general this will not be modular invariant, and further modifications are needed to achieve modular invariance. For perturbative monodromies, the shift in the circle coordinate $z$ must be accompanied by a shift in the coordinate of the T-dual circle \cite{Narain:1986qm,Narain:1990mw,Gautier:2019qiq}. Put differently, the quotient introduces phases dependent on both the momentum and the winding number on the circle, and on the charges of the state under the action of $\cal M$ \cite{Gautier:2019qiq}. For non-perturbative monodromies, the arguments of \cite{Ferrara:1995yx,Gautier:2019qiq} lead to phases dependent on other brane wrapping numbers.

Quantum effects 
can lead to corrections to the coefficients of the five-dimensional Chern-Simons terms $A\wedge F\wedge F$ at the two-derivative level and $A\wedge R\wedge R$ at the four-derivative level. There have been indications in the literature (see e.g. \cite{Bonetti:2013cza}) that the $A\wedge R\wedge R$ term can be supersymmetrized in the $\mathcal{N}=4$ $(0,2)$ theory but not in the $\mathcal{N}=4$ $(1,1)$ theory, nor in the $\mathcal{N}=6$ theory. In the $\mathcal{N}=2$ theory the supersymmetrization is known \cite{Hanaki:2006pj}. Our results are in agreement with these claims. That is, we find corrections only in the cases where supersymmetric Chern-Simons terms are expected.
The corrections to the Chern-Simons coefficients modify the black hole solutions in supergravity \cite{Castro:2007hc,deWit:2009de} and therefore also the entropy. We compute the quantum 
corrections to the Chern-Simons coefficients and the resulting modifications to the black hole entropies from
 supergravity and the Kaluza-Klein  modes from the circle compactification, using results in the literature for 5D $\mathcal{N}=2$ supergravity \cite{Castro:2007hc,deWit:2009de}. Similar calculations have been done in different setups, see \cite{Grimm:2018weo,Couzens:2019wls}.

As a by-product of our analysis, we present in detail the supergravity reduction of type IIB on $T^4$. While the general techniques and results are known in the literature \cite{Tanii:1984zk},    the explicit relation between the 10D and 6D fields has not been given, to the best of our knowledge. We present this calculation in \autoref{sec:covform}; the results are relevant for understanding which black holes survive which twist in subsequent sections.

In \autoref{sec:scherk-schwarz-reduction}, we perform the Scherk-Schwarz reduction to five dimensions, starting from the maximally supersymmetric 6D $(2,2)$ supergravity. We construct mass matrices and decompose the 5D field content into massless and massive multiplets. By simply truncating to the massless sector, we can embed known BPS black holes in these five dimensional theories. In \autoref{sec:5dblackholes}, we work out which choices of Scherk-Schwarz twist preserve the D1-D5-P black hole, the 
 F1-NS5-P black hole and the D3-D3-P black hole.
 By tuning the mass parameters, we can find twists that preserve more than  one black hole solution (e.g. 
 we find the twists that preserve both 
 the D1-D5-P and the F1-NS5-P black holes). In \autoref{sec:radiativecorrections}, we study one-loop effects by integrating out the massive fields. We compute the corrections to the Chern-Simons terms and to the entropy of 5D BPS black holes. Finally, in \autoref{sec:embeddingstringtheory}, we discuss how to embed our supergravity model in string theory and discuss the quantization conditions on the parameters $m_i$ of the twist.

\section{Duality invariant formulation of IIB supergravity on \texorpdfstring{$\boldsymbol{T^4}$}{T4} }\label{sec:covform}

Reducing type IIB supergravity on a four-torus gives six-dimensional maximal supergravity. This theory has $\mathcal{N}=(2,2)$ supersymmetry and a $\sofivefive$ duality symmetry group. The goal of this section is to write this supergravity theory in a form in which both the type IIB origin of the six-dimensional fields and the Spin$(5,5)$ symmetry  are manifest. We do this explicitly for the scalar and tensor fields.

\subsection{Ans\"atze for reduction to 6D}
\label{sec:ansatzereductionto6D}

We start from type IIB supergravity. Written in Einstein frame, the bosonic terms in the Lagrangian read
\begin{equation}
\begin{aligned}
\label{typeIIB}
\lagr_\text{\fontsize{6}{0}\selectfont {IIB}}=\;&\left(R^{(10)}-\frac{1}{2}|\text{d}\Phi|^2-\frac{1}{2}\,e^{-\Phi}\,\big|H^{(10)}_3\big|^2-\frac{1}{2}\,e^{2\Phi}\,|\text{d}a|^2-\frac{1}{2}\,e^{\Phi}\,\big|F^{(10)}_3\big|^2-\frac{1}{4}\big|{F}^{(10)}_5\big|^2\right) \ast 1\\[4pt]
&-\frac{1}{2}\:{C}^{(10)}_4\wedge H^{(10)}_3\wedge F^{(10)}_3\,,
\end{aligned}
\end{equation}
where the field strengths are given by
\begin{equation}
\label{field strengths 10d}
\begin{aligned}
&H^{(10)}_3=\diff B^{(10)}_2\,,\\[6pt]
&F^{(10)}_3=\diff C^{(10)}_2-a\,\diff B^{(10)}_2\,,\\[2pt]
&{F}^{(10)}_5=\diff {C}^{(10)}_4-\frac{1}{2}C^{(10)}_2\wedge\diff B^{(10)}_2+\frac{1}{2}B^{(10)}_2\wedge\diff C^{(10)}_2\,.
\end{aligned}
\end{equation}
The superscripts (10) indicate that the fields live in 10 dimensions. The field equations are supplemented by the self-duality constraint
\begin{equation}
{F}^{(10)}_5=*{F}^{(10)}_5\ .
\end{equation}

In our compactification to six dimensions, the coordinates split up as $X^M=(\hat{x}^{\hat{\mu}},y^m)$ with $M=0,\ldots,9$, $\hat{\mu}=0,\ldots,5$ and $m=1,\ldots,4$. We now present the ans\"atze that we use in our reduction. In order to arrive in Einstein frame in 6D, we decompose the ten-dimensional metric as
\begin{eqnarray}
\label{compansatz}
g_{M\!N}=\begin{pmatrix}
g_4^{-1/4}\,g_{\hat{\mu}\hat{\nu}}+g_{mn}\,\mathcal{A}^m_{\hat{\mu}}\mathcal{A}^n_{\hat{\nu}}\; & \;\:\:g_{mn}\,\mathcal{A}^m_{\hat{\mu}}\:\\[5pt]
g_{mn}\,\mathcal{A}^n_{\hat{\nu}} & \;\:g_{mn}
\end{pmatrix}\,,
\end{eqnarray}
where $g_4=\det(g_{mn})$. The compact part of the metric, $g_{mn}$, we parametrize in terms of scalar fields $\phi_i$ ($i=1,\ldots,4$) and $A_{mn}$ ($m<n$) by
\begin{eqnarray}
\label{torusmetric}
g_{mn}=\begin{cases}
\begin{aligned}
\;&e^{\vec{b}_m\cdot\vec{\phi}}+\sum\limits_{k<m}^{}e^{\vec{b}_k\cdot\vec{\phi}}\,(A_{km})^2 && \quad\text{for }\,m=n\\[6pt]
&e^{\vec{b}_m\cdot\vec{\phi}}\,A_{mn}+\sum\limits_{k<m}e^{\vec{b}_k\cdot\vec{\phi}}\,A_{km}A_{kn} && \quad\text{for }\,m<n\\[3pt]
&g_{nm}\phantom{\sum} && \quad\text{for }\,m>n\,.
\end{aligned}
\end{cases}
\end{eqnarray}
Here $\vec{\phi}=(\phi_1,\phi_2,\phi_3,\phi_4)$ and the vectors $\vec{b}_m$ are given by
\begin{eqnarray}
\label{dilatonvectors}
\begin{aligned}
\vec{b}_1&=(-\tfrac{1}{\sqrt{2}},-\tfrac{1}{\sqrt{2}},-\tfrac{1}{\sqrt{2}},\tfrac{1}{2})\,,\\
\vec{b}_2&=(-\tfrac{1}{\sqrt{2}},\tfrac{1}{\sqrt{2}},\tfrac{1}{\sqrt{2}},\tfrac{1}{2})\,,\\
\vec{b}_3&=(\tfrac{1}{\sqrt{2}},\tfrac{1}{\sqrt{2}},-\tfrac{1}{\sqrt{2}},\tfrac{1}{2})\,,\\
\vec{b}_4&=(\tfrac{1}{\sqrt{2}},-\tfrac{1}{\sqrt{2}},\tfrac{1}{\sqrt{2}},\tfrac{1}{2})\,.
\end{aligned}
\end{eqnarray}
From this, it can be computed that $g_4=e^{2\phi_4}$, so the scalar $\phi_4$ parametrizes the volume of the $T^4$.

We reduce the 10D form-valued fields by simply splitting into  components with different numbers of indices on the torus. For example, the Kalb-Ramond field $B_2^{(10)}$ decomposes as
\begin{equation}
\label{splitforms}
\begin{aligned}
B^{(10)}_2&=\frac{1}{2}\,B_{M\!N}\,\diff X^M\wedge\diff X^N\\
&=\frac{1}{2}\,B_{\hat{\mu}\hat{\nu}}\,\diff x^{\hat{\mu}}\wedge\diff x^{\hat{\nu}}+B_{\hat{\mu} m}\,\diff x^{\hat{\mu}}\wedge\diff y^m+\frac{1}{2}\,B_{mn}\,\diff y^m\wedge\diff y^n\\
&=B_2^{(6)}+B^{(6)}_{1,m}\wedge\diff y^m+\frac{1}{2}\,B_{mn}\,\diff y^m\wedge\diff y^n\,,
\end{aligned}
\end{equation}
where $B_2^{(6)}$, $B^{(6)}_{1,m}$ and $B_{mn}$ are 2, 1 and 0-forms defined on the six-dimensional non-compact space. The ten-dimensional scalars are simply equal to their six-dimensional descendants, e.g. $\Phi^{(10)}=\Phi^{(6)}=\Phi$. For this reason, we usually drop the superscript (D) for scalar fields.

\paragraph{Reduction of the self-dual five-form field strength.}

To find the fields that descend from the RR four-form ${C}_4^{(10)}$ we need to be a bit careful, since it has a self-dual field strength: $\ast{F}^{(10)}_5={F}^{(10)}_5$. Because of this self-duality, the action \eqref{typeIIB} does not properly describe the dynamics of the RR four-form. So instead of reducing the action, we should reduce the corresponding field equations along with the self-duality constraint. The action \eqref{typeIIB} with field strengths \eqref{field strengths 10d} yields the following equation of motion and Bianchi identity
\begin{align}
\text{d}\big(\!\ast F^{(10)}_5\big)&=\text{d}B^{(10)}_2 \wedge \text{d}C_2^{(10)} \,,\\[2pt]
\text{d}{F}^{(10)}_5&=\text{d}B^{(10)}_2 \wedge \text{d}C_2^{(10)} \,.\label{bianchiF5}
\end{align}
We see that, because of the self-duality of $F_5^{(10)}$, these two equations are identical, so we only have to reduce one of them. In what follows, we choose to reduce the Bianchi identity \eqref{bianchiF5}. Subsequently, we reduce the self-duality equation and use it to rewrite the six-dimensional Bianchi identities to a system of Bianchi identities and equations of motion. By integrating this system of equations to an action, we find the proper result of the reduction of ${C}_4^{(10)}$. Below, we work out this reduction in detail for the scalars and the two-forms.

First, we consider the scalars. In 6D, massless four-forms can be dualized to scalars, so we need to consider the components of $F_5^{(10)}$ that have either zero or four legs on the torus. The Bianchi identities for these components following from \eqref{bianchiF5} read
\begin{equation}
\begin{aligned}\label{bianchiP5}
\text{d}P_{1}^{(6)}&= \frac{1}{2!\,2!}\,\varepsilon^{mnpq}\,\text{d}B_{mn} \wedge \text{d} C_{pq} \,, \\[2pt]
\text{d}P_{5}^{(6)}&= \text{d}B^{(6)}_2 \wedge \text{d}C_2^{(6)} \,.
\end{aligned}
\end{equation}
Here we have introduced the notation $P_1^{(6)}=\frac{1}{4!}\,\varepsilon^{mnpq}\,F_{1,mnpq}^{(6)}$ and $P_5^{(6)}=F_5^{(6)}$. Next, we write down the relevant components that follow from the reduction of the self-duality constraint. By using the metric ansatz \eqref{compansatz}, and ignoring  interactions with the graviphotons $\mathcal{A}_{\hat{\mu}}^m$, we find
\begin{eqnarray}\label{selfdualityP5}
P_5^{(6)} = \frac{1}{g_4}\ast P_1^{(6)} \,.
\end{eqnarray}
We now use this constraint to eliminate $P_5^{(6)}$ from \eqref{bianchiP5}. In this way, we find the following Bianchi identity and equation of motion for the one-form field strength $P_1^{(6)}$
\begin{equation}
\begin{aligned}
\text{d}P_{1}^{(6)}&= \frac{1}{2!\,2!}\,\varepsilon^{mnpq}\,\text{d}B_{mn} \wedge \text{d} C_{pq} \,, \\[2pt]
\text{d}\big(e^{-2\phi_4}\ast P_{1}^{(6)}\big)&= \text{d}B^{(6)}_2 \wedge \text{d}C_2^{(6)} \,.
\end{aligned}
\end{equation}
From the first equation, we can find an expression for $P_1^{(6)}$ in terms of the corresponding scalar field that we denote by $b$. The second equation can be integrated to an action that contains both the kinetic term for $b$ and interaction terms between $b$ and other scalar and two-forms fields. These expressions can be found in \eqref{d=6scalars} and \eqref{d=6scalarfieldstrengths}.

Next, we look at the two-forms coming from $C_4^{(10)}$. We are interested in the action for the six-dimensional two-form fields and their interactions with scalar fields. We will ignore interactions with six-dimensional one-forms. The relevant components that follow from the reduction of \eqref{bianchiF5} read
\begin{equation}
\label{bianchiF3mn}
\begin{aligned}
\diff F_{3,mn}^{(6)}&=\diff B_{mn} \wedge \diff C_2^{(6)}+\diff B^{(6)}_2 \wedge \diff C_{mn}\\[2pt]
&=\diff \big(B_{mn}\;\diff C_2^{(6)}-C_{mn}\;\diff B^{(6)}_2\big)\,.
\end{aligned}
\end{equation}
These are Bianchi identities for six tensors in six dimensions. We want to eliminate half of these fields in exchange for equations of motion for the residual ones. We choose to retain the components $F_{3,mn}^{(6)}$ for $mn=12,13,14$ and to eliminate the ones with indices $mn=23,24,34$. For this, we again use the reduced self-duality constraint. The relevant components are
\begin{equation}
\label{selfdualitytensors}
F_{3,mn}^{(6)}=\frac{1}{2}\sqrt{g_4}\:\varepsilon_{mnpq}\:g^{pr}g^{qs}\ast F_{3,rs}^{(6)}\,.
\end{equation}
Due to the summations over the $r$ and $s$ indices, each component of this equation contains a linear combination of all the dual field strengths $\ast F_{3,rs}^{(6)}$ (recall that the metric on $T^4$ is given by \eqref{torusmetric}). Consequently, solving \eqref{selfdualitytensors} for three of the six field strengths results in unwieldy expressions. We choose not to write down these expressions here, but instead to give a step-by-step outline of the way we use them to find an action for the 6D tensors.

First, we introduce a new notation for the field strengths that we plan on retaining: $P_{3;\,1}^{(6)}=F_{3,12}^{(6)}$, $P_{3;\,2}^{(6)}=F_{3,14}^{(6)}$ and $P_{3;\,3}^{(6)}=F_{3,13}^{(6)}$. Here the first subscript indicates that these are three-forms, and the second subscript labels the three distinct field strengths (we will sometimes drop this label when we are talking about all three of them). The expressions for these field strengths in terms of the corresponding two-form fields can be deduced from \eqref{bianchiF3mn}. For example,
\begin{equation}
P_{3;\,1}^{(6)}=\diff R_{2;\,1}^{(6)}+B_{12}\;\diff C_2^{(6)}-C_{12}\;\diff B^{(6)}_2\, ,
\end{equation}
where $R_{2;\,1}^{(6)}$ is then one of the two-forms that arise from compactifying the ten-dimensional 4-form.
Similar expressions can be found for $P_{3;\,2}^{(6)}$ and $P_{3;\,3}^{(6)}$ in terms of fields that we call $R_{2;\,2}^{(6)}$ and $R_{2;\,3}^{(6)}$ respectively.

Next, we solve the six equations in \eqref{selfdualitytensors} for $F_{3,mn}^{(6)}$ and $\ast F_{3,mn}^{(6)}$ (for $mn=23,24,34$) in terms of the field strengths $P_{3}^{(6)}$ and their duals $\ast P_{3}^{(6)}$. By substituting these expressions in the components of \eqref{bianchiF3mn} for $mn=23,24,34$, we find the equations of motion for the tensor fields $R_{2}^{(6)}$ purely in terms of the (dual) field strengths $P_{3}^{(6)}$ and $\ast P_{3}^{(6)}$, and fields that don't descend from the RR four-form $C_4^{(10)}$. These field equations are quite unwieldy, but with some careful bookkeeping they can be integrated to an action. We will not write down this awkward version of the action here. Instead, we write down a more elegant version of the action for the six-dimensional tensor fields and their interactions with scalar fields in \autoref{tensors}.

\subsection{6D scalars}
\label{sec:scalars}

The field content of maximal six-dimensional supergravity contains 25 scalars. In terms of their origin in type IIB, these are $\Phi$, $\phi_i$, $A_{mn}$, $B_{mn}$, $C_{mn}$, $a$ and $b$. We find the action for these scalar fields by using the methods and ans\"atze described in the previous section. This yields
\begin{equation}
\label{d=6scalars}
\begin{aligned}
e^{-1}_{(6)}\,\lagr_\text{s} \;=\;& -\frac{1}{2}\,|\diff\Phi|^2 -\frac{1}{4}\,|\diff\phi_4|^2-\frac{1}{2}\,|\diff g_{mn}|^2 -\frac{1}{2}\,e^{-\Phi}\,\big|H^{(6)}_{1,mn}\big|^2 \\[4pt]
& -\frac{1}{2}\,e^{2{\Phi}}\,|\text{d}a|^2 -\frac{1}{2}\,e^{\Phi}\,\big|F^{(6)}_{1,mn}\big|^2 -\frac{1}{2}\,e^{-2{\phi_4}}\,\big|P_{1}^{(6)}\big|^2 \,.
\end{aligned}
\end{equation}
Note that the absolute values apply both to the 6D Lorentz indices and to the indices on the torus. For example, $|H^{(6)}_{1,mn}|^2 = \frac{1}{2!}\,H_{\hat{\mu}mn}H^{\hat{\mu}mn} = \frac{1}{2!}\,H_{\hat{\mu}mn}\,g^{mp}H^{\hat{\mu}}_{\;\;\,pq}\,g^{pn}$. The field strengths in \eqref{d=6scalars} are given by
\begin{eqnarray}
\label{d=6scalarfieldstrengths}
\notag {H}_{1,mn}^{(6)}&=&\diff {B}_{mn} \,,\\[6pt]
{F}_{1,mn}^{(6)}&=&\diff {C}_{mn}-a\,\diff {B}_{mn} \,,\\[3pt]
\notag P_{1}^{(6)}&=&\diff {b}+\frac{1}{8}\,\varepsilon^{mnpq}\left({B}_{mn}\,\diff {C}_{pq}-{C}_{mn}\,\diff {B}_{pq}\right) \,.
\end{eqnarray}
These 25 scalar fields together parametrize the coset $\sofivefive/(\spinfive \times \spinfive)$ \cite{Tanii:1984zk}. The action above has a global Spin$(5,5)$ and a local Spin$(5)$ $\!\times\!$ Spin$(5)$ symmetry. In its current form, these symmetries are not visible, so we will now write this action in a form that makes both symmetries manifest.

In order to do this, we construct a generalized vielbein (or coset representative) $\viel$ from the scalar fields. This vielbein is an element of $\sofivefive$ and it transforms as $\viel \rightarrow U\,\viel \, W(\hat{x})$, with $U \in \sofivefive$ and $W(\hat{x}) \in \spinfive \times \spinfive$. We now define the $\spinfive \times \spinfive$ invariant field $\mathcal{H} = \viel \, \viel^T$, that transforms as $\mathcal{H} \rightarrow U\,\mathcal{H}\,U^T$ under global $\sofivefive$ transformations\footnote{In this section we suppress $\sofivefive$ indices, but we will need them later on. With indices, $\mathcal{H}$ is written as $\mathcal{H}_{AB}$ and it transforms as $\mathcal{H}_{AB} \,\rightarrow\, U_A^{\;\;\,C} \, \mathcal{H}_{CD} \, \big(U^T\big)^D_{\;\;\,B\,}$. The inverse of $\mathcal{H}$ is written with upper indices: $\mathcal{H}^{-1}=\mathcal{H}^{AB}$.}. We can now write the scalar Lagrangian in terms of $\mathcal{H}$ as
\begin{equation}
\label{d=6scalarsSO55}
e^{-1}_{(6)}\,\lagr_\text{s} \:=\: \frac{1}{8}\,\text{Tr}\big[\partial_{\hat{\mu}}\mathcal{H}^{-1}\partial^{\hat{\mu}}\mathcal{H}\big] \,.
\end{equation}
In this formulation, the Lagrangian is manifestly invariant under the U-duality group $\sofivefive$.

We now specify the way we build $\viel$ from the 25 scalar fields so that the two Lagrangians \eqref{d=6scalars} and \eqref{d=6scalarsSO55} are equal to one another. We choose to build $\viel\in\sofivefive$ in $\tau$-frame, i.e. it satisfies $\viel^T\tau\,\viel=\tau$ (for the definition of $\tau$, see Appendix \ref{sec:groupappendix}). The exact construction is as follows:
\begin{equation}
\label{Scalarvielbein}
\begin{aligned}
\viel\:=\:&\exp\!\big[b \, T^{b}\big] \times \exp \!\bigg[ \sum_{1\leq m<n\leq4}\left(B_{mn} \, T^B_{mn} +C_{mn} \, T^C_{mn}\right) \! \bigg] \times \exp\!\big[a \, T^{a}\big] \\
&\times \bigg(\prod_{1\leq m<n\leq4} \exp \left[A_{mn} \, T^A_{mn}\right]\bigg) \times \exp\!\bigg[\Phi \, H_0 + \sum_{i=1}^{4} \phi_i \, H_i\bigg] \,.
\end{aligned}
\end{equation}
Here the $T$'s and the $H$'s are generators of $\mathfrak{so}(5,5)$ that span the subspace of $\mathfrak{so}(5,5)$ that generates the coset $\sofivefive/(\spinfive \times \spinfive)$. The precise expressions for these generators are given in Appendix \ref{sec:groupappendix}. All the scalar fields appear under the same name as in \eqref{d=6scalars}.

Because we construct our vielbein \eqref{Scalarvielbein} in $\tau$-frame\footnote{For the convenience of the reader, it might be useful to mention how this convention is related to those of other authors. The following relations hold: $\viel=X\,U_{\text{[Tanii]}}X$ where $U_{\text{[Tanii]}}$ is the vielbein that is used in \cite{Tanii:1984zk}, and $\viel=\viel_{\text{[BSS]}}X$ where $\viel_{\text{[BSS]}}$ is the vielbein that is used in \cite{Bergshoeff:2007ef}. The matrix $X$ is a conjugation matrix that is defined in Appendix \ref{sec:groupappendix}.}, the transformation matrices $U$ and $W$ are also written in $\tau$-frame. Henceforth, we use this frame whenever $\sofivefive$ and $\spinfive\times\spinfive$ groups appear (unless mentioned otherwise).

\subsection{6D tensors}
\label{tensors}

The field content of maximal supergravity in six dimensions contains five 2-form tensor gauge fields. Collectively, we denote these fields by $A^{(6)}_{2,a}$ ($a=1,\ldots,5$), and their field strengths by $G^{(6)}_{3,a}=\diff A^{(6)}_{2,a}$. The Lagrangian for these fields reads \cite{Tanii:1984zk,Bergshoeff:2007ef}
\begin{equation}
\lagr_{\text{t}}=-\frac{1}{2}\,K^{ab}\,G^{(6)}_{3,a} \wedge\ast\, G^{(6)}_{3,b}-\frac{1}{2}\,L^{ab}\,G^{(6)}_{3,a} \wedge G^{(6)}_{3,b}\,.
\end{equation}
Here $K^{ab}$ and $L^{ab}$ are functions of the scalar fields. We define a set of dual field strengths $\tilde{G}^{(6)a}_{3}=K^{ab}\ast G^{(6)}_{3,b}+L^{ab}\,G^{(6)}_{3,b}$ so that we can write the Lagrangian in the more compact form
\begin{equation}
\label{lagrtensors}
\lagr_{\text{t}}=-\frac{1}{2}\,G^{(6)}_{3,a} \wedge \tilde{G}^{(6)a}_{3}\,.
\end{equation}
In this notation, we write the Bianchi identities and the equations of motion as $\diff G^{(6)}_{3,a}=0$ and $\diff\tilde{G}^{(6)a}_{3}=0$. We can combine these in the more compact notation $\diff G^{(6)}_{3,A}=0$, where $G^{(6)}_{3,A}$ is defined as
\begin{equation}\label{tensorvectordefinition}
G^{(6)}_{3,A}=\begin{pmatrix}
G^{(6)}_{3,a}\\[2pt]
\tilde{G}^{(6)a}_{3}
\end{pmatrix}\,.
\end{equation}
The $\sofivefive$ duality symmetry acts on this ten-component vector as
\begin{equation}\label{tensortransformation6d}
G^{(6)}_{3,A} \;\:\rightarrow\;\:
U_A^{\;\;\,B} \: G^{(6)}_{3,B}\,,
\qquad\quad U_A^{\;\;\,B}\in\,\sofivefive\,.
\end{equation}
Only the subgroup $\glfive\subset\sofivefive$ is a symmetry of the action. The full symmetry group is only manifest on the level of the field equations.

When we decompose our coset representative in $5\times5$ blocks as $\viel=\begin{psmallmatrix}a&b\\c&d\end{psmallmatrix}$, we can write the matrices $K^{ab}$ and $L^{ab}$ as
\begin{equation}
K=\tfrac{1}{2}((c+d)(a+b)^{-1}-(c-d)(a-b)^{-1})\,, \quad L=\tfrac{1}{2}((c+d)(a+b)^{-1}+(c-d)(a-b)^{-1})\,.
\end{equation}
Now, by making the identification
\begin{equation}
A^{(6)}_{2,a}=\big(R_{2;\,1}^{(6)}\,,\,R_{2;\,2}^{(6)}\,,\,R_{2;\,3}^{(6)}\,,\,C^{(6)}_2,\,-B^{(6)}_2\big)\,,
\end{equation}
the Lagrangian \eqref{lagrtensors} is exactly equal to the one that we find by explicit reduction from type IIB supergravity using the ans\"{a}tze given in \autoref{sec:ansatzereductionto6D}. The advantage of \eqref{lagrtensors} is that we have made the duality symmetry manifest.

\paragraph{Doubled formalism.}

It is a common feature of supergravity actions in even dimensions that only a subgroup of the duality group is a symmetry of the action. In such cases, one can use the so-called doubled formalism \cite{Cremmer:1997ct} to construct an action that realizes the full symmetry group. In order to do this, one needs to introduce twice the original amount of form-valued fields as well as a constraint that makes sure that the doubled theory does not contain more degrees of freedom than the original theory.

We apply this formalism to our 6D tensor fields. We promote the $\tilde{G}^{(6)a}_{3}$ to field strengths that correspond to the doubled fields, i.e. we write them as $\tilde{G}^{(6)a}_{3}=\diff\tilde{A}^{(6)a}_{2}$. These doubled fields $\tilde{A}^{(6)a}_{2}$ are now treated as independent fields. We write down the doubled Lagrangian as
\begin{equation}
\label{lagrtensorsdoubled}
\lagr^{\text{(doubled)}}_{\text{t}}=-\frac{1}{4}\,\mathcal{H}^{AB}\,G^{(6)}_{3,A}\wedge\ast\,G^{(6)}_{3,B}\,.
\end{equation}
In this formulation we have ten field strengths $G^{(6)}_{3,A}$ that satisfy the Bianchi identities $\diff G^{(6)}_{3,A}=0$ and the equations of motion $\diff\big(\mathcal{H}^{AB}\ast G^{(6)}_{3,B}\big)=0$. Furthermore, these fields are subject to the self-duality constraint
\begin{equation}
\label{constraintdoubled}
G^{(6)}_{3,A}=\tau_{AB}\,\mathcal{H}^{BC}\ast G^{(6)}_{3,C}\,.
\end{equation}
By imposing this constraint on the field equations, we see that they reduce to the ones that correspond to the undoubled action. Thus we have found a proper doubled version of \eqref{lagrtensors}. Both the action \eqref{lagrtensorsdoubled} and the constraint \eqref{constraintdoubled} are invariant under the full $\sofivefive$ duality group. This can be seen directly from the way that these transformations work on the fields:
\begin{equation}
\label{transformations6dfields}
\mathcal{H}^{AB} \;\rightarrow\; \big(U^{-T}\big)^A_{\;\;\:C} \, \mathcal{H}^{CD} \, \big(U^{-1}\big)_D^{\;\;\;\,B}\,,
\qquad
G^{(6)}_{3,A} \;\rightarrow\; U_A^{\;\;\,B} \, G^{(6)}_{3,B}\,,
\qquad
U_A^{\;\;\,B}\in\,\sofivefive\,,
\end{equation}
where we use the notation $U^{-T}=(U^{-1})^T$.

\section{Scherk-Schwarz reduction to five dimensions}
\label{sec:scherk-schwarz-reduction}

In a Scherk-Schwarz reduction, one considers a $(D+1)$-dimensional supergravity theory with a global symmetry given by a Lie group $G$ that is compactified to $D$ dimensions. The difference between `ordinary' Kaluza-Klein and Scherk-Schwarz reduction lies in the compactification ansatz. 
Consider a field $\hat{\psi}$ in the $(D+1)$-dimensional theory that transforms as $\hat{\psi}\rightarrow g\hat{\psi}$ with $g\in G$ (for scalars, this is typically a non-linear realization, while some fields such as the metric in Einstein frame will be invariant). The Scherk-Schwarz ansatz then gives $\hat{\psi}$  a dependence on the  coordinate  $z$ on the circle, which has periodicity $z\simeq z+2\pi R$, given by
\begin{equation}
\label{SSansatz}
\hat{\psi}(x^{\mu},z) = \exp\left({\frac{M z}{2\pi R}}\right)\, \psi(x^{\mu}) \,,
\end{equation}
where $M$ lies in the Lie algebra of $G$. This ansatz is not periodic around the circle, but picks up a monodromy $\mathcal{M}=e^{M} \in G$. The Lie algebra element $M$ is sometimes called the mass matrix because it  appears in mass terms in the $D$-dimensional theory. For more details, see \cite{Scherk:1978ta,Scherk:1979zr,Cremmer:1979uq,Kaloper:1999yr,Dabholkar:2002sy,Hull:2003kr,Andrianopoli:2004xu,Hull:2005hk,Hull:2007jy,Hull:2009sg,Hull:2017llx,Gautier:2019qiq,Ozer:2003} and references therein. A conjugate mass matrix 
\begin{equation}
M'=gMg^{-1} \,,
\end{equation}
with $g\in G$, gives a conjugate monodromy
\begin{equation}
\mathcal{M}'=g\mathcal{M} g^{-1} \,.
\end{equation}
This conjugated monodromy gives a massive theory that is related to the one for the monodromy $\mathcal{M}$ by a field redefinition, so that it defines an equivalent theory. Thus the possible Scherk-Schwarz reductions are classified by the conjugacy classes of the duality group \cite{Dabholkar:2002sy}.

In our case, we reduce from 6D to 5D on a circle with a Scherk-Schwarz twist. We denote the coordinates on the five-dimensional Minkowski space by $x^\mu$ and the coordinate on the circle by $z$. The compact coordinate is periodic with periodicity $z\simeq z+2\pi R$.
The metric (in Einstein frame) is inert under the duality group, 
so we choose  the conventional  Kaluza-Klein  metric ansatz:
\begin{equation}
\label{KK-ansatz general}
g_{\hat{\mu}\hat{\nu}}=
\begin{pmatrix}
e^{-\sqrt{1/6}\,\phi_5}\,g_{\mu\nu}+e^{\sqrt{3/2}\,\phi_5}\,\mathcal{A}^5_\mu\mathcal{A}^5_\nu & \;\;\; e^{\sqrt{3/2}\,\phi_5}\,\mathcal{A}^5_\mu \\[6pt]
e^{\sqrt{3/2}\,\phi_5}\,\mathcal{A}^5_\nu & \;\;\; e^{\sqrt{3/2}\,\phi_5}
\end{pmatrix}\,.
\end{equation}
The factors in the exponents are chosen so that we arrive in Einstein frame in five dimensions and the scalar field $\phi_5$ is canonically normalized \cite{pope2000kaluza}.

The result of our reduction is a gauged $\mathcal{N}=8$ supergravity theory in five dimensions in which a non-semi-simple subgroup of $\sofivefive$ is gauged. The gauge group contains an important $\uone$ subgroup for which $\mathcal{A}^5_\mu$ is the corresponding gauge field. For each twist, the theory has a vacuum (partially) breaking the supersymmetry where it can be described by an $\mathcal{N}<8$ effective field theory. This reduction from 6D to 5D has been considered previously in \cite{Hull:2003kr,Andrianopoli:2004xu}. An important feature is that reducing self-dual 2-form gauge fields in 6D can result in massive self-dual 2-form fields in 5D \cite{Hull:2003kr}. See \cite{Ozer:2003,Hull:2007jy} for further details.

\subsection{Monodromies and masses}

In six dimensions the global symmetry is $G=\sofivefive$, so in principle we can choose the mass matrix  to be  any  element of the Lie algebra of $G$. However, our goal is to obtain a Minkowski vacuum with partially broken supersymmetry, so, as discussed in the introduction, we restrict our twist to be conjugate to an element of the R-symmetry group
\begin{equation}
\label{R-symmetry group}
\uspfour_\text{L}\times \uspfour_\text{R} \; =\; \spinfive_\text{L}\times \spinfive_\text{R}  \,,
\end{equation}
that preserves the identity in $\sofivefive$. We take then a monodromy
\begin{equation}
\label{monocon}
\mathcal{M}=g \tilde {\mathcal{M} }g^{-1} \,, \qquad g\in \sofivefive \,, \qquad  \tilde{ \mathcal{M}} \in \uspfour_\text{L}\times \uspfour_\text{R} \subset \sofivefive \,.
\end{equation}
By a further conjugation, we can bring  $ \tilde{ \mathcal{M}}$ to an element $ \bar {\mathcal{M}}$ of a maximal torus $\mathbb{T}=\text{U}(1)^4$ of the R-symmetry group $\uspfour_\text{L}\times \uspfour_\text{R} $
\begin{equation}
\label{monocon2}
 \tilde{ \mathcal{M}}=h \bar {\mathcal{M} }h^{-1} \,, \qquad h\in \uspfour_\text{L}\times \uspfour_\text{R} \,, \qquad  \bar{ \mathcal{M}} \in \mathbb{T} \subset
 \uspfour_\text{L}\times \uspfour_\text{R}  \,.
\end{equation}
The element $ \bar {\mathcal{M} }$ of a maximal torus $\mathbb{T}=\text{U}(1)^4$ is then specified by four angles, which we denote $m_1,m_2,m_3,m_4$;
we  take $0\le m_i < 2\pi$. 
Writing 
\begin{equation}
\label{monocon3}
\bar {\mathcal{M} } = (\mathcal{M}_\text{L}^{\mathfrak{usp}(4)}, \mathcal{M}_\text{R}^{\mathfrak{usp}(4)} ) \,, \qquad
\mathcal{M}_\text{L}^{\mathfrak{usp}(4)} \in \uspfour_\text{L} \,, \qquad
\mathcal{M}_\text{R}^{\mathfrak{usp}(4)} \in \uspfour_\text{R} \,,
\end{equation}
we can take the monodromies to be in the $\sutwo\times \sutwo$ subgroup of $\text{USp}(4)$ for both the left and right factors (note that $\sutwo \cong \text{USp}(2)$):
\begin{equation}
\label{subgroupRsymm}
\sutwo _{\text{L}_1} \times \sutwo_{\text{L}_2} \times \sutwo_{\text{R}_1} \times \sutwo_{\text{R}_2} \;\subset\; \uspfour_\text{L} \times \uspfour_\text{R} \,.
\end{equation}
We can then take, for example, 
\begin{equation}
\label{monoex}
\mathcal{M}_\text{L}^{\mathfrak{usp}(4)} = e^{m_1 \sigma _3} \otimes e^{m_2 \sigma _3} \,, \qquad
\mathcal{M}_\text{R}^{\mathfrak{usp}(4)} = e^{m_3 \sigma _3} \otimes e^{m_4 \sigma _3} \,,
\end{equation}
where $\sigma _3$ is the usual Pauli matrix.
Other choices of the monodromy are related to this by $\uspfour_\text{L} \times \uspfour_\text{R}$ conjugation.

The six-dimensional supergravity fields fit into the following representations under the R-symmetry group (see e.g. \cite{Tanii:1984zk,Andrianopoli:2004xu}):
\begin{equation}
\label{sugra representation}
\begin{alignedat}{2}
\text{scalars}:& \qquad\; (\textbf{5},\textbf{5}) 
\,,\\[3pt]
\text{vectors}:& \qquad\; (\textbf{4},\textbf{4}) 
\,,\\[3pt]
\text{tensors}:& \;\:(\textbf{5},\textbf{1})+(\textbf{1},\textbf{5}) 
\,,\\[3pt]
\text{gravitini}:& \;\:(\textbf{4},\textbf{1})+(\textbf{1},\textbf{4})
 \,,\\[3pt]
\text{dilatini}:& \;\:(\textbf{5},\textbf{4})+(\textbf{4},\textbf{5})
 \,.
\end{alignedat}
\end{equation}
We have an equal number of self-dual and anti-self-dual 2-form tensor fields, and an equal number of fermions of positive and negative chirality. In terms of the R-symmetry representations above, the self-dual tensors $B^+_2$ transform in the $(\textbf{5},\textbf{1})$ and the anti-self-dual tensors $B^-_2$ transform in the $(\textbf{1},\textbf{5})$. The positive chiral gravitini $\psi^+_\mu$ and dilatini $\chi^+$ transform in the $(\textbf{4},\textbf{1})$ and $(\textbf{5},\textbf{4})$ respectively, and the negative chiral gravitini $\psi^-_\mu$ and dilatini $\chi^-$ transform in the $(\textbf{1},\textbf{4})$ and $(\textbf{4},\textbf{5})$.

These representations determine the charges $(e_1,e_2,e_3,e_4)$ of each field under $\text{U}(1)^4\subset  \uspfour_\text{L}\times \uspfour_\text{R}$.
A field with charges $(e_1,e_2,e_3,e_4)$ will then be an eigenvector of the mass matrix with eigenvalue $i \mu$ and will have $z$-dependence $e^{i \mu z/2\pi R}$ where
\begin{equation}
\label{muis}
\mu =   \sum _{i=1}^4  e_im_i \ .
\end{equation}
The resulting mass for the field will turn out to be $|\mu |/2\pi R$.

\subsection{Supersymmetry breaking and massless field content}
\label{sec:supersymmetrybreaking}

The R-symmetry representations \eqref{sugra representation} decompose into the following representations under the  $\sutwo ^4$ subgroup \eqref{subgroupRsymm}:
\begin{equation}
\label{representation branching}
\begin{alignedat}{2}
\text{scalars}:& \qquad\; (\textbf{5},\textbf{5}) \;&&\rightarrow\; (\textbf{2},\textbf{2},\textbf{2},\textbf{2})+(\textbf{2},\textbf{2},\textbf{1},\textbf{1})+(\textbf{1},\textbf{1},\textbf{2},\textbf{2})+(\textbf{1},\textbf{1},\textbf{1},\textbf{1})\,,\\[3pt]
\text{vectors}:& \qquad\; (\textbf{4},\textbf{4}) \;&&\rightarrow\; (\textbf{2},\textbf{1},\textbf{2},\textbf{1})+(\textbf{2},\textbf{1},\textbf{1},\textbf{2})+(\textbf{1},\textbf{2},\textbf{2},\textbf{1})+(\textbf{1},\textbf{2},\textbf{1},\textbf{2})\,,\\[3pt]
\text{tensors}:& \;\:(\textbf{5},\textbf{1})+(\textbf{1},\textbf{5}) \;&&\rightarrow\; (\textbf{2},\textbf{2},\textbf{1},\textbf{1})+(\textbf{1},\textbf{1},\textbf{2},\textbf{2})+2\,(\textbf{1},\textbf{1},\textbf{1},\textbf{1})\,,\\[3pt]
\text{gravitini}:& \;\:(\textbf{4},\textbf{1})+(\textbf{1},\textbf{4}) \;&&\rightarrow\; (\textbf{2},\textbf{1},\textbf{1},\textbf{1})+(\textbf{1},\textbf{2},\textbf{1},\textbf{1})+(\textbf{1},\textbf{1},\textbf{2},\textbf{1})+(\textbf{1},\textbf{1},\textbf{1},\textbf{2})\,,\\[3pt]
\text{dilatini}:& \;\:(\textbf{5},\textbf{4})+(\textbf{4},\textbf{5}) \;&&\rightarrow\; (\textbf{2},\textbf{2},\textbf{2},\textbf{1})+(\textbf{2},\textbf{2},\textbf{1},\textbf{2})+(\textbf{2},\textbf{1},\textbf{2},\textbf{2})+(\textbf{1},\textbf{2},\textbf{2},\textbf{2})\\[2pt]
& &&\phantom{\rightarrow}\;\;\;+(\textbf{2},\textbf{1},\textbf{1},\textbf{1})+(\textbf{1},\textbf{2},\textbf{1},\textbf{1})+(\textbf{1},\textbf{1},\textbf{2},\textbf{1})+(\textbf{1},\textbf{1},\textbf{1},\textbf{2})\,.
\end{alignedat}
\end{equation}
This then determines the four charges $e_i$ under the $\text{U}(1)^4$ subgroup: each doublet gives charges $\pm 1$ and each singlet gives charge $0$.
For example, the sixteen vector fields in the
\begin{equation}
(\textbf{4},\textbf{4}) \;\rightarrow\; (\textbf{2},\textbf{1},\textbf{2},\textbf{1})+(\textbf{2},\textbf{1},\textbf{1},\textbf{2})+(\textbf{1},\textbf{2},\textbf{2},\textbf{1})+(\textbf{1},\textbf{2},\textbf{1},\textbf{2})
\end{equation}
have charges
\begin{equation}
(e_1,e_2,e_3,e_4) = (\pm 1,{0},\pm 1,{0})+({\pm 1},{0},{0},{\pm 1})+({0},{\pm 1},{\pm 1},{0})+({0},{\pm 1},{0},{\pm 1}) \,.
\end{equation}
These charges then determine the masses through \eqref{muis}. The eight gravitini (symplectic Weyl spinors) in the $(\textbf{4},\textbf{1})+(\textbf{1},\textbf{4})$ representation of the R-symmetry group $\uspfour_\text{L} \times \uspfour_\text{R}$ decompose into four pairs, each of which has a different mass $|m_i| / 2\pi R$, with $i = 1,2,3,4$. 
The number $\mathcal{N}$ of unbroken supersymmetries is then given by the number of massless gravitini, which is $\mathcal{N}= 2r$
where $r$ is the number of parameters $m_i$ that are zero.
The different values of $r$ give
 rise to 5D supergravities with $\mathcal{N}=8, 6,4,2,0$ Minkowski vacua, corresponding to twisting in $4-r$ of the $\sutwo$ factors in
\eqref{subgroupRsymm}.

In general, all fields that are charged, with at least one of the $e_i \neq 0$ corresponding to an $m_i
\neq 0$, become massive in 5D. Below we give the massless field content of reductions with twists that preserve $\mathcal{N}=8,6,4,2,0$ supersymmetry in the Minkowski vacuum and check that they fit into the relevant supermultiplets of 5D supergravities \cite{Cremmer:1980gs}.

\begin{itemize}
\item
$\mathcal{N}=8$

We start with the untwisted case, $m_i=0$, where all fields remain massless. Apart from the 5D graviton, the spectrum contains 8 gravitini, 27 vectors, 48 dilatini and 42 scalars (all massless). As expected, these fields make up a single gravity multiplet of maximal 5D supergravity.
\item
$\mathcal{N}=6$

In order to end up with $\mathcal{N}=6$ supergravity, we  take only one of the four mass parameters to be non-zero so that we twist in only one of the four  $\sutwo$ subgroups. The massless spectrum from such a reduction contains a graviton, 6 gravitini, 15 vectors, 20 dilatini and 14 scalars. These fields form the gravity multiplet of the $\mathcal{N}=6$ theory.

\item
$\mathcal{N}=4$

We obtain $\mathcal{N}=4$ supergravity by twisting in two $\sutwo$ groups, with two mass parameters zero. This can be done in two qualitatively different ways: either with a chiral twist, say in $\sutwo_{\text{R}_1}$ and $\sutwo_{\text{R}_2}$ with $m_1=m_2=0$, or with a non-chiral twist, for example in  $\sutwo_{\text{L}_2}$ and $\sutwo_{\text{R}_2}$ with $m_1=m_3=0$. Both types of twists result in the same massless spectrum: the graviton, 4 gravitini, 7 vectors, 8 dilatini and 6 scalars, although as we shall see, they result in different massive spectra.

In the $\mathcal{N}=4$ theory, the gravity multiplet contains the graviton, 4 gravitini, 6 vectors, 4 dilatini and a single scalar field, and the vector multiplet contains 1 vector, 4 dilatini and 5 scalars \cite{Awada:1985ep}. We see that our massless spectrum consists of the gravity multiplet coupled to one vector multiplet.

\item
$\mathcal{N}=2$

We end up with minimal 5D supergravity by twisting in three of the four $\sutwo$ subgroups, with just one of the mass parameters zero. The massless field content after such a twist contains the graviton, 2 gravitini, 3 vectors, 4 dilatini and 2 scalars.

For $\mathcal{N}=2$ supersymmetry, the gravity multiplet contains the graviton, 2 gravitini, and 1 vector field, and the vector multiplet contains 1 vector, 2 dilatini and 1 scalar field. Thus, the field content that we find from this reduction forms a gravity multiplet coupled to two vector multiplets.

\item
$\mathcal{N}=0$

By twisting in all four $\sutwo$ groups, with all four mass parameters non-zero, we break all supersymmetry. The only fields that are not charged under such a twist are the graviton and the singlets 
which are completely uncharged, with all $e_i=0$.
As a result, the massless spectrum in 5D consists of the graviton, 3 vectors and 2 scalars. Note that all fermions become massive. 
\end{itemize}

\subsection{Massive field content}
\label{sec: massivespectra}

The charges $(e_1,e_2,e_3,e_4)$ following from \eqref{representation branching} determine the massive spectrum for the reduced theory in five dimensions. This spectrum is summarized in \autoref{SpectrumTable}. The spectrum of \autoref{SpectrumTable} has been previously derived from Scherk-Schwarz reduction in \cite{Andrianopoli:2004xu} and corresponds to a gauging of $\mathcal{N}=8$ five-dimensional supergravity.

\begin{table}[ht!]
\centering
\begin{tabular}{c|c|c}
\thinspace Fields \rule{0pt}{2.7ex}\thinspace & \thinspace Representation \thinspace & \thinspace  $|\mu|=$ Mass (multiplied by $2\pi R$)   \thinspace\\[1pt] \hline\hline
Scalars \rule{0pt}{3ex} & $(\textbf{5,5})$ & $\;\;\big|\!\pm m_1\pm m_2\pm m_3 \pm m_4 \big|\;\;$ \\[2pt]
\rule{0pt}{3ex} &  & $\big|\!\pm m_1\pm m_2\big|$ \\[2pt]
\rule{0pt}{3ex} &  & $\big|\!\pm m_3\pm m_4 \big|$ \\[2pt]
\rule{0pt}{3ex} &  & $0$ \\[3pt] \hline
Vectors\rule{0pt}{3ex} & $(\textbf{4},\textbf{4})$ & $\big|\!\pm m_{1,2}\pm m_{3,4}\big|$ \\[3pt] \hline
Tensors\rule{0pt}{3ex} & $(\textbf{5},\textbf{1})$ & $\big|\!\pm m_1\pm m_2\big|\,$, $0$ \\[2pt]
\rule{0pt}{3ex} & $(\textbf{1},\textbf{5})$ & $\big|\!\pm m_3\pm m_4\big|\,$, $0$ \\[3pt] \hline
\;Gravitini\;\rule{0pt}{3ex} & $(\textbf{4},\textbf{1})$ & $\big|\!\pm m_{1,2}\big|$ \\[2pt]
\rule{0pt}{3ex} & $(\textbf{1},\textbf{4})$ & $\big|\!\pm m_{3,4}\big|$ \\[3pt] \hline
Dilatini\rule{0pt}{3ex} & $(\textbf{5},\textbf{4})$ & $\big|\!\pm m_1\pm m_2\pm m_{3,4}\big|$ \\[2pt]
\rule{0pt}{3ex} &  & $\big|\!\pm m_{3,4}\big|$ \\[2pt]
\rule{0pt}{3ex} & $(\textbf{4},\textbf{5})$ & $\big|\!\pm m_{1,2} \pm m_3\pm m_4\big|$ \\[2pt]
\rule{0pt}{3ex} &  & $\big|\!\pm m_{1,2}\big|$ \\[3pt]
\end{tabular}
\captionsetup{width=.9\linewidth}
\caption{\textit{This table gives the value of $|\mu(m_i)|$ for the 5D fields coming from the different types of 6D fields. The mass of the field is then $|\mu (m_i)|/2\pi R$. The notation $m_{i,j}$ indicates that both $m_i$ and $m_j$ occur. There is no correlation between the $\pm$ signs and the $_{ij}$ indices, so that e.g. $(\pm \, m_1\pm m_2)$ denotes 4 different combinations of mass parameters, and $(\pm \, m_{1,2}\pm m_{3,4})$ denotes 16 different combinations.
For example, the 5 tensors in the $(\boldsymbol{5},\boldsymbol{1})$ representation consist of two with mass
$| m_1+ m_2|\,$, two with mass $|  m_1- m_2|\,$ and one with mass $0$.
 }}
\label{SpectrumTable}
\end{table}

We now give the supermultiplet structure of the massive spectra that follow from the various twists preserving different amounts of supersymmetry. All fields that acquire mass also become charged under the graviphoton $\mathcal{A}_1^5$ with covariant derivatives of the form
\begin{equation}
D_\mu \,=\, \partial_\mu - i q \, g \, \mathcal{A}_\mu^5 \,.
\end{equation}
Here the gauge coupling is $g=1/R$, and the charge $q$ of each 5D field is equal to $1/g = R$ times its mass. Because the massive fields are charged, the real fields that follow from the reduction have to combine into complex fields.
In the spectra that we give below, we list the number of complex fields (unless stated otherwise). Furthermore, when we give the mass of a field or collection of fields we only write down $|
\mu|$. In order to find the actual mass, this needs to be divided by $2\pi R$.

The massive multiplets we find are all BPS multiplets in five dimensions; these multiplets were analyzed and classified in  \cite{Hull:2000cf}
and are labeled by two integers $(p,q)$. For $ \mathcal{N}$ supersymmetries in five dimensions (with $\mathcal{N}$ even), the R-symmetry is $\text{USp}(\mathcal{N})$.
For a $(p,q)$ massive multiplet, the choice of central charge breaks the R-symmetry to a subgroup $\text{USp}(2p)\times \text{USp}(2q)$, i.e. the subgroup of $\text{USp}(\mathcal{N})$ preserving the central charge, where $2p+2q=\mathcal{N}$. The nomenclature was chosen such that a 
massless supermultiplet of $(p,q)$ supersymmetry in six-dimensions has, after reducing on a circle, Kaluza-Klein modes that fit into $(p,q)$ massive supermultiplets in five dimensions. 
The physical states of a $(p,q)$ massive multiplet in five dimensions then fit into representations of
\begin{equation}
\sutwo\times \sutwo\times 
\text{USp}(2p)\times \text{USp}(2q) \ ,
\end{equation}
where
$\sutwo\times \sutwo\sim \text{SO}(4)$ is the little group for massive representations in five dimensions.
The representations of
the little group $\sutwo\times \sutwo$ that arise include
$(3,2)$ and $(2,3)$ for massive gravitini and $(2,2)$ for massive vector fields.
The representation $(3,1)$ corresponds to a massive self-dual two-form field satisfying the five-dimensional duality condition
\begin{equation}\label{tensorproperty}
\diff B_2 = -i m \ast B_2\ ,
\end{equation}
while the $(1,3)$ representation corresponds to the anti-self dual case with $\diff B_2=im\ast B_2$.
In the following, we consider the cases in which the Scherk-Schwarz reduction breaks the supersymmetry to $\mathcal{N}=6,4,2$. The massless states are in the $\mathcal{N}$  supersymmetry representations given in the previous subsection, and we now give the $\mathcal{N}$  supersymmetry representations of the massive fields.
It was already pointed out in \autoref{sec:supersymmetrybreaking} that there are two qualitatively different twists that result in a theory with $\mathcal{N}=4$ supersymmetry: a chiral one and a non-chiral one. Both theories have the same massless spectrum (see \autoref{sec:supersymmetrybreaking}), but their massive spectra are different. 
The non-chiral twist gives massive fields fitting into $(1,1)$ multiplets and we will refer to this as the $(1,1)$ theory. The chiral twist leads to $(0,2)$ supermultiplets and we will refer to this as the $(0,2)$ (or $(2,0)$) theory.

\begin{itemize}
\item
$\mathcal{N}=6$

In order to break to $\mathcal{N}=6$, we twist with just  one of the four mass parameters non-zero. Without loss of generality, we take $m_1\neq0$ and  the other three parameters  equal to zero. The physical states will then fall into representations of
\begin{equation}
\sutwo\times \sutwo\times \usptwo \times \uspfour \,.
\end{equation}
The massive field content from such a twist contains 1 gravitino, 2 self-dual tensors, 4 vectors, 13 dilatini and 10 scalars. All these fields are complex, and their mass is equal to $|{m_1}|$. This is a $(1,2)$ BPS supermultiplet with the representations
\begin{equation}
(3,2;1,1)+(3,1;2,1)+(2,2;1,4)+(1,2;1,5)+(2,1;2,4)+(1,1;2,5) \,.
\end{equation}

\item
$\mathcal{N}=4$ $(0,2)$

We obtain the $(0,2)$ theory by taking chiral twist with $m_1,m_2 \neq 0$ and $m_3,m_4 = 0$. The physical states will then fall in representations of
\begin{equation}
\sutwo\times \sutwo\times 
\uspfour \ .
\end{equation}
From the reduction we find two massive $(0,2)$ spin-$\tfrac{3}{2}$ multiplets,  one with mass $|m_1|$, and the other with mass $|m_2|$. Each consists of 1 gravitino, 4 vectors and 5 dilatini, which are in the representations
\begin{equation}
(3,2;1)+(2,2;4)+(1,2;5) \ .
\end{equation}
 Furthermore, we find two massive $(0,2)$ tensor multiplets with masses $|m_1+m_2|$ and $|m_1-m_2|$. Each of these contains one self-dual 2-form satisfying \eqref{tensorproperty}, 4 dilatini and 5 scalars \cite{Hull:2000cf}, fitting in the representations
 \begin{equation}
(3,1;1)+(2,1;4)+(1,1;5) \ .
\end{equation}
We note at this point that a part of the massive spectrum above can be made massless by tuning the mass parameters. That is, if we choose $m_1=\pm \, m_2$, one of the two (complex) tensor multiplets becomes massless. This gives two additional real vector multiplets in the massless sector of the $\mathcal{N}=4$ theory (see \autoref{sec:supersymmetrybreaking}).

\item
$\mathcal{N}=4$ $(1,1)$

For the non-chiral twist, we choose $m_1,m_3 \neq 0$ and $m_2,m_4 = 0$ in order obtain the $(1,1)$ theory. 
There are two massive $(1,1)$ vector multiplets, one with mass $|m_1+m_3|$ and one with mass $|m_1-m_3|$.
Each consists of 1 vector, 4 dilatini and 4 scalars \cite{Hull:2000cf} corresponding to a representation of 
\begin{equation}
\sutwo\times \sutwo\times 
\usptwo\times \usptwo \,,
\end{equation}
given by
\begin{equation}
(2,2;1,1)+(2,1;2,1)+(1,2;1,2)+(1,1;2,2) \,.
\end{equation}
In addition, there are two massive $(1,1)$ spin-$\tfrac{3}{2}$ multiplets, one with
mass $|m_1|$, and one with mass $|m_3|$. Each consists of 1 gravitino, 2 (anti-)self-dual tensors, 2 vectors, 5 dilatini and 2 scalars. The one with mass $|m_1|$ is in the representation
\begin{equation}
(3,2;1,1)+(3,1;2,1) +(2,2;1,2)+
(1,2;1,1)+(2,1;2,2)+(1,1;2,1) \,,
\end{equation}
and the one with mass $|m_3|$ is in the representation
\begin{equation}
(2,3;1,1)+(1,3;1,2) +(2,2;2,1)+
(2,1;1,1)+(1,2;2,2)+(1,1;1,2) \,.
\end{equation}
As in the $(0,2)$ theory, we can tune the mass parameters in such a way that a part of this spectrum becomes massless. For $m_1=\pm \, m_3$, one of the massive vector multiplets becomes massless, and so we get two more real vector multiplets in the massless sector of the theory (again see \autoref{sec:supersymmetrybreaking}). Note that, even though the massive tensor multiplet of the $(0,2)$ theory and the massive vector multiplet of the $(1,1)$ theory contain different fields, they give the same field content in the massless limit.

\item
$\mathcal{N}=2$

We choose $m_1,m_2,m_3 \neq 0$ and $m_4=0$ to obtain the $\mathcal{N}=2$ case with massive $(0,1)$ multiplets in representations of
\begin{equation}\label{N=2group}
\sutwo\times \sutwo\times \usptwo \ .
\end{equation}
There are
four massive hypermultiplets   with masses $|m_1 \pm m_2 \pm m_3|$ consisting of 1 complex dilatino and 2 complex scalars in the 
\begin{equation}\label{N=2dilatino2scalar}
(2,1;1)+ (1,1;2)
\end{equation}
representation.
The four vector multiplets with masses $|m_{1,2} \pm m_3|$ consist of 1 vector and 2 dilatini in the 
\begin{equation}
(2,2;1)+ (1,2;2)
\end{equation}
representation.
Furthermore, we find two tensor multiplets (1 self-dual tensor, 2 dilatini, 1 scalar) with masses $|m_1 \pm m_2|$ in the following representation of \eqref{N=2group}:
\begin{equation}
(3,1;1)+ (2,1;2) +(1,1;1) \ .
\end{equation}
There are also two spin-$\tfrac{3}{2}$ multiplets, one with mass $|m_1|$ and one with mass $|m_2|$, containing 1 gravitino, 2 vectors and 1 dilatino in the
\begin{equation}
(3,2;1)+ (2,2;2) +(1,2;1)
\end{equation}
representation. We also find another multiplet containing a spin-$\tfrac{3}{2}$ field: 1 gravitino, 2 anti-self-dual tensors, 1 dilatino and 2 scalars with mass $|m_3|$. This is reducible, giving one massive hypermultiplet consisting of 1 dilatino and 2 scalars with the representation \eqref{N=2dilatino2scalar} and one multiplet consisting of 1 gravitino and 2 anti-self-dual tensors in the representation:
\begin{equation}
(2,3;1)+ (1,3;2) \ .
\end{equation}
As for the $\mathcal{N}=4$ theories, we can  tune the mass parameters in order to obtain extra massless fields. Choosing $m_1=\pm \, m_2$ or $m_{1,2}=\pm \, m_3$ would make either a tensor multiplet or a vector multiplet massless. Both of these would give two real massless vector multiplets. Another choice would be to set $m_1=\pm \, m_2 \pm m_3$ so that one of the massive hypermultiplets becomes massless. 

\end{itemize}

\subsection{Mass matrices }
\label{sec: massmat}

The monodromies $\mathcal{M}_\text{L}^{\mathfrak{usp}(4)} \in \uspfour_\text{L}$ and $\mathcal{M}_\text{R}^{\mathfrak{usp}(4)} \in \uspfour_\text{R}$ in \eqref{monocon3} are the exponentials of mass matrices in the Lie algebra of $\uspfour$:
\begin{equation}
\mathcal{M}_\text{L}^{\mathfrak{usp}(4)} 
=\exp ({{M}_\text{L}^{\mathfrak{usp}(4)} })
\,, \qquad
\mathcal{M}_\text{R}^{\mathfrak{usp}(4)} =\exp ({{M}_\text{R}^{\mathfrak{usp}(4)} )
} \,.
\end{equation}
For the monodromies \eqref{monoex}, the mass matrices are given by
\begin{equation}
\label{monoexal}
 {M}_\text{L}^{\mathfrak{usp}(4)} = {m_1 \sigma _3} \oplus {m_2 \sigma _3} \,, \qquad
 {M}_\text{R}^{\mathfrak{usp}(4)} = {m_3 \sigma _3} \oplus {m_4 \sigma _3} \,.
\end{equation}
By conjugating, as in \eqref{monocon2}, by an element $h$ of the $\sutwo^4$ subgroup \eqref{subgroupRsymm}, we can bring this to the form
\begin{equation}
\label{monoexalp}
 {M}_\text{L}^{\mathfrak{usp}(4)} = 
 {m_1 (n_1 \cdot \sigma )}
  \oplus {m_2 (n_2 \cdot \sigma )} \,, \qquad
 {M}_\text{R}^{\mathfrak{usp}(4)} = {m_3 (n_3 \cdot \sigma )} \oplus
  {m_4(n_4 \cdot \sigma )} \,,
 \end{equation}
for any four unit 3-vectors $n_i$. Here $\sigma$ is the 3-vector of Pauli matrices.

The Lie algebra of $\uspfour$ consists of anti-hermitian $4\times 4 $ matrices $M_A{}^B$ ($M^{\dagger} = - M $)
such that $M^{AB} = \Omega^{AC} M_C{}^B$ is symmetric ($M^{AB}=M^{BA}$), where $\Omega^{AB}=-\Omega^{BA}$ is the symplectic invariant; see Appendix \ref{appendix: isomorphism} for more details.
In a basis in which
$\Omega= \sigma _2 \oplus \sigma _2$ and the subgroup \eqref{subgroupRsymm} is block diagonal,
we have the $4\times 4 $ matrix representation
\begin{equation}
{M}_\text{L}^{\mathfrak{usp}(4)} = 
 \begin{pmatrix}
{m_1 (n_1 \cdot \sigma )} &0\\
0 & {m_2 (n_2 \cdot \sigma )}
\end{pmatrix}
\,, \qquad
\Omega^{AB} = \begin{pmatrix}
\sigma _2 &0\\
0 & \sigma _2
\end{pmatrix} \,.
\end{equation} 
However, for our purposes, it will be useful to have mass matrices in a basis in which
\begin{equation}
\Omega^{AB} = \begin{pmatrix}
0 & \;\mathbbm{1}_{2} \\
- \mathbbm{1}_{2} & \;0
\end{pmatrix} \ .
\end{equation} 
In this basis, we can take for example
\begin{eqnarray}
\label{massusp}
M_\text{L}^{\mathfrak{usp}(4)}&=&\begin{pmatrix}
0 & 0 & -m_1 & 0 \\
0 & 0 & 0 & -m_2 \\
\,m_1\, & 0 & 0 & 0 \\
0 & \,m_2\, & 0 & 0
\end{pmatrix}\,,
\end{eqnarray}
and a similar expression for $M_\text{R}^{\mathfrak{usp}(4)}$ that can be found by replacing $m_1 \rightarrow m_3$ and $m_2 \rightarrow m_4$. The monodromy for the above mass matrix is given by
\begin{equation}
\mathcal{M}_\text{L}^{\mathfrak{usp}(4)}=\begin{pmatrix}
\cos(m_1) & 0 & -\sin(m_1) & 0 \\
0 & \cos(m_2) & 0 & -\sin(m_2) \\
\sin(m_1)\, & 0 & \cos(m_1) & 0 \\
0 & \sin(m_2)\, & 0 & \cos(m_2)
\end{pmatrix}\,,
\end{equation}
and there is a similar expression for $\mathcal{M}_\text{R}^{\mathfrak{usp}(4)}$. We can use the isomorphism $\mathfrak{usp}(4)\cong\mathfrak{so}(5)$ to map \eqref{massusp} to the corresponding generator in the Lie algebra of $\sofive$. This yields
\begin{equation}
\label{leftmassmatrixd1d5}
M_\text{L}=\begin{pmatrix}
0 & -(m_1+m_2) & 0 & \:0\: & 0 \\
m_1+m_2 & 0 & 0 & \:0\: & 0 \\
0 & 0 & 0 & \:0\: & -(m_1-m_2) \\
0 & 0 & 0 & \:0\: & 0 \\
0 & 0 & m_1-m_2 & \:0\: & 0
\end{pmatrix} \,,
\end{equation}
and a similar expression for $M_\text{R}$ where we replace $m_1 \rightarrow m_3$ and $m_2 \rightarrow m_4$ (see Appendix \ref{appendix: isomorphism} for more information on the isomorphism $\mathfrak{usp}(4)\cong\mathfrak{so}(5)$). The corresponding $\sofive$ monodromy is given by
\begin{equation}
\mathcal{M}_\text{L}=\begin{pmatrix}
\cos (m_1+m_2) & -\sin(m_1+m_2) & 0 & \:0\: & 0 \\
\sin(m_1+m_2) & \cos(m_1+m_2) & 0 & \:0\: & 0 \\
0 & 0 & \cos(m_1-m_2) & \:0\: & -\sin(m_1-m_2) \\
0 & 0 & 0 & \:1\: & 0 \\
0 & 0 & \sin(m_1-m_2) & \:0\: & \cos(m_1-m_2)
\end{pmatrix} \,.
\end{equation}
The $\uspfour$ monodromy is of course a double cover of the $\sofive$ monodromy: taking e.g. $m_1=m_2=\pi$ gives $\mathcal{M}_\text{L}=\mathbbm{1}$ but $\mathcal{M}_\text{L}^{\mathfrak{usp}(4)}=-\mathbbm{1}$.

We can use the mass matrices $M_\text{L}$ and ${M}_\text{R}$ in the algebras of $\sofive_\text{L}$ and $\sofive_\text{R}$ to create an $\mathfrak{so}(5,5)$ mass matrix. In the basis in which the $\text{SO}(5,5)$ metric takes the form
\begin{equation}
\label{taus}
\tau_{AB} = \begin{pmatrix}
0 & \;\mathbbm{1}_5   \\
\mathbbm{1}_5  & \;0
\end{pmatrix} \,,
\end{equation}
(see Appendix \ref{sec:groupappendix}) this $\mathfrak{so}(5,5)$ mass matrix is given by
\begin{equation}
\label{embedding monodromy}
\massmatr_A^{\;\;\,B}=\frac{1}{2}\begin{pmatrix}
(M_\text{L}+{M}_\text{R})_a^{\;\;\,b}\, & \,(M_\text{L}-{M}_\text{R})_{ab}\\[3pt]
(M_\text{L}-{M}_\text{R})^{ab}\, & \,(M_\text{L}+{M}_\text{R})^a_{\;\;\,b}
\end{pmatrix}
\;\in\; \mathfrak{so}(5,5) \,.
\end{equation}
It is this matrix that appears explicitly in the bosonic action, as we shall see in the following subsections.

In \autoref{sec:5dblackholes}, we consider various brane configurations that result in five-dimensional black holes. For each of these systems, we choose $M_\text{L}$ and ${M}_\text{R}$ in such a way that the fields that charge the black hole remain massless in 5D. All of these are conjugate to the ones given here. In particular, they all have the same eigenvalues and so give the same mass spectrum.

\subsection{5D scalars}
\label{sec:SSscalars}

In this section, we go through the reduction of the 6D scalar fields in detail. The goal is to compute the mass that each of the 25 scalar fields obtains in 5D. For notational convenience, we set $R=\tfrac{1}{2\pi}$ here and in the next subsection where we reduce the 6D tensors. Consequently, the masses that we compute here carry an `invisible' factor $\tfrac{1}{2\pi R}$ that can be reinstated by checking   the mass dimensions.

The scalar Lagrangian in six dimensions reads (see \autoref{sec:scalars})
\begin{equation}
\label{lagrscalars6d}
e^{-1}_{(6)}\,\lagr_{\text{s}}=\frac{1}{8}\,\text{Tr}\big[\partial_{\hat{\mu}}\mathcal{H}^{-1}\partial^{\hat{\mu}}\mathcal{H}\big]\,.
\end{equation}
The global $\sofivefive$ transformations   act as $\mathcal{H} \;\rightarrow\; U \, \mathcal{H} \, U^{T}$ with $U\in\,\sofivefive$. This leads us to the following   Scherk-Schwarz ansatz:
\begin{equation}
\label{ssansatzscalars}
\mathcal{H}(\hat{x}^{\hat{\mu}})=e^{\massmatr z}\,\mathcal{H}(x^\mu)\,e^{\massmatr^Tz}\,,
\end{equation}
where $\massmatr$ is the mass matrix defined in \eqref{embedding monodromy}. By substituting this ansatz in \eqref{lagrscalars6d}, we find the five-dimensional Lagrangian  
\begin{equation}
\label{lagrscalars5d}
e^{-1}_{(5)}\,\lagr_{\text{s}}=\frac{1}{8}\,\text{Tr}\big[D_{{\mu}}\mathcal{H}^{-1}D^{{\mu}}\mathcal{H}\big]-V(\mathcal{H}) \,.
\end{equation}
Matter that is charged under the monodromy becomes charged under the $\uone$ symmetry corresponding to the graviphoton $\graviph_1^5$ in 5D. The covariant derivative on $\mathcal{H}$ is given by
\begin{equation}
D_{{\mu}}\mathcal{H} = \partial_\mu\mathcal{H} - \graviph_\mu^5\,\big(\massmatr\,\mathcal{H}+\mathcal{H}\,\massmatr^T\big) \,.
\end{equation}
The potential in \eqref{lagrscalars5d} is given by
\begin{equation}
\label{SS-potential}
V(\mathcal{H})=\frac{1}{4}\,e^{-\sqrt{8/3}\,\phi_5}\,\Tr\!\big[\massmatr^2+\massmatr^T\mathcal{H}^{-1}\massmatr\,\mathcal{H}\big]\,.
\end{equation}
For an R-symmetry twist, such potentials must be non-negative \cite{Dabholkar:2002sy}; consequently, a global minimum can be found by solving $V=0$. We find such a minimum by putting all 25 scalar fields to zero, so that $\mathcal{H}=\mathbbm{1}$. By realizing that our mass matrix is anti-symmetric, $\massmatr^T=-\,\massmatr$, we immediately see that this gives $V=0$.

We now compute the masses of the scalar fields in this minimum. We denote the collection of all 25 scalar fields by $\sigma^i$, with $i=1,\ldots,25$, and compute the mass matrix as\footnote{The kinetic term of the sigma model is diagonal at the minimum of the potential, i.e. it takes the form $-\frac{1}{2}g_{ij}(\sigma^k)\,\partial_\mu\sigma^i\partial^\mu\sigma^j$ with $g_{ij}(0)=\delta_{ij}$.}
\begin{equation}
m_{ij}=\left.\frac{\partial^2V}{\partial\sigma^i\partial\sigma^j}\right|_{\sigma^k=0}\,.
\end{equation}
We diagonalize this mass matrix as $m_{ij}=Q_i^{\;\:k}\, m^{\text{diag}}_{kl}\,Q^l_{\;\,j}$, where $m^{\text{diag}}$ is a diagonal matrix and $Q$ is a conjugation matrix built from an orthonormal basis of eigenvectors. In this way, we find the mass that corresponds to each of the redefined fields $\tilde{\sigma}^i=Q^i_{\;\,j}\sigma^j$.

We have computed these masses explicitly for the mass matrices that preserve the various 6D black string configurations that we consider in \autoref{sec:5dblackholes}. Tables are provided in \autoref{appendix: masses}.

\subsection{5D tensors}
\label{sec:SStensors}

In this section we work out the reduction of the six-dimensional tensor fields in detail, 
following \cite{Hull:2003kr,Ozer:2003}. Just like in the previous subsection, we set $R=\tfrac{1}{2\pi}$ and neglect the Kaluza-Klein towers for notational convenience.

The Lagrangian for the six-dimensional tensor fields reads
\begin{equation}
\lagr^{\text{(doubled)}}_{\text{t}}=-\frac{1}{4}\,\mathcal{H}^{AB}\,G^{(6)}_{3,A}\wedge\ast\,G^{(6)}_{3,B}\,.
\end{equation}
The ten three-form field strengths $G^{(6)}_{3,A}$ transform as in \eqref{transformations6dfields}, so we choose our Scherk-Schwarz ansatz to be
\begin{equation}
\label{6dtensorssansatz}
G^{(6)}_{3,A}(\hat{x}^{\hat{\mu}})=\big(e^{\massmatr z}\big)_A^{\;\;\;B}\Big(G^{(5)}_{3,B}(x^\mu) \,+\, G^{(5)}_{2,B}(x^\mu)\wedge\big(\diff z+\mathcal{A}_1^5\big)\Big)\,,
\end{equation}
where $G^{(5)}_{3,A}$ and $G^{(5)}_{2,A}$ are five-dimensional field strengths that are independent of the circle coordinate $z$. As usual for self-dual tensor fields, we don't compactify the Lagrangian of the theory but rather its field equations. We start by reducing the six-dimensional Bianchi identities $\diff G^{(6)}_{3,A}=0$. We find
\begin{equation}
\begin{aligned}
\diff G^{(5)}_{3,A} + \diff\big( G^{(5)}_{2,A} \wedge \mathcal{A}_1^5 \big) &= 0 \,,\\[4pt]
\diff G^{(5)}_{2,A} - \massmatr_A^{\;\;\,B}\big(G^{(5)}_{3,B} + G^{(5)}_{2,B} \wedge \mathcal{A}_1^5\big) &= 0 \,.
\end{aligned}
\end{equation}
From these we deduce expressions for the five-dimensional field strengths in terms of the corresponding two-form and one-form potentials:
\begin{equation}
\label{fieldstrengths6d}
\begin{aligned}
G^{(5)}_{3,A} &= \diff A^{(5)}_{2,A} - G^{(5)}_{2,A} \wedge \mathcal{A}_1^5 \,, \\[4pt]
G^{(5)}_{2,A} &= \diff A^{(5)}_{1,A} + \massmatr_A^{\;\;\,B}\,A^{(5)}_{2,B}\,.
\end{aligned}
\end{equation}
Normally at this point, we would like to shift $A^{(5)}_{2,A} \,\rightarrow\, A^{(5)}_{2,A} - (\massmatr^{-1})_A^{\;\;\,B} \, \diff A^{(5)}_{1,B}\,$ so that the field strengths in \eqref{fieldstrengths6d} would lose their dependence on $A^{(5)}_{1,A}$. This is not possible, however, because our mass matrix $\massmatr_A^{\;\;\,B}$ is not invertible. We therefore need to diagonalize $\massmatr_A^{\;\;\,B}$ and split the indices that correspond to zero and non-zero eigenvalues. In the most general case where the combinations $m_1\pm m_2$ and $m_3\pm m_4$ are non-zero, this splitting goes like $A\rightarrow(\alpha,\dot{\alpha})$ with $\dot{\alpha}\in\{i,i+5\}$, where $i$ is the index that corresponds to the row and column that we set to zero in $M_\text{L}$ and $M_\text{R}$. For example, for the reduction of the D1-D5 system (see \eqref{mass matrices SO(5)}) we have $\dot{\alpha}\in\{4,9\}$. The index $\alpha$ takes the other eight values of the original index $A$. The second equation in \eqref{fieldstrengths6d} now separates into
\begin{equation}
\begin{aligned}
G^{(5)}_{2,\alpha} &= \diff A^{(5)}_{1,\alpha} + \massmatr_\alpha^{\;\;\,\beta}\,A^{(5)}_{2,\beta} \,, \\[4pt]
G^{(5)}_{2,\dot{\alpha}} &= \diff A^{(5)}_{1,\dot{\alpha}}\,.
\end{aligned}
\end{equation}
The matrix $\massmatr_\alpha^{\;\;\,\beta}$ is invertible, so now we can shift $A^{(5)}_{2,\alpha} \,\rightarrow\, A^{(5)}_{2,\alpha} - (\massmatr^{-1})_\alpha^{\;\;\,\beta} \, \diff A^{(5)}_{1,\beta}$. After this shift, the five-dimensional field strengths read
\begin{equation}
\label{fieldstrenghts6dshifted}
\begin{aligned}
G^{(5)}_{3,\alpha} &= \diff A^{(5)}_{2,\alpha} - G^{(5)}_{2,\alpha} \wedge \mathcal{A}_1^5 \,, \qquad\quad & G^{(5)}_{3,\dot{\alpha}} &= \diff A^{(5)}_{2,\dot{\alpha}} - G^{(5)}_{2,\dot{\alpha}} \wedge \mathcal{A}_1^5 \,, \\[4pt]
G^{(5)}_{2,\alpha} &= \massmatr_\alpha^{\;\;\,\beta}\,A^{(5)}_{2,\beta} \,, & G^{(5)}_{2,\dot{\alpha}} &= \diff A^{(5)}_{1,\dot{\alpha}} \,.
\end{aligned}
\end{equation}
The six-dimensional field strengths are subject to the self-duality constraint
\begin{equation}
\label{6dselfduality}
G^{(6)}_{3,A}=\tau_{AB}\,\mathcal{H}^{BC}\ast G^{(6)}_{3,C} \,.
\end{equation}
We now compactify this constraint. First, we need to reduce the six-dimensional Hodge star to five dimensions. By using the metric decomposition \eqref{KK-ansatz general}, we find
\begin{equation}
\begin{aligned}
\ast^{(6)} G^{(6)}_{3,A} &\,=\, \big(e^{\massmatr z}\big)_A^{\;\;\;B} \ast^{(6)}\! \big(G^{(5)}_{3,B} \,+\, G^{(5)}_{2,B}\wedge\big(\diff z+\mathcal{A}_1^5\big)\big) \\[4pt]
&\,=\, \big(e^{\massmatr z}\big)_A^{\;\;\;B} \,\big(e^{\sqrt{2/3}\,\phi_5}\,\ast^{(5)}\! G^{(5)}_{3,B}\wedge\big(\diff z+\mathcal{A}_1^5\big) \,-\, e^{-\sqrt{2/3}\,\phi_5}\,\ast^{(5)}\!G^{(5)}_{2,B}\big) \,.
\end{aligned}
\end{equation}
This result allows us to write down the 5D self-duality constraint that follows from \eqref{6dselfduality} as
\begin{equation}
\label{5dselfduality}
G^{(5)}_{3,A}=-\,e^{-\sqrt{2/3}\,\phi_5} \, \tau_{AB}\,\mathcal{H}^{BC} \ast G^{(5)}_{2,C} \,.
\end{equation}
Recall for the derivation of this result that $\mathcal{H}^{AB}$ with raised indices is the inverse of the matrix $\mathcal{H}$ as defined in \autoref{sec:scalars}. Consequently, we use the inverse of \eqref{ssansatzscalars} as Scherk-Schwarz ansatz.

\paragraph{Mass spectrum.}

In order to find the mass spectrum of the fields that descend from $G^{(6)}_{3,A}$, we put all other fields in \eqref{5dselfduality} to zero. In particular, this means that $\mathcal{H}^{AB}=\delta^{AB}$. We find
\begin{equation}
\label{selfduality5d}
\diff A^{(5)}_{2,\alpha}=-\,\tau_\alpha^{\;\;\,\beta} \, \massmatr_\beta^{\;\;\,\gamma} \ast A^{(5)}_{2,\gamma} \,, \qquad\quad \diff A^{(5)}_{2,\dot{\alpha}}=-\,\tau_{\dot{\alpha}}^{\;\;\,\dot{\beta}} \ast \diff A^{(5)}_{1,\dot{\beta}} \,,
\end{equation}
where we use the notation $\tau_\alpha^{\;\;\,\beta}=\tau_{\alpha\gamma}\,\delta^{\gamma\beta}$ and an analogous expression for the dotted indices. These are massive and massless five-dimensional self-duality conditions. From these, we can deduce the equations of motion for the corresponding fields (following \cite{Townsend:1983xs}). They read
\begin{equation}
\diff\big(\!\ast\diff A^{(5)}_{2,\alpha}\,\big)=-\,\big(\tau \massmatr \tau \massmatr\big)_\alpha^{\;\;\,\beta}\ast A^{(5)}_{2,\beta} \,, \qquad\quad \diff\big(\!\ast\diff A^{(5)}_{1,\dot{\alpha}}\,\big)=0 \,.
\end{equation}
So in 5D, we end up with eight massive tensors and two massless vectors (again, this is for the case where $m_1\pm m_2$ and $m_3\pm m_4$ are non-zero). The self-duality constraint \eqref{5dselfduality} eliminates the massless tensors $A^{(5)}_{2,\dot{\alpha}}$ and makes sure that the massive tensors $A^{(5)}_{2,\alpha}$ carry only half their usual degrees of freedom. The masses of the fields $A^{(5)}_{2,\alpha}$ are determined by the mass matrix $-\,\big(\tau \massmatr \tau \massmatr\big)_\alpha^{\;\;\,\beta}$. By diagonalizing this matrix, we find the mass corresponding to each field.

Just as for the scalar fields, we have computed these masses explicitly for the mass matrices that we use for the reduction of the D1-D5 system and the dual brane configurations in \autoref{sec:5dblackholes}. These masses can be found in \autoref{appendix: masses}.

\paragraph{Graviphoton interactions.}

We now pay some extra attention to the interactions between the graviphoton and the vector and tensor fields that we find in this subsection. They will prove to be very important in \autoref{sec:radiativecorrections}. As it turns out, there is a difference in the result that we find for the reduction of a self-dual 6D tensor and an anti-self-dual 6D tensor. We illustrate this difference with two simple examples.

Consider a six-dimensional (anti-)self-dual tensor field $\hat{B}_2$ with field strength $\hat{H}_3=\hat{\diff} \hat{B}_2$ (here hats denote 6D quantities). The field equations and self-duality constraint for this field read
\begin{equation}
\hat{\diff} \hat{H}_3 = 0 \,, \qquad\quad \hat{\ast} \, \hat{H}_3 = \pm \, \hat{H}_3 \,.
\end{equation}
By decomposing this field (strength) as $\hat{H}_3 = H_3 + H_2 \wedge (\diff z + \graviph_1)$, and by using straightforward reduction techniques and the conventions of this paper, we find the following 5D Lagrangian:
\begin{equation}
\lagr = - \, \tfrac{1}{2} \, H_2 \wedge \ast \, H_2 \, \pm \, \tfrac{1}{2} \, \graviph_1 \wedge H_2 \wedge H_2 \,,
\end{equation}
with $H_2 = \diff B_1$. We see that a self-dual and an anti-self-dual tensor give a Chern-Simons interaction term with the graviphoton with an opposite sign.

Now take a real doublet of (anti-)self-dual tensor fields, that we Scherk-Schwarz reduce from 6D to 5D with the ansatz
\begin{equation}
\hat{H}_3 = \exp{\left[
{\left(\begin{matrix} 0 & -m\\m & 0 \end{matrix}\right)\,{z}} 
\right]}
\, \big( H_3 + H_2 \wedge (\diff z + \graviph_1) \big),
\end{equation}
(apart from the ansatz and the fact that we are considering a doublet this set-up is similar to the previous one). Going through this reduction gives a complex massive tensor $B_2$ in five dimensions subject to the self-duality equation
\begin{equation}
\label{selfdualityequation5Dgeneral}
\diff B_2 \,-\, i m \, \graviph_1 \wedge B_2 \,\pm\, i m \ast B_2 = 0 \,.
\end{equation}
Again, the $\pm$ sign indicates the difference between the result for the reduction of a self-dual and an anti-self-dual tensor from six dimensions. Now, this sign is not in front of the interaction with the graviphoton, but we can still flip it by redefining $\graviph_1 \rightarrow - \graviph_1$. To see this, recall that the field $B_2$ is complex so that we also have the complex conjugate of \eqref{selfdualityequation5Dgeneral}. By flipping the sign of the graviphoton, we effectively switch the particle and the anti-particle $B_2 \leftrightarrow \bar{B}_2$ in order to protect the sign in the covariant derivative. The $\pm$ sign in the mass term of the equation for $\bar{B}_2$ is flipped with respect to \eqref{selfdualityequation5Dgeneral}, and so we see that redefining $\graviph_1 \rightarrow - \graviph_1$ effectively interchanges the result for a self-dual and an anti-self-dual tensor.

\subsection{Conjugate monodromies}
\label{Conjugate Monodromies}

We have so far considered monodromies in the R-symmetry group $\spinfive_\text{L}\times \spinfive_\text{R} $ preserving the identity in the coset $\sofivefive/\spinfive_\text{L}\times \spinfive_\text{R} $, which is  the point in the moduli space at which all scalar fields vanish.
Then this point  in  moduli space is a fixed point under the action of the $\sofivefive$ transformation  $\psi \to {\cal M}\psi$ given by the monodromy, and  as we have seen this point is  a  minimum of the Scherk-Schwarz potential giving a Minkowski vacuum.
Conjugating by an element of the R-symmetry group
\begin{equation}
\label{monto}
{\mathcal{M}}\to h \bar {\mathcal{M} }h^{-1} \,, \qquad h\in \uspfour_\text{L}\times \uspfour_\text{R} 
\end{equation}
will then preserve the fixed point in the moduli space and the minimum will remain at the origin.

However, for the embedding in string theory (see \autoref{sec:quantization mass parameters}), we will need to consider monodromies that are related to an R-symmetry transformation by conjugation by an element of $\sofivefive$
\begin{equation}
\label{monoconqq}
\mathcal{M}=g \tilde {\mathcal{M} }g^{-1} \,, \qquad g\in \sofivefive \,, \qquad  \tilde{ \mathcal{M}} \in \uspfour_\text{L}\times \uspfour_\text{R} \subset \sofivefive \,.
\end{equation}
This change of monodromy can be thought of as the result of acting on the theory twisted with monodromy $\tilde {\mathcal{M} }$ by a transformation
$\psi \to g \psi$. For the supergravity theory,  this is just a  field redefinition giving an equivalent theory, but as we shall see later this has consequences for the embedding in string theory.
The fixed point is now at the coset containing $g$, $[g]= \{gh \:|\: h\in \spinfive_\text{L}\times \spinfive_\text{R} \}$, and this is now the location of the  minimum of the potential \cite{Dabholkar:2002sy}.
At this point, the kinetic terms of the various fields are not conventionally normalized. On bringing these to standard form, the masses become precisely the ones given earlier for the theory with monodromy
$\tilde {\mathcal{M} }$. This was of course to be expected: a field redefinition cannot change physical parameters such as masses.

\subsection{Gauged supergravity and gauge group}
\label{sec:GaugedSupergravityandGaugeGroup}

The result of the Scherk-Schwarz reduction is a gauged $\mathcal{N}=8$ supergravity theory in which a subgroup of the E$_6$ duality symmetry
of the ungauged 5D theory is promoted to a gauge symmetry. 
In this subsection we discuss this gauged supergravity and its gauge group. 

We start with the case in which the twist is a T-duality transformation in the T-duality subgroup $\spinfourfour$ of $\sofivefive$.
Consider first the bosonic NS-NS sector of the ten-dimensional supergravity theory, consisting of the metric, B-field and dilaton.
Compactifying on $T^4$ gives a 6D theory with $\sofourfour$ symmetry.
There is a 6D metric, B-field and dilaton, together with 8 vector fields $A_{\hat \mu}^A$ in the {\bf 8} of $\sofourfour$ (with $A=1,\dots, 8$ labelling the vector representation of $\sofourfour$)
and scalars in the coset space $\sofourfour/\sofour\times \sofour$.
The Scherk-Schwarz compactification of this on a circle with an $\sofourfour$ twist with mass matrix $N_A{}^B$ was given in detail in \cite{Hull:2007jy}.
In 5D, there are then 10 gauge fields: eight 
 $A_{ \mu}^A$ arising from the 6D vector fields, the graviphoton vector field  $\mathcal{A}^5_\mu$ from the metric and a vector field
 $\mathcal{B}^5_\mu$ from the reduction of the 6D B-field.
 Then $(A_{ \mu}^A, \mathcal{A}^5_\mu, \mathcal{B}^5_\mu)$
  are the gauge fields for a gauge group with 10 generators $T_A, T_z, T_{\tilde z}$ respectively.
  After the field redefinitions given in \cite{Hull:2007jy}  to obtain tensorial fields transforming covariantly under duality transformations, the gauge algebra is \cite{Hull:2007jy} 
  \begin{equation}
[T_z, T_A]= N_A{}^B \, T_B \,, \qquad [T_A, T_B]= N_{AB} \, T_{\tilde z} \,,
\end{equation}
with all other commutators vanishing. Here $N_{AB}=N_A{}^C\eta _{CB}$ where $\eta _{AB}$  is the $\sofourfour $-invariant metric, so that $N_{AB}=-
N_{BA}$ as the mass matrix is in the Lie algebra of  $\sofourfour $.
This then represents a gauging of a 10-dimensional subgroup of   $\sofourfour $, which has a $\uone^2$ subgroup generated by $T_z, T_{\tilde z}$.
A further $\uone$ factor can be obtained by dualising the 2-form $b_{\mu \nu}$ to give an extra gauge field and the generator $t$ of this $\uone$ factor commutes with all other generators.

Next, consider reintroducing the R-R sector. In six dimensions, there are a further 8 one-form gauge fields
$C_{\hat{\mu}}^\alpha$ transforming as a Weyl spinor of $\spinfourfour$  ($\alpha=1,\dots ,8$), which combine with the 8 NS-NS one-form gauge fields to form the {\bf 16} of $\sofivefive$.
There are also a further 4 two-form gauge fields, which split into four self-dual ones and four anti-self dual ones that transform as an {\bf 8} of $\sofourfour $.
These combine with the degrees of freedom of the NS-NS 2-form to form the {\bf 10} of $\sofivefive$.
The mass matrix acts on the spinor representation through 
$N_{\alpha}{}^ \beta$ which is given as usual by $N_{\alpha \beta}=\frac 1 4 N_{AB}( \gamma ^{AB})_{\alpha \beta}$
where  $N_{\alpha \beta}=N_{\alpha}{}^ \gamma \eta _{\gamma\beta}$ and $\eta _{\alpha\beta}$ is the symmetric charge conjugation matrix.
The structure in the spinor representation is related to that in the vector representation by $\sofourfour$ triality.
The gauge algebra then gains the terms
  \begin{equation}
[T_z, T_\alpha]= N_\alpha{}^\beta \, T_\beta \,, \qquad [T_\alpha, T_\beta]= N_{\alpha \beta} \, T_{\tilde z} \,,
\end{equation}
to give an 18-dimensional gauge group. This corresponds to gauging an 18-dimensional subgroup of E$_6$.
For generic values of the  parameters $m_i$, the two-form gauge fields in the {\bf 8} of $\sofourfour $ become massive, while the 5D NS-NS two-form remains massless and can again be dualized to give a further $\uone$ factor with generator  $t$. For special values of the parameters, some of the two-forms in the {\bf 8} of $\sofourfour$ can become invariant under the twist and so become massless as well. These can be dualized to give further $\uone$ factors.

The gauge algebra can now be written
\begin{equation}
[T_z, T_a]= M_a{}^b \, T_b \,, \qquad [T_a, T_b]= M_{ab} \, T_{\tilde z} \,,
\end{equation}
with all other commutators vanishing, where $T_a= (T_A,T_\alpha)$ and
\begin{equation}
M_a{}^b=
 \begin{pmatrix}
{N_A{}^B} &0\\
0 & {N_{\alpha}{}^{\beta}}
\end{pmatrix} \,.
\end{equation}
There is a $\uone^3$ subgroup generated by $t$ (if the NS-NS two-form is dualized) with possible further $\uone$ factors coming in if some of the R-R two-forms remain massless.

In the generic case in which $M_a{}^b$ has no zero eigenvalues, then the vector fields $A^a$ corresponding to the generators $T_a$ all become massive, while the gauge fields corresponding to the generators $T_z,T_{\tilde z},t$ remain massless. Then the gauge group is spontaneously broken to the $U(1)^3$ subgroup generated by $T_z,T_{\tilde z},t$. For special values of the parameters $m_i$ such that $M_a{}^b$ has some zero eigenvalues, there will be more massless gauge fields and the unbroken gauge group will be larger.

In \autoref{sec:dualbraneconfigurations}, we will consider a twist of this kind in the compact $\spinfour\times \spinfour$ subgroup of the $\spinfourfour$  T-duality group. The other twists we will consider are all related to this one by conjugation (see \autoref{Conjugate Monodromies} and \ref{sec:dualbraneconfigurations}) and will give isomorphic gauge groups.

One can argue what part of the matter content is charged under each of these generators of the gauge group by Scherk-Schwarz reducing the 6D gauge transformations and seeing how the 5D fields transform under these reduced transformations. The 6D gauge transformations can be found in \cite{Tanii:1984zk,Bergshoeff:2007ef}. The generators $T_a$ come from the gauge transformations corresponding to the 16 vector fields in 6D. These transform the 6D vectors and the 6D tensors, so the 5D descendants of these fields can become charged under generators $T_a$. The generators $T_{\tilde z}$ and $t$ come from the gauge transformation that correspond to the 6D tensor field that is a singlet under the twist. This transformation acts only on this tensor field, so after reduction no matter becomes charged under the resulting 5D transformations $T_{\tilde z}$ and $t$. The generator $T_z$ (corresponding to the graviphoton $\mathcal{A}_1^5$) comes from 6D diffeomorphisms in the circle direction. By explicit reduction of these diffeomorphisms, we find that all fields that become massive in 5D carry $\uone$ charge under~$T_z$.

\subsection{Kaluza-Klein towers}
\label{sec:kaluzakleintowers}

In the previous subsections, we have constructed 5D theories with both massless and massive fields 
from Scherk-Schwarz reduction. However, this is not the whole story: if we consider a compactification on $S^1$, then each field picks up an infinite Kaluza-Klein tower\footnote{Of course, even this is not the whole story. There are also Kaluza-Klein modes that come from the reduction from 10D to 6D on the four-torus, plus stringy degrees of freedom. We will return to these in \autoref{sec:embeddingstringtheory}, where we discuss the full string theory.}. We choose Scherk-Schwarz ans\"{a}tze including Kaluza-Klein towers on the $S^1$ of the form
\begin{equation}
\psi(x^{\mu},z) = \exp\left({\frac{{{M}} z}{2\pi R}}\right)\, \sum_{n\in\mathbb{Z}}\,e^{inz/R} \, \psi_n(x^{\mu}) \,,
\end{equation}
where we use $\psi$ as a schematic notation for any field in the theory that transforms in some representation of the R-symmetry group. 
Then if $\psi $ has charges $e_i$, it is an eigenvector of ${{M}}$ with eigenvalue $i \mu$ given by
\eqref{muis}, ${{M}}\psi=i\mu\,\psi$, so that
\begin{equation}
\label{KKm}
\psi(x^{\mu},z) =  \sum_{n\in\mathbb{Z}}\,\exp {\left( i \left(
\frac {\mu}{2\pi}+n
\right)\frac{z}{R} \right) }
 \, \psi_n(x^{\mu}) \, .
\end{equation}
Clearly, shifting $\tfrac{\mu}{2\pi}$ by an integer $r$ can be absorbed into a shift $n\to n-r$ and so corresponds to changing the $n$'th Kaluza-Klein mode to the $(n-r)$'th one while leaving the sum unchanged.
From \autoref{SpectrumTable}, we see that shifting the $m_i$ by $2\pi r_i$ for any integers $r_i$ shifts all the $\tfrac{\mu}{2\pi}$ by an integer and so leaves the above sum \eqref{KKm} unchanged. For this reason, there is no loss of generality in taking $m_i\in [0,2\pi)$.

Without loss of generality, we can restrict the $m_i$'s further by realizing that all eigenvalues $i\mu$ appear with a $\pm$ sign in front of them. By taking into account two towers of the form \eqref{KKm}, one of them with a minus sign in front of $\mu$, we see that the \textit{combination} of these towers is unchanged under $\tfrac{\mu}{2\pi} \rightarrow 1 - \tfrac{\mu}{2\pi}$. Consequently, we can take $m_i\in [0,\pi]$ without loss of generality.

The Scherk-Schwarz ansatz is a truncation of \eqref{KKm} to the $n=0$ mode. This gives a consistent truncation to a gauged five-dimensional supergravity theory, which is sufficient for e.g.  determining which twists preserve which brane configuration in \autoref{sec:5dblackholes}. The full string theory requires keeping all these modes, together with stringy modes and degrees of freedom from branes wrapping the internal space.

The mass of the $n$'th KK-mode is given by
\begin{equation}
\label{massesKKtower}
\left|  \frac{\mu}{2\pi R}  + \frac {n}{R} \right| \,, \qquad\quad n\in\mathbb{Z} \,,
\end{equation}
and the value of $\mu(m_i)$ for each field can be read off from \autoref{SpectrumTable}.
 As an example, we check this for the reduction of the 6D tensors. If the whole tower is taken into account, the Scherk-Schwarz ansatz \eqref{6dtensorssansatz} is extended to
\begin{equation}
\label{6dtensorsansatzincludingKK}
G^{(6)}_{3}(\hat{x}^{\hat{\mu}})=\exp\left(\frac{\massmatr z}{2\pi R}\right)\,\sum_{n\in\mathbb{Z}}\:e^{inz/ R}\,\Big(G^{(5)}_{3,n}(x^\mu) \,+\, G^{(5)}_{2,n}(x^\mu)\wedge\big(\diff z+\mathcal{A}_1^5\big)\Big)\,.
\end{equation}
Note that we have restored the circle radius $R$ in this ansatz; from now on, we will keep it manifest in all our equations. Furthermore, in \eqref{6dtensorsansatzincludingKK} the $\sofivefive$ indices are suppressed for clarity. It can be seen directly that this extended ansatz essentially changes the mass matrix $\mathbbm{M}$ as we used it in \autoref{sec:SStensors} to
\begin{equation}
\left(
 \frac{\mathbbm{M}} {2\pi R} +  \frac{in\mathbbm{1}}{R}
 \right) \,, \qquad\quad n\in\mathbb{Z} \,.
\end{equation}
We can now use  that the eigenvalue of $ {\mathbbm{M}} $ is $i\mu$ with  $\mu$ given by \eqref{muis} to see that the masses of the Kaluza-Klein modes are given by \eqref{massesKKtower}.

Note that the modes with $n=0$ that are kept in the Scherk-Schwarz reduction are not necessarily the lightest modes in the tower. In particular, if the parameters $m_i$ are chosen so that $\tfrac{\mu(m_i)}{2\pi}$ is an integer, $\tfrac{\mu(m_i)}{2\pi} = N$, then the mode with $n=-N$ will be massless. As an example of this, we can choose
\begin{equation}
m_{1}=m_{2}=\tfrac{\pi}{2} \,, \; m_3 = \pi \,, \; m_4=0 \,.
\end{equation}
By using \autoref{SpectrumTable}, we can see which additional massless fields arise. In this case, there are four scalars and two spin-$\tfrac{1}{2}$ fermions that become massless, which form a hypermultiplet of $\mathcal{N}=2$ supergravity. For further discussion of such accidental massless modes, see \cite{Dabholkar:2002sy,Hull:2017llx}.

\section{Five-dimensional black hole solutions}
\label{sec:5dblackholes}

In this section, we consider several 10D brane configurations that we compactify to give  black holes in 5D. 
We do this in two steps. First, we reduce the brane configuration to a black string solution of $(2,2)$ supergravity in six dimensions. This solution will not be invariant under the whole $\sofivefive$ duality group, but will be preserved by a stabilizing subgroup. If we then do a standard (untwisted) compactification of  this on a circle with the black string wrapped along the circle, we obtain a BPS black hole solution of ${\cal N}=8$ supergravity in five dimensions. This reduction can be modified by including a duality twist on the circle. If the duality twist is in the stabilizing subgroup, the same black hole solution will remain a solution of the gauged supergravity resulting from the Scherk-Schwarz reduction, and of its truncation to an effective ${\cal N}<8$ supergravity describing the massless sector. This is because the only fields that become massive as a result of the Scherk-Schwarz twist are the ones that are trivial (zero) in the black hole solution. As a consequence, the black hole will also be BPS and preserving (at least) four supercharges. Indeed, it descends from a BPS black string solution in six dimensions, and the duality twist preserves the supercharges and Killing spinors of the truncated theory that has the black hole as a solution.

Primarily, we focus on the D1-D5 system, but later in this section we also consider the dual F1-NS5 and D3-D3 systems.

\subsection{The D1-D5-P system}
\label{sec:d1/d5}

The D1-D5 system, sometimes more accurately called the D1-D5-P system, consists of D1-branes, D5-branes and waves carrying momentum. The ten-dimensional configuration is as follows:
\renewcommand{\arraystretch}{1.1}
\begin{table}[ht!]
\centering
	\begin{tabular}{c|c|cccc|c|cccc}
		\multicolumn{1}{c}{} & \multicolumn{5}{c}{$\mathbb{R}^{1,4}$} & \multicolumn{1}{c}{$S^1$} & \multicolumn{4}{c}{$T^4$} \\[-8pt]
		\multicolumn{1}{c}{} & \multicolumn{5}{c}{\downbracefill} & \multicolumn{1}{c}{\!\downbracefill\!} & \multicolumn{4}{c}{\downbracefill} \\
		& $t$ & $r$ & $\theta$ & $\varphi_1$ & $\varphi_2$ & $z$ & $y_1$ & $y_2$ & $y_3$ & $y_4$ \\ \hline
		\multicolumn{1}{c|}{$\,\,\,\text{D1}^{\phantom{\hat{1}}}$} & $-$ & $\cdot$ & $\cdot$ & $\cdot$ & $\cdot$ & $-$ & $\cdots$ & $\cdots$ & $\cdots$ & $\cdots$ \\ \hline
		\multicolumn{1}{c|}{$\,\,\,\text{D5}^{\phantom{\hat{1}}}$} & $-$ & $\cdot$ & $\cdot$ & $\cdot$ & $\cdot$ & $-$ & $-$ & $-$ & $-$ & $-$ \\ \hline
		\multicolumn{1}{c|}{$\,\,\,\text{P}^{\phantom{\hat{1}}}$} & $-$ & $\cdot$ & $\cdot$ & $\cdot$ & $\cdot$ & $-$ & $\cdots$ & $\cdots$ & $\cdots$ & $\cdots$
	\end{tabular} 
\end{table}
\renewcommand{\arraystretch}{1}

Here a line ($-$) denotes an extended direction, a dot ($\,\cdot\,$) denotes a pointlike direction, and multiple dots ($\cdots$) denote a direction in which the brane or wave is smeared.

We start from the ten-dimensional solution and reduce it to 5D with the ans\"atze that are given in previous sections. The D1-D5 solution of type IIB supergravity in Einstein frame reads
\begin{equation}
	\label{10dsolution}
	\begin{cases}
		\begin{aligned}
			\diff s^2_{(10)}=\;&H_1^{-\frac{3}{4}}H_5^{-\frac{1}{4}}\big[\!-\diff t^2+\diff z^2+K(\diff t-\diff z)^2\big]+H_1^{\frac{1}{4}}H_5^{\frac{3}{4}}\big[\diff r^2+r^2\diff\Omega_3^2\big]\\
			&+H_1^{\frac{1}{4}}H_5^{-\frac{1}{4}}\big[\diff y_1^2+\diff y_2^2+\diff y_3^2+\diff y_4^2\big]\\[4pt]
			e^{\Phi}=\;&H_1^{\frac{1}{2}}H_5^{-\frac{1}{2}}\\[4pt]
			C^{(10)}_2=\;&(H_1^{-1}-1)\,\diff t\wedge\diff z+Q_5\cos^2\!\theta\:\diff\varphi_1\wedge\diff\varphi_2\,,
		\end{aligned}
	\end{cases}
\end{equation}
where $\diff\Omega_3^2=\diff\theta^2+\sin^2\!\theta\:\diff\varphi_1^2+\cos^2\!\theta\:\diff\varphi_2^2$ is the metric on the three-sphere written in Hopf coordinates. The harmonic functions corresponding to the D1-branes, the D5-branes and the momentum modes can be written in terms of their total charges as
\begin{equation}
	H_1=1+\frac{Q_1}{r^2}\,,\qquad\quad H_5=1+\frac{Q_5}{r^2}\,,\qquad\quad H_K=1+K=1+\frac{Q_K}{r^2}\,.
\end{equation}

\paragraph{Reduction to six dimensions.}

We compactify the metric in \eqref{10dsolution} to 6D using the ansatz \eqref{compansatz}. The metric on the torus, $g_{mn}$, is diagonal in \eqref{10dsolution} so we find that
\begin{equation}
e^{\vec{b}_m\cdot\vec{\phi}}=H_1^{\frac{1}{4}}H_5^{-\frac{1}{4}} \,, \qquad m=1,\ldots,4 \,.
\end{equation}
By using the expressions for the vectors $\vec{b}_m$ given in \eqref{dilatonvectors}, we can solve for the individual scalar fields $\phi_i$. We find that only one of them is non-zero in the 6D solution:
\begin{equation}
e^{\phi_4}=H_1^{\frac{1}{2}}H_5^{-\frac{1}{2}}\,, \qquad\quad \phi_1=\phi_2=\phi_3=0\,.
\end{equation}
The rest of the reduction is straightforward. The six-dimensional Einstein frame metric is related to the ten-dimensional one by a Weyl rescaling with $g_4^{1/4}=H_1^{\frac{1}{4}}H_5^{-\frac{1}{4}}$, which is incorporated in the ansatz \eqref{compansatz}. The dilaton $\Phi$ and the R-R two-form $C_2^{(10)}$ have no non-zero components on the torus, so they reduce trivially. The result reads
\begin{equation}
	\label{6dsolution}
	\begin{cases}
		\begin{aligned}
			\diff{s}^2_{(6)}=\;&H_1^{-\frac{1}{2}}H_5^{-\frac{1}{2}}\big[\!-\diff t^2+\diff z^2+K(\diff t-\diff z)^2\big]+H_1^{\frac{1}{2}}H_5^{\frac{1}{2}}\big[\diff r^2+r^2\diff\Omega_3^2\big]\\[4pt]
			e^{\phi_+}=\;&H_1^{\sqrt{\frac{1}{2}}}H_5^{-\sqrt{\frac{1}{2}}}\\[4pt]
			C^{(6)}_2=\;&(H_1^{-1}-1)\,\diff t\wedge\diff z+Q_5\cos^2\!\theta\:\diff\varphi_1\wedge\diff\varphi_2\,.
		\end{aligned}
	\end{cases}
\end{equation}
Here, we have defined the scalar field $\phi_+=\frac{1}{\sqrt{2}}(\phi_4+\Phi)$. This solution describes a black string in six dimensions.

When we use the doubled formalism (see \autoref{tensors}) we can rewrite the solution above in terms of the doubled tensor fields. To derive the contributions of these doubled fields to the black string solution, recall that the dual field strengths are defined as $\tilde{G}^{(6)a}_{3}=K^{ab}\ast G^{(6)}_{3,b}+L^{ab}\,G^{(6)}_{3,b}$. By putting all scalar fields except   $\phi_+$ to zero, this reduces to $\tilde{G}^{(6)a}_{3}=K^{ab}\ast G^{(6)}_{3,b}$ with $K^{ab}=\text{diag}\,(1,1,1,e^{\sqrt{2}\,\phi_+},1)$. Hence, we find that  the doubled tensors to which  the black string solution couples are given by
\begin{equation}
		\begin{aligned}
			C^{(6)}_2=\;&(H_1^{-1}-1)\,\diff t\wedge\diff z+Q_5\cos^2\!\theta\:\diff\varphi_1\wedge\diff\varphi_2\, , \\[2pt]
			\tilde{C}^{(6)}_2=\;&(H_5^{-1}-1)\,\diff t\wedge\diff z+Q_1\cos^2\!\theta\:\diff\varphi_1\wedge\diff\varphi_2\,.
		\end{aligned}
\end{equation}
In the doubled formalism, the degrees of freedom of both these fields are halved by the self-duality constraint \eqref{constraintdoubled} so the total number of degrees of freedom of the fields that the black string couples to remain unchanged.

\paragraph{Scherk-Schwarz reduction to five dimensions.}

The last step is to Scherk-Schwarz reduce the six-dimensional black string solution, which results in a black hole in five dimensions. We choose the twist matrices 
to be in the stabilizing subgroup of the R-symmetry group, i.e. the subgroup of the R-symmetry that preserves the solution.
As a result, all the fields that are non-constant in the 
 black hole remain massless.

For the D1-D5 system, we choose the following $\mathfrak{usp}(4)$ mass matrices:
\begin{equation}
\label{mass matrices USp(4)}
M_\text{L}^{\mathfrak{usp}(4)}=\begin{pmatrix}
0 & 0 & -m_1 & 0 \\
0 & 0 & 0 & -m_2 \\
\,m_1\, & 0 & 0 & 0 \\
0 & \,m_2\, & 0 & 0
\end{pmatrix}\,,
\qquad\quad
{M}_\text{R}^{\mathfrak{usp}(4)}=\begin{pmatrix}
0 & 0 & -m_3 & 0 \\
0 & 0 & 0 & -m_4 \\
\,m_3\, & 0 & 0 & 0 \\
0 & \,m_4\, & 0 & 0
\end{pmatrix}\,.
\end{equation}
Here $m_1$, $m_2$, $m_3$ and $m_4$ are real mass parameters, each corresponding to one $\sutwo$ in the R-symmetry subgroup \eqref{subgroupRsymm}. The isomorphism $\mathfrak{usp}(4)\cong \mathfrak{so}(5)$ of Appendix \ref{appendix: isomorphism}  maps these to $\mathfrak{so}(5)$ mass matrices. We find
\begin{equation}
\label{mass matrices SO(5)}
\begin{aligned}
M_\text{L}&=\begin{pmatrix}
0 & -(m_1+m_2) & 0 & \:0\: & 0 \\
m_1+m_2 & 0 & 0 & \:0\: & 0 \\
0 & 0 & 0 & \:0\: & -(m_1-m_2) \\
0 & 0 & 0 & \:0\: & 0 \\
0 & 0 & m_1-m_2 & \:0\: & 0
\end{pmatrix}\,,\\[8pt]
{M}_\text{R}&=\begin{pmatrix}
0 & -(m_3+m_4) & 0 & \:0\: & 0 \\
m_3+m_4 & 0 & 0 & \:0\: & 0 \\
0 & 0 & 0 & \:0\: & -(m_3-m_4) \\
0 & 0 & 0 & \:0\: & 0 \\
0 & 0 & m_3-m_4 & \:0\: & 0
\end{pmatrix}\,.
\end{aligned}
\end{equation}
The embedding of these $\mathfrak{so}(5)$ matrices in the $\mathfrak{so}(5,5)$ mass matrix $\massmatr_A^{\;\;\,B}$ is given in \eqref{embedding monodromy}.

We can now follow the techniques of \autoref{sec:SSscalars} and \ref{sec:SStensors} to determine the masses that each of the scalar and tensor fields acquires due to this twist. The results of these calculations for
the mass matrices \eqref{mass matrices SO(5)} are presented in Appendix \ref{app: masses d1d5}. In particular, we find that the fields that appear in the six-dimensional black string solution (${\phi_+}$, $C^{(6)}_2$ and $\tilde{C}^{(6)}_2$)  do not become massive in this Scherk-Schwarz reduction, as required. This means that the reduction of the solution \eqref{6dsolution} to a 5D black hole is the same as in the untwisted case.

The two self-dual tensors that charge the black string solution, $C^{(6)}_2$ and $\tilde{C}^{(6)}_2$, yield two tensors and two vector fields in 5D. We denote these by $C^{(5)}_2$, $C^{(5)}_1$, $\tilde{C}^{(5)}_2$, $\tilde{C}^{(5)}_1$. We now consider the self-duality conditions for these fields from \eqref{5dselfduality}, where we only take along fields that are non-zero in the 5D black hole solution. We find that they are pairwise dual by the relations
\begin{equation}
\diff C_1^{(5)} = \,e^{\sqrt{2/3}\,\phi_5} \, e^{-\sqrt{2}\,\phi_+} \ast \diff \tilde{C}_2^{(5)} \,, \qquad\quad \diff \tilde{C}_1^{(5)} = \,e^{\sqrt{2/3}\,\phi_5} \, e^{\sqrt{2}\,\phi_+} \ast \diff C_2^{(5)} \,.
\end{equation}
We use these to write the contributions of $C_2^{(5)}$ and $\tilde{C}_2^{(5)}$ to the black hole solution in terms of the dual one-forms. In doing so, we move to  an undoubled formalism. The full five-dimensional black hole solution is then given by
\begin{equation}
	\label{5dsolution}
	\begin{cases}
		\begin{aligned}
			\diff{s}^2_{(5)}=\;&-(H_1 H_5 H_K)^{-\frac{2}{3}}\,\diff t^2 + (H_1 H_5 H_K)^{\frac{1}{3}}\, \big[\diff r^2+r^2\diff\Omega_3^2\big]\\[3pt]
			e^{\phi_+}=\;&H_1^{\sqrt{\frac{1}{2}}}H_5^{-\sqrt{\frac{1}{2}}}\\[2pt]
			e^{\phi_5}=\;&H_1^{-\sqrt{\frac{1}{6}}} H_5^{-\sqrt{\frac{1}{6}}} H_K^{\,\sqrt{\frac{2}{3}}}\\[4pt]
			C^{(5)}_1=\;&(H_1^{-1}-1)\,\diff t\\[4pt]
			\tilde{C}^{(5)}_1=\;&(H_5^{-1}-1)\,\diff t\\[4pt]
			\mathcal{A}^5_1=\;&(H_K^{-1}-1)\,\diff t \,.
		\end{aligned}
	\end{cases}
\end{equation}
Here $C^{(5)}_1$ and $\tilde{C}^{(5)}_1$ are full vector fields, meaning that they are not subject to a self-duality constraint and carry the usual number of degrees of freedom. Note that this compactification can be generalized by adding angular momentum in directions transverse to the 10D branes to give a rotating black hole in five dimensions.

This three-charge black hole has been well studied in the literature. Its charges are quantized as $Q_i=c_i N_i$, where $N_i$ are integers and the the basic charges are given by \cite{Maldacena:1996ky}
\begin{equation}\label{chargecoefficients}
c_1 = \frac{4 G_N^{(5)}R}{\pi \alpha' g_s} \, , \qquad c_5 = \alpha' g_s \, , \qquad c_K = \frac{4 G_N^{(5)}}{\pi R} \,.
\end{equation}
The entropy of this black hole can be computed with the Bekenstein-Hawking formula, which yields
\begin{equation}\label{5dentropy}
S_\text{BH} \,=\, \frac{A}{4G_N^{(5)}} \,=\, \frac{\pi^2}{2G_N^{(5)}} \sqrt{Q_1Q_5Q_K} \,=\, 2 \pi \sqrt{N_1N_5N_K} \ .
\end{equation}

\subsection{Dual brane configurations}
\label{sec:dualbraneconfigurations}

\paragraph{The F1-NS5-P system.}

We now study the F1-NS5-P system, which consists of F1 and NS5-branes arranged as follows:\renewcommand{\arraystretch}{1.1}
\begin{table}[ht!]
\centering
	\begin{tabular}{c|c|cccc|c|cccc}
		& $t$ & $r$ & $\theta$ & $\varphi_1$ & $\varphi_2$ & $z$ & $y_1$ & $y_2$ & $y_3$ & $y_4$ \\ \hline
		\multicolumn{1}{c|}{$\,\,\,\text{F1}^{\phantom{\hat{1}}}$} & $-$ & $\cdot$ & $\cdot$ & $\cdot$ & $\cdot$ & $-$ & $\cdots$ & $\cdots$ & $\cdots$ & $\cdots$ \\ \hline
		\multicolumn{1}{c|}{$\,\,\,\text{NS5}^{\phantom{\hat{1}}}$} & $-$ & $\cdot$ & $\cdot$ & $\cdot$ & $\cdot$ & $-$ & $-$ & $-$ & $-$ & $-$
	\end{tabular} 
\end{table}
\renewcommand{\arraystretch}{1}

Again there are  waves with momentum in the $z$-direction. This system is related to the D1-D5 system via S-duality. As in the previous case, we start by considering the supergravity solution in ten dimensions. It can be written in Einstein frame as
\begin{equation}
	\label{10dsolutionNS5F1}
	\begin{cases}
		\begin{aligned}
			\diff s^2_{(10)}=\;&H_F^{-\frac{3}{4}}H_N^{-\frac{1}{4}}\big[\!-\diff t^2+\diff z^2+K(\diff t-\diff z)^2\big]+H_F^{\frac{1}{4}}H_N^{\frac{3}{4}}\big[\diff r^2+r^2\diff\Omega_3^2\big]\\
			&+H_F^{\frac{1}{4}}H_N^{-\frac{1}{4}}\big[\diff y_1^2+\diff y_2^2+\diff y_3^2+\diff y_4^2\big]\\[4pt]
			e^{\Phi}=\;&H_F^{-\frac{1}{2}}H_N^{\frac{1}{2}}\\[4pt]
			B^{(10)}_2=\;&(H_F^{-1}-1)\,\diff t\wedge\diff z+Q_N\cos^2\!\theta\:\diff\varphi_1\wedge\diff\varphi_2\,,
		\end{aligned}
	\end{cases}
\end{equation}
where we have the harmonic functions
\begin{equation}
	H_F=1+\frac{Q_F}{r^2}\,,\qquad\quad H_N=1+\frac{Q_N}{r^2}\,,
\end{equation}
and $H_K$ is as before. Note that this solution can be obtained from the D1-D5 solution \eqref{10dsolution} by  an S-duality transformation, which sends $\Phi \rightarrow -\,\Phi$ and $C_2^{(10)} \rightarrow B_2^{(10)}$. After reduction on $T^4$, we obtain a very similar six-dimensional solution, 
given by
 \eqref{6dsolution}  with  the replacements $\phi_+ \rightarrow \phi_-=\frac{1}{\sqrt{2}}(\phi_4-\Phi)$ and $C_2^{(6)} \rightarrow B_2^{(6)}$. In the doubled formalism, the
black string couples to the two-forms
\begin{equation}
		\begin{aligned}
			B^{(6)}_2=\;&(H_F^{-1}-1)\,\diff t\wedge\diff z+Q_N\cos^2\!\theta\:\diff\varphi_1\wedge\diff\varphi_2\, , \\[2pt]
			\tilde{B}^{(6)}_2=\;&(H_N^{-1}-1)\,\diff t\wedge\diff z+Q_F\cos^2\!\theta\:\diff\varphi_1\wedge\diff\varphi_2\,.
		\end{aligned}
\end{equation}
Again these fields carry only half their usual degrees of freedom due to the self-duality constraint \eqref{constraintdoubled}. 

To reduce to five dimensions, we need to specify   the mass matrices. 
We choose the Scherk-Schwarz twist to be in the stabilizing subgroup of the R-symmetry group. 
Since the F1-NS5 system couples to $B_2$ instead of $C_2$, the twist is chosen to preserve $B_2$. 
We choose the $\mathfrak{so}(5)$ matrices
\begin{equation}
\label{mass matrices SO(5) NS5F1}
\begin{aligned}
M_\text{L}&=\begin{pmatrix}
0 & -(m_1+m_2) & 0 & 0 & \:0\, \\
m_1+m_2 & 0 & 0 & 0 & \:0\, \\
0 & 0 & 0 & -(m_1-m_2) & \:0\, \\
0 & 0 & m_1-m_2 & 0 & \:0\, \\
0 & 0 & 0 & 0 & \:0\,
\end{pmatrix}\,,
\end{aligned}
\end{equation}
and ${M}_\text{R}$ similar with $m_1 \rightarrow m_3$ and $m_2 \rightarrow m_4$. By using the isomorphism in Appendix \ref{appendix: isomorphism}, these map to $\mathfrak{usp}(4)$ generators of the form
\begin{equation}
\label{mass matrices USp(4) NS5F1}
M_\text{L}^{\mathfrak{usp}(4)}=\begin{pmatrix}
0 & 0 & -\frac{m_1+m_2}{2} & \frac{m_1-m_2}{2} \\
0 & 0 & \frac{m_1-m_2}{2} & -\frac{m_1+m_2}{2}  \\
\frac{m_1+m_2}{2} & \frac{-m_1+m_2}{2} & 0 & 0 \\
\frac{-m_1+m_2}{2} & \frac{m_1+m_2}{2}  & 0 & 0
\end{pmatrix}\,.
\end{equation}
The masses of the scalar and tensor fields that follow from the reduction with these mass matrices are given in Appendix \ref{appendix:NS5F1masstables}.

The resulting five-dimensional black hole is given by \eqref{5dsolution} with the field redefinitions $\phi_+ \rightarrow \phi_-$, $C_2^{(5)} \rightarrow B_2^{(5)}$ and $\tilde{C}_2^{(5)} \rightarrow \tilde{B}_2^{(5)}$. It is not surprising that these black holes are related by field redefinitions. After all, the D1-D5 and F1-NS5 systems are related by U-duality, and the corresponding mass matrices are related by conjugation
\begin{eqnarray}\label{conjclassNS5}
\massmatr_\text{F1-NS5} = C \,  \massmatr_\text{D1-D5} \,  C^{-1} \,, \qquad C \in \sofivefive \,.
\end{eqnarray}
This conjugation matrix $C$ is given by
\begin{eqnarray}
 C=\begin{pmatrix}
 c\: & \:0 \\
 0\: & \:c 
 \end{pmatrix} \,, \qquad
c=\begin{pmatrix}
1 & \:0\: & 0 & \:0\: & 0 \\
0 & \:1\: & 0 & \:0\: & 0 \\
0 & \:0\: & 1 & \:0\: & 0 \\
0 & \:0\: & 0 & \:0\: & 1 \\
0 & \:0\: & 0 & \:1\: & 0
\end{pmatrix} \,.
\end{eqnarray}
Essentially this conjugation matrix interchanges the fourth and fifth row and column and the ninth and tenth row and column in the mass matrix (and monodromy).

\paragraph{The D3-D3-P systems.}

Finally, we consider the reduction of the D3-D3-P system of branes. We specify the brane configuration:
\renewcommand{\arraystretch}{1.1}
\begin{table}[ht!]
\centering
	\begin{tabular}{c|c|cccc|c|cccc}
		& $t$ & $r$ & $\theta$ & $\varphi_1$ & $\varphi_2$ & $z$ & $y_1$ & $y_2$ & $y_3$ & $y_4$ \\ \hline
		\multicolumn{1}{c|}{$\,\,\text{D3}^{\phantom{\hat{1}}}$} & $-$ & $\cdot$ & $\cdot$ & $\cdot$ & $\cdot$ & $-$ & $-$ & $-$ & $\cdots$ & $\cdots$ \\ \hline
		\multicolumn{1}{c|}{$\,\,\,\text{D3}'^{\phantom{\hat{1}}}$} & $-$ & $\cdot$ & $\cdot$ & $\cdot$ & $\cdot$ & $-$ & $\cdots$ & $\cdots$ & $-$ & $-$
	\end{tabular} 
\end{table}
\renewcommand{\arraystretch}{1}

As before, we also have momentum in the $z$-direction. We start with the supergravity solution in ten dimensions, in Einstein frame it can be written as
\begin{equation}
\label{10dsolutionD3}
\begin{cases}
\begin{aligned}
\diff s^2_{(10)}=\;&H_{3}^{-\frac{1}{2}}H_{3'}^{-\frac{1}{2}}\big[\!-\diff t^2+\diff z^2+K(\diff t-\diff z)^2\big]+H_{3}^{\frac{1}{2}}H_{3'}^{\frac{1}{2}}\big[\diff r^2+r^2\diff\Omega_3^2\big]\\
&+ H_{3}^{-\frac{1}{2}}H_{3'}^{\frac{1}{2}}\big[\diff y_1^2+\diff y_2^2\big] + H_{3}^{\frac{1}{2}}H_{3'}^{-\frac{1}{2}}\big[\diff y_3^2+\diff y_4^2\big]\\[4pt]
C^{(10)}_4=\;&(H_3^{-1}-1)\,\diff t\wedge\diff z\wedge\diff y_1\wedge\diff y_2 + (H_{3'}^{-1}-1)\,\diff t\wedge\diff z\wedge\diff y_3\wedge\diff y_4\,,
\end{aligned}
\end{cases}
\end{equation}
where the harmonic functions are given by
\begin{equation}
	H_3=1+\frac{Q_3}{r^2}\,,\qquad\quad H_{3'}=1+\frac{Q_{3'}}{r^2}\,.
\end{equation}
On compactifying to six dimensions on $T^4$ by taking the coordinates $y_1,\ldots ,y_4$ periodic, this brane configuration is related  to the D1-D5 system by T-duality. This means that the black string solution for the D3-D3 system can be obtained from that for the D1-D5 system \eqref{6dsolution} by a field redefinition. We find this field redefinition as $C_2^{(6)} \rightarrow R_{2;\,1}^{(6)}$ and $\phi_+ \rightarrow \phi_1$. In the doubled formalism, the six-dimensional black string arising from the D3-D3 system couples to the two-forms
\begin{equation}
\begin{aligned}
R^{(6)}_{2;\,1}=\;&(H_{3}^{-1}-1)\,\diff t\wedge\diff z+Q_{3'}\cos^2\!\theta\;\diff\varphi_1\wedge\diff\varphi_2 \,, \\[2pt]
\tilde{R}^{(6)}_{2;\,1}=\;&(H_{3'}^{-1}-1)\,\diff t\wedge\diff z+Q_{3}\cos^2\!\theta\;\diff\varphi_1\wedge\diff\varphi_2 \,.
\end{aligned}
\end{equation}
Different D3-D3 systems can be constructed by arranging the D3-branes differently on the torus. These would be charged under the  two-forms coming from the reduction of  $C^{(10)}_4$ in such systems. All of these systems are related by T-duality.

In the last step of the reduction we need to  ensure the fields that are non-trivial in the black hole  solution  remain massless in 5D. For this twisted reduction we choose $\mathfrak{so}(5)$ mass matrices of the form
\begin{equation}
\label{mass matrices SO(5) D3}
\begin{aligned}
M_\text{L}&=\begin{pmatrix}
\,0\: & 0 & 0 & 0 & 0 \\
\,0\: & 0 & 0 & -(m_1+m_2) & 0 \\
\,0\: & 0 & 0 & 0 & -(m_1-m_2) \\
\,0\: & m_1+m_2 & 0 & 0 & 0 \\
\,0\: & 0 & m_1-m_2 & 0 & 0
\end{pmatrix}\,,
\end{aligned}
\end{equation}
which results in $R_{2;\,1}^{(6)}$ remaining massless. In $\mathfrak{usp}(4)$ this mass matrix reads
\begin{equation}
\label{mass matrices USp(4) D3}
M_\text{L}^{\mathfrak{usp}(4)}=\begin{pmatrix}
0 & 0 & -\frac{m_1-m_2}{2} & \frac{i(m_1+m_2)}{2} \\
0 & 0 & \frac{i(m_1+m_2)}{2} & \frac{m_1-m_2}{2}  \\
\frac{m_1-m_2}{2}  & \frac{i(m_1+m_2)}{2} & 0 & 0 \\
 \frac{i(m_1+m_2)}{2}& -\frac{m_1-m_2}{2}  & 0 & 0
\end{pmatrix}\,.
\end{equation}
The scalar and tensor masses that follow from the reduction with these mass matrices are given in Appendix \ref{app: masses d3d3}. The resulting five-dimensional black hole is 
given by making the field redefinitions $\phi_+ \rightarrow \phi_1$, $C_2^{(5)} \rightarrow R_{2;\,1}^{(5)}$ and $\tilde{C}_2^{(5)} \rightarrow \tilde{R}_{2;\,1}^{(5)}$ in the solution \eqref{5dsolution}. The mass matrices are again conjugate to those of the dual D1-D5 and  F1-NS5 solutions. The relation is similar to the F1-NS5 result in \eqref{conjclassNS5}, except now the matrix $C$ switches the first and fourth rows and columns instead of the fourth and fifth ones.


\subsection{Preserving further black holes by tuning mass parameters}
\label{sec:preservingfurtherBHs}

In the previous subsection, we chose twist matrices with four arbitrary real parameters $m_i$. For each black hole solution (D1-D5, F1-NS5, D3-D3), we chose this matrix in such a way that the fields that source the black hole are left unchanged by the Scherk-Schwarz twist. Consequently, the black hole remains a valid solution of the 5D theory for all values of the mass parameters.

Here, we treat the special cases in which the mass parameters can be tuned in such a way that, in addition to the original black hole, other   black hole solutions are also preserved by the same twist. 
For example, we consider twists that preserve both the D1-D5 and F1-NS5 black holes.
As it turns out, this can only be done in the $\mathcal{N}=4$ $(0,2)$ theory and in the $\mathcal{N}=0$ theory. Since we are interested mostly in partial supersymmetry breaking, we treat an example of the $\mathcal{N}=4$ $(0,2)$ case in detail below.

\paragraph{Preserving D1-D5 with T-duality twist in \texorpdfstring{$\boldsymbol{\mathcal{N}=4$ $(0,2)}$}{N=4: (0,2)}.}

For this example, we consider mass matrices of the form given in \eqref{mass matrices SO(5) NS5F1} that preserve the F1-NS5 black hole solution. In order to twist to the $\mathcal{N}=4$ $(0,2)$ theory, we choose $m_1,m_2\neq0$ and $m_3,m_4=0$.

Suppose that, in addition to the F1-NS5 solution, we also want to preserve the D1-D5 solution with this twist. Then the fields
\begin{equation}
\label{fieldschargingD1D5}
\big\{ \phi_+=\tfrac{1}{\sqrt{2}}(\phi_4+\Phi)\,, \: C_2^{(5)}, \: \tilde{C}_2^{(5)} \big\}
\end{equation}
have to remain massless as well. The masses of these fields for this  twist matrix can be found in Appendix \ref{appendix:NS5F1masstables}. By setting $m_3,m_4=0$, we see that each field either becomes massive with mass $|m_1-m_2|$ or remains massless. It is therefore straightforward to tune the mass parameters in such a way that all of these fields remain massless by taking $m_1=m_2$.

We thus see that the D1-D5 solution can be preserved in a reduction to $\mathcal{N}=4$ $(0,2)$ with the twist matrix that was originally proposed to preserve the F1-NS5 solution, simply by taking the two mass parameters to be equal. This particular example offers some interesting possibilities. On the one hand, we note that the twist that preserves the F1-NS5 solution lies in the perturbative SO$(4,4)$ subgroup of the duality group. From the perspective of the full string theory this is a T-duality twist. Since T-duality is a perturbative symmetry, we can in principle work out the corresponding orbifold compactification of the perturbative string theory explicitly. On the other hand, the microscopic description of the D1-D5 black hole, the D1-D5 CFT, is understood reasonably well. Therefore, it should be possible to study this particular reduction thoroughly both from the perspective of the full string theory, and from the perspective of the black hole microscopics. We will return to this elsewhere.

\paragraph{Other possibilities in \texorpdfstring{$\boldsymbol{\mathcal{N}=4}$ $\boldsymbol{(0,2)}$}{N=4: (0,2)}.}

By taking $m_1=m_2$ in the example above, we managed to keep the fields that couple to the D1-D5 black hole massless, and so we could preserve this particular solution. It turns out, however, that this choice kept more fields massless than just the ones that charge the D1-D5 solution. In particular, the fields
\begin{equation}
\big\{\phi_3\,,\, R_{2;  3}^{(5)} \,,\, \tilde{R}_{2; 3}^{(5)}\big\}
\end{equation}
also remain massless (as can be checked from the tables in Appendix \ref{appendix:NS5F1masstables}). These are exactly the fields that are non-trivial in one of the three possible D3-D3 black holes. So not only the D1-D5 and F1-NS5 black holes, but also one of the D3-D3 black holes is preserved in this reduction.

Suppose now that, instead of $m_1=m_2$, we choose $m_1=-\,m_2$ in this reduction to $\mathcal{N}=4$ $(0,2)$. This choice does not preserve the D1-D5 solution and the D3-D3 solution charged under $R_{2;  3}^{(5)}$, but instead other solutions are preserved. Now the fields coupling to the two other D3-D3 black holes remain massless:
\begin{equation}
\big\{\phi_1\,,\, R_{2;  1}^{(5)} \,,\, \tilde{R}_{2; 1}^{(5)}\big\} \quad \text{and} \quad \big\{\phi_2\,,\, R_{2;  2}^{(5)} \,,\, \tilde{R}_{2; 2}^{(5)}\big\} \,.
\end{equation}
These are all the possibilities for preserving multiple black hole solutions with a T-duality twist to the $\mathcal{N}=4$ $(0,2)$ theory. The same game can be played, however, with the other twist matrices given in \autoref{sec:d1/d5} and \ref{sec:dualbraneconfigurations}. For each case, we find that taking either $m_1=m_2$ or $m_1=-\,m_2$ in the reduction to $\mathcal{N}=4$ $(0,2)$ results in the preservation of two additional black hole solutions.

\paragraph{Preserving further black holes in \texorpdfstring{$\boldsymbol{\mathcal{N}=0}$}{N=0}.}

The only other theory in which we can preserve several black hole solutions by tuning mass parameters is the one in which we break all supersymmetry: the $\mathcal{N}=0$ case. Now all four mass parameters are non-zero. As an example, let's consider the geometric F1-NS5 twist \eqref{mass matrices SO(5) NS5F1} again. If we take $m_1=m_2$ and $m_3=m_4$ in this reduction, all fields \eqref{fieldschargingD1D5} that are non-trivial in the D1-D5 black hole solution remain massless. Consequently, the D1-D5 solution is preserved.  Other examples can be worked out for similar reductions to $\mathcal{N}=0$.

\section{Quantum corrections}
\label{sec:radiativecorrections}

So far, we have considered five-dimensional supergravity theories with both massless and massive fields. For the purpose of finding black hole solutions in these theories, we truncated (consistently) to the $n=0$ modes of the Kaluza-Klein towers and identified black hole solutions in the massless sector after this truncation.

Under certain conditions, which we discuss in the next section, the black hole solutions we have been considering lift to solutions of the full string theory.
In the string theory, 
the effective supergravity theory receives quantum corrections. In particular, there are quantum corrections to the coefficients of the 5D Chern-Simons terms which in turn lead to modifications of the black hole solutions and hence to quantum corrections to their entropy.

 In this section, we consider corrections to the coefficients of the 5D Chern-Simons terms that result from integrating out the massive spectrum. 
 It is a little unusual that it is {\it massive} fields that contribute to these parity-violating terms. This is because in five dimensions massive fields can be in chiral
 representations of the  little group $\sutwo\times \sutwo$
  and so can contribute to the parity-violating Chern-Simons terms.
 First, we consider these quantum  corrections in a general setting and then discuss their origins and consequences for the entropy of the black holes solutions of \autoref{sec:5dblackholes}. Subsequently, we 
 compute the quantum
 corrections to the Chern-Simons
  terms from integrating out massive supergravity fields. This is of course not the full story: there are in principle further corrections from stringy modes; these will be considered elsewhere.

\subsection{Corrections to Chern-Simons terms}
\label{sec:corrections chern-simons}

In five dimensions massive fields can be {\it chiral} as  they are in representations $(s,s')$ of the little group $\sutwo\times \sutwo$, and we will refer to them as chiral if $s\ne s'$.
In the supergravity theory we have been discussing, the chiral massive field content consists of the gravitino in the $(3,2)$ representation, the self-dual two-form field in the  $(3,1)$ representation and the spin-half dilatino in the $(2,1)$ representation (together with their anti-chiral counterparts  
$(2,3)$, $(1,3)$ and $(1,2)$). As we have seen in \autoref{sec: massivespectra}, these massive fields fit into $(p,q)$ BPS supermultiplets.
 By integrating out this chiral matter, we can obtain corrections to the 5D Chern-Simons terms \cite{Bonetti:2013ela}. In principle, one would need to integrate out the entire chiral massive spectrum; the fields that we found in our supergravity calculation, as well as massive stringy modes. We focus on the supergravity fields here.

From the fields that we obtain in our duality-twisted compactification of 6D supergravity, only the self-dual tensors, gravitini (spin-$\tfrac{3}{2}$ fermions) and dilatini (spin-$\tfrac{1}{2}$ fermions) contribute to the Chern-Simons terms. Integrating out other types of massive fields does not yield Chern-Simons couplings \cite{Bonetti:2013ela}. The origin of this lies in parity: since the Chern-Simons terms violate parity, they can only be generated by integrating out parity-violating fields.

The non-abelian gauge symmetry of the 5-dimensional gauged supergravity is spontaneously broken to an abelian subgroup with massless abelian gauge field one-forms $A^I$ with field strengths $F^I=\diff A^I$, with the index $I$ running over the number of massless vector fields in the theory.
The pure gauge and the mixed gauge-gravitational Chern-Simons terms involving these fields are of the form
\begin{eqnarray}
\label{chernsimonsterms}
S_{AFF} = \frac{-\,g^3}{48\pi^2} \int k_{IJK} A^I \wedge F^J \wedge F^K \,, \quad\;\;\; S_{ARR} = \frac{-\,g}{48\pi^2} \int k_{I} A^I \wedge \tr \, (R \wedge R) 
\end{eqnarray}
for some coefficients $k_{IJK}$, $k_I$.
Here  $g$ denotes the  gauge coupling and $R$ denotes the curvature two-form.
Integrating out the chiral massive fields yields quantum corrections to 
the coefficients $k_{IJK}$, $k_I$.

Consider first the Chern-Simons terms in the  classical 5D supergravity obtained by Scherk-Schwarz reduction from maximal 6D supergravity.
By explicit reduction, we find that there are no $A\wedge R\wedge R$ terms. There are $A\wedge F\wedge F$ terms present however. For example, in the reduction with the Scherk-Schwarz twist that preserves the D1-D5 black hole, we find the term
\begin{equation}
\label{05K CS term}
\frac{1}{2\kappa^2_{(5)}} \int \mathcal{A}_1^5 \wedge \diff {C}_1^{(5)} \wedge \diff \tilde{C}_1^{(5)} \,,
\end{equation}
so that we have $k_{IJK}=-\tfrac{4\pi^2}{\kappa^2_{(5)} g^3}$ for the indices $I,J,K$ corresponding to the three gauge fields in \eqref{05K CS term}. This Chern-Simons term (and other similar terms) can be found from the reduction of the 6D tensor fields (following \autoref{sec:SStensors}). There are also Chern-Simons terms coming from the reduction of the 6D vectors.

Quantum corrections to the Chern-Simons terms are only allowed for certain amounts of unbroken supersymmetry. The coefficients of the $A\wedge F\wedge F$ term are fixed for $\mathcal{N}>2$ supersymmetry, so corrections to this terms are only allowed in the $\mathcal{N}=2$ (and 0) theories. For $\mathcal{N}=2$, the supersymmetric completion of the $A\wedge R\wedge R$ term exists and is known \cite{Hanaki:2006pj}, but this is not the case for theories with more supersymmetry. However, in the chiral $\mathcal{N}=4$ $(0,2)$ theory a  $A\wedge R\wedge R$ term  
is generated by quantum corrections, leading to the
   conjecture that a supersymmetric completion of this term should exist \cite{Bonetti:2013cza}. There is no such quantum $A\wedge R\wedge R$ term  for the non-chiral $\mathcal{N}=4$ $(1,1)$ theory, nor for the $\mathcal{N}=8,6$ theories. We will see in \autoref{sec:sugracalculationchernsimons} that the corrections that we find from integrating out the massive fields that come from our duality-twisted compactification of 6D supergravity (including the Kaluza-Klein towers from the circle compactification) are in agreement with the above: we find corrections to the $A\wedge F\wedge F$ term only for  $\mathcal{N}=2$ supersymmetry and a quantum $A\wedge R\wedge R$ term   is induced only for $\mathcal{N}=2$ and the chiral $\mathcal{N}=4$ $(0,2)$ theory.

For our purposes, we will focus on  the Chern-Simons  terms
$\mathcal{A}^5 \wedge \diff\mathcal{A}^5 \wedge \diff\mathcal{A}^5$ and $\mathcal{A}^5\wedge R\wedge R$
 that
 involve the graviphoton $\mathcal{A}^5$. This is because the black holes that we consider couple only to the graviphoton and to vectors descending from the 6D tensors (see \autoref{sec:5dblackholes}). The chiral massive field content that we find from duality-twisted compactification is not charged under the gauge symmetries corresponding to the vectors that descend from 6D tensors, so for the purposes of studying corrections to the black hole solutions we only need to consider couplings of this chiral matter to the graviphoton; these then lead to    corrections to the coefficients of the
 $\mathcal{A}^5 \wedge \diff\mathcal{A}^5 \wedge \diff\mathcal{A}^5$ and $\mathcal{A}^5\wedge R\wedge R$ terms.
 
We introduce the notation $k_{AFF}$ for the coefficient of the $\mathcal{A}^5 \wedge \diff\mathcal{A}^5 \wedge \diff\mathcal{A}^5$ term and $k_{ARR}$ for the coefficient of the $\mathcal{A}^5\wedge R\wedge R$ term. Neither of these terms are present in the classical theory -- there is no $\mathcal{A}^5 \wedge \diff\mathcal{A}^5 \wedge \diff\mathcal{A}^5$ term for the graviphoton.
As a result, both  $k_{AFF}$ and  $k_{ARR}$ have no classical contributions and arise only from quantum corrections.

\subsection{Corrections to black hole entropy}

We now study the effect that the corrections to the Chern-Simons terms have on the black holes that we studied in \autoref{sec:5dblackholes}. As it turns out, both the coefficients $k_{AFF}$ and $k_{ARR}$ affect the black hole solutions. In particular, the entropy of these black holes is modified by the corrections to these coefficients.

In \cite{Castro:2007hc,deWit:2009de} general   BPS black hole solutions
were found for $\mathcal{N}=2$ supergravity
with both pure gauge and gauge-gravitational Chern-Simons terms \eqref{chernsimonsterms}. These general results  then give BPS  black hole solutions for our $\mathcal{N}=2$ supergravity models, with the specific values of the Chern-Simons coefficients obtained in the next subsection. In particular, these   BPS black holes  are preserved by four supersymmetries, and these are the black holes for which we compute the entropy.

We can also apply this to the black holes in the $\mathcal{N}=4$ $(0,2)$ theory. As discussed in the previous subsection, the
$\mathcal{N}=2$
and $\mathcal{N}=4$ $(0,2)$ theories
 are the only ones for which corrections to the Chern-Simons coefficients are allowed, and so these are the only theories in which we find corrected black hole solutions.

Consider the $\mathcal{N}=4$ $(0,2)$ theory. By integrating out the massive field content we obtain a non-zero coefficient $k_{ARR}$ for the gauge-gravitational Chern-Simons term. In order to compute corrected BPS black hole solutions in this theory, we use the framework of \cite{Castro:2007hc,deWit:2009de} for $\mathcal{N}=2$ supergravity. 
We can  consistently truncate this $\mathcal{N}=4$ theory to an $\mathcal{N}=2$ theory
by decomposing fields into representations of an $\usptwo\times\usptwo$ subgroup of the
  R-symmetry group  and removing all fields that transform non-trivially under one of these $\usptwo$'s.
  For each of the black hole solutions we have considered, we make a corresponding choice of the embedding of the $\usptwo\times\usptwo$ subgroup so that all the fields that are non-trivial in the black hole solution survive the truncation.
As a result, the  black hole solutions of  the effective theory with an $A\wedge R\wedge R$ term  given in \cite{Castro:2007hc,deWit:2009de} will also be solutions of the quantum-corrected $\mathcal{N}=4$ $(0,2)$ theory that we have been considering here.

We now briefly review the procedure to compute the entropy of BPS black holes in these quantum corrected theories. It is given by the formula \cite{Castro:2007hc, deWit:2009de}
\begin{equation}
\label{bheq}
S = \frac{\pi}{6} \, k_{IJK} \, X^I X^J X^K \, ,
\end{equation}
where $X^I$ are the (rescaled) moduli corresponding to the three gauge fields $A^I$ that couple to the black hole charges
and $ k_{IJK}$ are the Chern-Simons coefficients from \eqref{chernsimonsterms}.
The values of  these moduli in the solution are found by  solving the attractor equation, which 
 in the near-horizon limit is
\begin{equation}
\label{attractoreq}
- \frac{1}{2} \, k_{IJK} \, X^J X^K \,=\, \frac{\pi}{2 g \, G_N^{(5)}} \, Q_I + 2 \, k_I \, ,
\end{equation}
More comprehensive studies of these solutions can be found in \cite{Castro:2007hc,deWit:2009de}.

We now apply this to our setup. When we solve \eqref{attractoreq} and compute \eqref{bheq} for general coefficients $k_{AFF}$ and $k_{ARR}$ (to the Chern-Simons terms that contain the graviphoton $\mathcal{A}^5$), we find the entropy of the corrected D1-D5-P black hole solution in terms of its three charges to be
\begin{equation}
\label{correctedentropy}
S_\text{BH}  \:=\: \frac{2 \pi^2}{4G_N^{(5)}} \, \sqrt{Q_1 Q_5 \hat{Q}_K \: \frac{\,2\left(1+\sqrt{1+k_{AFF}\,\frac{4 G_N^{(5)}}{\pi R^3}\,\frac{Q_1 Q_5\phantom{^i}}{\hat{Q}_K^2}}+k_{AFF}\,\frac{4 G_N^{(5)}}{3\pi R^3}\,\frac{Q_1 Q_5\phantom{^i}}{\hat{Q}_K^2}\right)^2}{\left(1+\sqrt{1+k_{AFF}\,\frac{4 G_N^{(5)}}{\pi R^3}\,\frac{Q_1 Q_5\phantom{^i}}{\hat{Q}_K^2}}\,\right)^3} } \;.
\end{equation}
Here the charge arising from momentum in the $z$ direction is shifted
\begin{equation}
\hat{Q}_K \,=\, Q_K + \frac{4 G_N^{(5)}}{\pi R} \, k_{ARR} \,.
\end{equation}
It can easily be checked that for $k_{AFF}=k_{ARR}=0$ this expression for the black hole entropy reduces to the uncorrected result
\begin{equation}\label{uncorrectedentropy}
S_\text{BH} \,=\, \frac{\pi^2}{2G_N^{(5)}} \, \sqrt{Q_1Q_5Q_K} \:.
\end{equation}
Just as was done for the uncorrected expression for the entropy, we can express the three charges in terms of integers $N_i$  times the basic charges as $Q_i=c_i N_i$ with the basic charges $c_i$ as given in \eqref{chargecoefficients}. This yields
\begin{equation}
\label{correctedentropyd1d5}
S_\text{BH} \:=\: 2 \pi \sqrt{N_1 N_5 \hat{N}_K \: \frac{\,2\left(1+\sqrt{1+k_{AFF}\,\frac{N_1 N_5\phantom{^i}}{\hat{N}_K^2}}+\frac{1}{3}\,k_{AFF}\,\frac{N_1 N_5}{\hat{N}_K^2}\right)^2}{\left(1+\sqrt{1+k_{AFF}\,\frac{N_1 N_5\phantom{^i}}{\hat{N}_K^2}}\,\right)^3} } \;,
\end{equation}
where the shifted momentum charge number is given by
\begin{equation}
\hat{N}_K \,=\, N_K + k_{ARR} \,.
\end{equation}
The expression \eqref{correctedentropyd1d5} can be expanded for small $k_{AFF}$ as
\begin{equation}
S_\text{BH} \, = \, 2 \pi \sqrt{N_1 N_5 \hat{N}_K} + \frac{\pi}{12} \, k_{AFF} \left(\frac{N_1 N_5}{\hat{N}_K}\right)^{\frac{3}{2}} + \mathcal{O} \big( k_{AFF}^2\big) \,.
\end{equation}
The first term is equal to the uncorrected black hole entropy \eqref{5dentropy} and the second term is the correction to first order  in $k_{AFF}$.

\subsection{One-loop calculation of the Chern-Simons coefficients}
\label{sec:sugracalculationchernsimons}

In this section we compute the contributions to $k_{AFF}$ and $k_{ARR}$ that come from integrating out chiral massive fields arising from the  duality-twisted compactification of 6D supergravity. While this is a well-defined calculation, some caution is needed since there will also be contributions from the chiral spectrum of stringy modes to the Chern-Simons coefficients. The coefficients that we compute here come purely from the supergravity modes.

\begin{figure}
\centering
\includegraphics[scale=1]{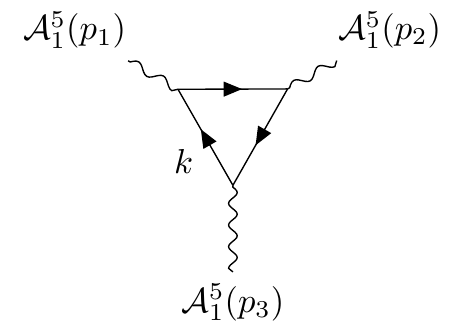}
\captionsetup{width=.9\linewidth}
\caption{\textit {This diagram generates corrections to the $A\wedge F\wedge F$ Chern-Simons coupling. The external lines represent the graviphoton whilst the solid internal lines represent a massive self-dual tensor, gravitino or dilatino running in the loop.}}
\label{fig:kaffdiagram}
\end{figure}

Contributions are only obtained from integrating out massive self-dual tensors, gravitini (spin-$\tfrac{3}{2}$ fermions) and dilatini (spin-$\tfrac{1}{2}$ fermions). The relevant diagrams for corrections to the couplings \eqref{chernsimonsterms} have been computed in \cite{Bonetti:2013ela}. As an example, we show the diagram that contributes to the $A\wedge F\wedge F$ term in \autoref{fig:kaffdiagram}. The diagrams that contribute to the $A\wedge R\wedge R$ term can be found in \cite{Bonetti:2013ela}. The results of these computations are shown in the table below.

\renewcommand\arraystretch{1.3}
\begin{table}[ht!]
\centering
\begin{tabular}{c|ccc}
& \;self-dual tensor $B_2$\;\; & gravitino $\psi_{\mu}$ & \;\;dilatino $\chi$\; \\ \hline\hline
$k_{AFF}$ & $-4\,c_B\,q^3$ & $5\,c_\psi\,q^3$ &$c_\chi\,q^3$ \\[.15cm]
$k_{ARR}$ & $c_B\,q$  & $-\frac{19}{8}\,c_\psi\,q$ & $\frac{1}{8}\,c_\chi\,q$ \\
\end{tabular}
\end{table}
\renewcommand\arraystretch{1}

We see that the contribution of a massive field to each of the Chern-Simons couplings consists of three parts: a prefactor that depends on the field type, a constant $c_{\text{field}}$ (equal to $\pm 1$) that depends on the field's representation under the massive little group, and the field's $\uone$ charge $q$ under the graviphoton $\mathcal{A}_1^5$.

In order to find the corrections to the Chern-Simons terms that are induced by the massive spectra of our 5D theories, we need to know two things about each of the massive fields: the sign of $c_{\text{field}}$ and the charge $q$. We always take $q\geq0$ and absorb any minus signs into the corresponding $c_\text{field}$.

The conventions in this work are such that 5D tensors that descend from 6D self-dual tensors and 5D fermions that descend from 6D positive chiral fermions have $c_{\text{field}}=-1$, while tensors descending from 6D anti-self-dual tensors and fermions descending from 6D negative chiral fermions have $c_{\text{field}}=+1$. In terms of the six-dimensional R-symmetry representations, the signs of $c_{\text{field}}$ of the corresponding five-dimensional massive fields are
\begin{alignat}{4}
\notag & {(\textbf{5},\textbf{1})}: \quad c_B\:&=-1 \,, \qquad\qquad\quad & {(\textbf{1},\textbf{5})}: \quad c_B\:&=+1 \,, \\[2pt]
& {(\textbf{4},\textbf{1})}: \quad c_{\psi}\:&=-1 \,, \qquad\qquad\quad & {(\textbf{1},\textbf{4})}: \quad c_{\psi}\:&=+1 \,, \\[2pt]
\notag & {(\textbf{5},\textbf{4})}: \quad c_{\chi}\:&=-1 \,, \qquad\qquad\quad & {(\textbf{4},\textbf{5})}: \quad c_{\chi}\:&=+1 \,.
\end{alignat}

We know from \autoref{sec:kaluzakleintowers} that each 6D field produces a Kaluza-Klein tower of 5D fields for which the sum of the charges is given by
\begin{equation}
\sum\limits_{n=-\infty}^{\infty}\left| \frac{\mu (m_i)}{2\pi} + {n}\, \right| \, .
\end{equation}
We need to regularize such sums (and similar sums in which we take the sum of the cube of the charges). Following \cite{Grimm:2013oga},  the regularized expressions are
\begin{alignat}{3}
\label{functions1}
s_1[m]=&\sum\limits_{n=-\infty}^{\infty}\left|\frac{m}{2\pi} + {n} \right| &&=&&\: \left|\frac{m}{2\pi}\right|\,(2k +1)-k(k+1) - \frac{1}{6} \,, \\[4pt]
s_3[m]=&\sum\limits_{n=-\infty}^{\infty}\left|\frac{m}{2\pi} + {n} \right|^3 &&=&&\: \left|\frac{m}{2\pi}\right|^3  (2k+1)-3\left(\frac{m}{2\pi}\right)^2  \left(k(k+1)+\frac{1}{6}\right) \nonumber\\
& \, && &&\:+  3\left|\frac{m}{2\pi}\right|\,\left(\frac{k(k+1)(2k+1)}{3}\right) - \frac{k^2(k+1)^2}{2} + \frac{1}{60} \label{functions3}\,.
\end{alignat}
Here we use the notation 
$$k\equiv \left \lfloor \left|\tfrac{m}{2\pi}\right| \right \rfloor \,,$$
 where  
$\left \lfloor x \right \rfloor$ is the integer part of $x$.

We now have all the information that we need to compute the corrections to the Chern-Simons terms \eqref{chernsimonsterms} that are generated by integrating out our massive five-dimensional spectra. Now, for a general twist (i.e. all twist parameters are turned on) we find the correction to the pure gauge term 
\begin{equation}
\label{correctionAFF}
\begin{aligned}
k_{AFF} \,=\, &\: 4 \big(\! -s_3[m_1]- s_3[m_2]+ s_3[m_3]+ s_3[m_4] \\
&\phantom{\: 4 \big(}\!+ s_3[m_1+m_2]+ s_3[m_1-m_2]- s_3[m_3+m_4]- s_3[m_3-m_4] \,\big) \\[2pt]
& - s_3[m_1+m_2+m_3]- s_3[m_1+m_2-m_3]- s_3[m_1-m_2+m_3] \\
&- s_3[m_1-m_2-m_3] - s_3[m_1+m_2+m_4]- s_3[m_1+m_2-m_4]\\
&- s_3[m_1-m_2+m_4]- s_3[m_1-m_2-m_4] + s_3[m_1+m_3+m_4]\\
&+ s_3[m_1+m_3-m_4]+ s_3[m_1-m_3+m_4]+ s_3[m_1-m_3-m_4] \\
&+ s_3[m_2+m_3+m_4]+ s_3[m_2+m_3-m_4]+ s_3[m_2-m_3+m_4]\\
&+ s_3[m_2-m_3-m_4] \,,
\end{aligned}
\end{equation}
and the correction to the mixed gauge-gravitational term 
\begin{equation}
\label{correctionARR}
\begin{aligned}
k_{ARR} \,=\, &\:\tfrac{5}{2} \big( s_1[m_1]+ s_1[m_2]- s_1[m_3]- s_1[m_4] \,\big)\\[2pt]
&- s_1[m_1+m_2]- s_1[m_1-m_2]+ s_1[m_3+m_4]+ s_1[m_3-m_4]\\[2pt]
& +\tfrac{1}{8} \big( -s_1[m_1+m_2+m_3]-s_1[m_1+m_2-m_3]-s_1[m_1-m_2+m_3] \\
&\phantom{\;\,+\tfrac{1}{8} \big(} -s_1[m_1-m_2-m_3]-s_1[m_1+m_2+m_4]-s_1[m_1+m_2-m_4] \\
&\phantom{\;\,+\tfrac{1}{8} \big(} -s_1[m_1-m_2+m_4]-s_1[m_1-m_2-m_4]+s_1[m_1+m_3+m_4] \\
&\phantom{\;\,+\tfrac{1}{8} \big(} +s_1[m_1+m_3-m_4]+s_1[m_1-m_3+m_4]+s_1[m_1-m_3-m_4] \\
&\phantom{\;\,+\tfrac{1}{8} \big(} +s_1[m_2+m_3+m_4]+s_1[m_2+m_3-m_4]+s_1[m_2-m_3+m_4] \\
&\phantom{\;\,+\tfrac{1}{8} \big(} +s_1[m_2-m_3-m_4] \,\big) \,.
\end{aligned}
\end{equation}

The above formulae give the contributions from summing over all Kaluza Klein modes arising from the reduction from 6D to 5D.
The Scherk-Schwarz reduction to  5D supergravity keeps only the  $n=0$ modes and not the whole KK-towers, and on restricting to the  $n=0$ modes the
functions $s_1$ and $s_3$ reduce to 
\begin{equation}
\label{sugrmods}
s_1[m] = \left|\tfrac{m}{2\pi}\right| \,, \qquad s_3[m] = \left|\tfrac{m}{2\pi}\right|^3 \,.
\end{equation}
Then the quantum corrections to the Chern-Simons coefficients $k_{AFF}$ and $k_{ARR}$ from integrating out only the massive modes of the 5D supergravity that arises from Scherk-Schwarz reduction 
are  given by \eqref{correctionAFF} and \eqref{correctionARR} with the simpler expressions \eqref{sugrmods} for $s_1,s_3$.

The expressions  \eqref{correctionAFF} and \eqref{correctionARR}  are the quantum corrections for general values of the mass parameters.
The results for twists that preserve supersymmetry can be found by taking certain parameters in \eqref{correctionAFF} and \eqref{correctionARR} equal to zero. We work out some interesting cases below.

\begin{itemize}
\item
$\mathcal{N}=8$, $\mathcal{N}=6$ and $\mathcal{N}=4$ $(1,1)$

By twisting to any of these cases we find that $k_{AFF}=0$ and $k_{ARR}=0$, as can be checked straightforwardly by setting the appropriate mass parameters equal to zero in \eqref{correctionAFF} and \eqref{correctionARR}. This is consistent with expectations based on supersymmetry and chirality, as was explained earlier in this section.

\item
$\mathcal{N}=4$ $(0,2)$

For the case where we choose a chiral twist to the $\mathcal{N}=4$ theory, say with $m_1,m_2\neq0$ and $m_3=m_4=0$, we find that $k_{AFF}$ vanishes but $k_{ARR}$ does not. For such a twist, we find the correction from the $n=0$ modes to be
\begin{equation}
k_{ARR} \,=\, \tfrac{1}{2\pi} \left( 3 \,|m_1| +3\, |m_2| - \tfrac{3}{2}\,|m_1+m_2| - \tfrac{3}{2}\,|m_1-m_2|\right)\,,
\end{equation}
and by taking into account the Kaluza-Klein towers as well we find
\begin{equation}
k_{ARR} \,=\, \tfrac{1}{2}+ 3 \,s_1[m_1] +3\, s_1[m_2] - \tfrac{3}{2}\,s_1[m_1+m_2] - \tfrac{3}{2}\,s_1[m_1-m_2]\,  \,.
\end{equation}

\item
$\mathcal{N}=2$ $(0,1)$

In the minimal $\mathcal{N}=2$ theory corrections to both the Chern-Simons coefficients are allowed, and the supersymmetric extension of the $A\wedge R\wedge R$ term is known \cite{Hanaki:2006pj}. The coefficients $k_{AFF}$ and $k_{ARR}$ can be computed from the general formulas \eqref{correctionAFF} and \eqref{correctionARR} by taking $m_4=0$ and the other parameters non-zero. The general expressions are quite unwieldy, but if we take $m_1=m_2=m_3=m$ they simplify substantially. For this choice of mass parameters the corrections due to the $n=0$ modes are
\begin{equation}
k_{AFF} \,=\, 36 \,\left|\frac{m}{2\pi}\right|^3 \,, \qquad\quad
k_{ARR} \,=\, \frac{9}{4} \, \left|\frac{m}{2\pi}\right| \,,
\end{equation}
and the corrections due to both the $n=0$ modes and the Kaluza-Klein towers read
\begin{align}
\label{n=2chernsimonskaff}
k_{AFF} \,&=\, \tfrac{1}{6} - 15\,s_3[m] + 6\,s_3[2m] - s_3[3m] \,, \\[6pt]
\label{n=2chernsimonskarr}
k_{ARR} \,&=\, \tfrac{13}{24} + \tfrac{33}{8}\,s_1[m] - \tfrac{3}{4}\,s_1[2m] - \tfrac{1}{8}\,s_1[3m] \,.
\end{align}

\end{itemize}

The expressions for the coefficients $k_{AFF}$ and $k_{ARR}$ that we found in this subsection are computed from the supergravity fields that come from the duality-twisted compactification. A more thorough calculation would be needed to include all the stringy modes as well. The embedding into string theory is     discussed in the next section. The full string theory calculation of the coefficients $k_{AFF}$ and $k_{ARR}$, however, is beyond the scope of this paper and left for future study.

\section{Embedding in string theory}
\label{sec:embeddingstringtheory}

So far we have considered a supergravity setup in which we studied BPS black holes in a Scherk-Schwarz reduced theory. In some cases, Scherk-Schwarz reductions can be lifted to string theory as compactifications with a duality twist. We study such lifts in this section.

\subsection{Quantization of the twist parameters}
\label{sec:quantization mass parameters}

The Scherk-Schwarz reductions we have been considering  have lifts to string theory (or M-theory) only for special values of the  parameters $m_i$. We now investigate the  lifts of Scherk-Schwarz reductions to full string theory constructions.
The supergravity duality symmetry is broken to the discrete U-duality symmetry 
$ \text{Spin}(5,5;\mathbb{Z})$ 
\cite{Hull:1994ys}, and
the Scherk-Schwarz monodromy has to be restricted to be in this discrete subgroup \cite{Hull:1998vy,Dabholkar:2002sy}.

We then have three  conditions on the monodromy, similar to the three conditions in \cite{Hull:2017llx,Gautier:2019qiq}.

\begin{enumerate}
\item The monodromy is a U-duality
\begin{equation}
{\cal{M}}\in \text{Spin} (5,5;\mathbb{Z}) \,.
\end{equation}

\item The monodromy is conjugate to an R-symmetry

\begin{equation}
\label{monoconzz}
\mathcal{M}=g \tilde {\mathcal{M} }g^{-1}, \qquad g\in \sofivefive, \qquad  \tilde{ \mathcal{M}} \in \uspfour_\text{L}\times \uspfour_\text{R} \subset \sofivefive \,.
\end{equation}
This ensures that there is a Minkowski vacuum and implies that the monodromy is in fact conjugate to an element of a maximal torus \eqref{monocon2} parameterised by four angles $m_i$. Note that the conjugation is by an element $g$ of the {\it continuous group} $\sofivefive$.

\item At least one of the parameters $m_i$ is zero, so that the  monodromy is conjugate to a subgroup of the R-symmetry
\begin{equation}
 \tilde{ \mathcal{M}} \in \sutwo \times \uspfour \in \uspfour \times \uspfour  \,.
\end{equation}
This condition ensures that some supersymmetry is preserved.

\end{enumerate}

Conditions (1) and (2) imply that $\mathcal{M}$ satisfies $\mathcal{M}^p=\mathbbm{1}$ for some integer $p$, so that $\mathcal{M}$ 
 generates a cyclic group $\mathbb{Z}_p$  \cite{Dabholkar:2002sy}.
As a a result, the phases $e^{i m_i}$ are all $p$'th roots of unity, so that
\begin{equation}
m_i= \frac {2\pi n_i} p \, , \quad \qquad i=1,\ldots , 4 \,,
\end{equation}
for some integers $n_i$. This can be thought of as a quantization of the parameters $m_i$.

The point in the moduli space given by the coset $[g]$ of the group element $g\in \sofivefive$ in \eqref{monoconzz} is a fixed point under the action of the $\mathbb{Z}_p$ generated by $\mathcal{M}$, and this is the point at which the scalar potential has its minimum \cite{Dabholkar:2002sy}. The corresponding low energy supergravity description is as described in \autoref{Conjugate Monodromies}.

The general solution to these three requirements is not known.
Consider, however, the special case in which 
\begin{equation}
{\cal{M}}\in 
\text{SL}(2;\mathbb{Z}) \times
\text{SL}(2;\mathbb{Z}) \times
\text{SL}(2;\mathbb{Z}) \times
\text{SL}(2;\mathbb{Z}) 
\subset
\text{Spin} (5,5;\mathbb{Z}) \,.
\end{equation}
This subgroup arises from considering
\begin{equation}
\text{Spin}(2,2)
\times \text{Spin}(2,2)
\subset \text{Spin}(4,4)
\subset \text{Spin}(5,5) \,,
\end{equation}
and the isomorphism
\begin{equation}
\text{Spin}(2,2) \,\cong\,
\text{SL}(2;\mathbb{R}) \times
\text{SL}(2;\mathbb{R})  \,.
\end{equation}
Then taking
\begin{equation}
{\cal{M}} = M_1\times 
M_2\times
M_3\times
M_4
\in 
\text{SL}(2;\mathbb{Z}) \times
\text{SL}(2;\mathbb{Z}) \times
\text{SL}(2;\mathbb{Z}) \times
\text{SL}(2;\mathbb{Z}) \ ,
\end{equation}
there are solutions in which each $M_i \in \text{SL}(2;\mathbb{Z})$ is an element of an elliptic conjugacy class of  $\text{SL}(2;\mathbb{Z})$  \cite{Dabholkar:2002sy}. 
Each $M_i$ is then conjugate to a rotation:
\begin{equation}
M_i= k_i \, R(m_i) \, k_i^{-1} \, ,
\end{equation}
where
\begin{equation}
k_i\in SL(2;\mathbb{R}) \, , \quad\qquad
R(m_i) = \begin{pmatrix}
\,\cos m_i\: & -\sin m_i \\[2pt]
\,\sin m_i\: & \cos m_i
\end{pmatrix} \,.
\end{equation}
The angles $m_i$ must each take one of the values
\begin{equation}
\label{massquants}
m_i \in \big\{ 0,\tfrac{\pi}{3},\tfrac{\pi}{2},\tfrac{2\pi}{3} ,\pi \big\} \,,
\end{equation}
and each $M_i$ generates a $\mathbb{Z}_{n_i}$ subgroup of $\text{SL}(2;\mathbb{Z})$ with each  $n_i$ being one of $1,2,3,4,6$ (the lowest number such that $R(m_i)^{n_i} = \mathbbm{1}$). The monodromy ${\cal M}$ then generates a $\mathbb{Z}_p$ where $p$ is
the least common multiple of the  $n_i$ ($i=1,\dots, 4$) and so
 is equal to $2,3,4,6$ or $12$ (excluding the trivial case ${\cal M}=\mathbbm{1}$). 
 
 The quantization  condition on the parameters $m_i$ then provides a condition on the corrections to the coefficients of the Chern-Simons terms. We have checked that for the values of the $m_i$
 given by \eqref{massquants}, the corrections to the coefficients of the Chern-Simons terms satisfy the  appropriate quantization conditions.

\subsection{Orbifold picture and modular invariance}

The point in moduli space at which there is a minimum of the scalar potential  is a fixed point under the action of the $\mathbb{Z}_p$ generated by
the U-duality transformation $\mathcal{M}$.
At this point, the construction can be realized as a generalized orbifold of IIB string theory compactified on $T^4\times S^1$  \cite{Dabholkar:2002sy}.
The full string construction is then IIB string theory on $T^4\times S^1$  quotiented by the $\mathbb{Z}_p$
generated by the monodromy $\mathcal{M}$ combined with a shift on the $S^1$ given by
 $z\rightarrow z+2\pi R/p$.

The T-duality subgroup of the U-duality group is a particular embedding of
 $\text{Spin} (4,4;\mathbb{Z})\subset \text{Spin} (5,5;\mathbb{Z})$, and when the monodromy is a T-duality, this orbifold construction  becomes a conventional asymmetric orbifold \cite{Narain:1986qm,Narain:1990mw}.
However, this asymmetric orbifold is not 
 modular invariant in general. The remedy is straightforward \cite{Narain:1986qm,Narain:1990mw,Gautier:2019qiq}:
  modular invariance can be achieved if the shift in the circle coordinate $z$ is accompanied by a shift in the coordinate of the T-dual circle. 
  The T-dual circle has radius $\alpha' /R$, and its  coordinate $\tilde z$ undergoes a shift $\tilde z\to \tilde z + 2\pi n \alpha' / p R$ for a particular integer $n$ which is determined as in \cite{Narain:1990mw,Gautier:2019qiq}.  
 This can also be understood in momentum space. The quotient introduces phases dependent on both the momentum and the winding number on the circle, and  dependent on the charges $e_i$ under the action of $\cal M$; see  \cite{Gautier:2019qiq} for further discussion.
This then gives an exact conformal field theory formulation  
 of the duality twisted theory in its Minkowski vacuum  \cite{Dabholkar:2002sy}.

Acting on this asymmetric orbifold with a U-duality transformation will take the monodromy to a conjugate U-duality monodromy that will in general not be a T-duality.
It will then take the phase depending on the winding number to a phase depending on brane wrapping numbers, giving a non-perturbative construction similar to the ones given in
\cite{Ferrara:1995yx,Gautier:2019qiq}.

\section{Conclusion}

In this paper we have studied duality twists and their effect on black holes in string theory. Our set-up was type IIB string theory  compactified on $T^4$  and then further compactified on $ S^1$ with a duality twist along the circle.
If the twist is with a diffeomorphism of $T^4$, this gives $T^4$ bundle over $S^1$, but for a U-duality twist this gives a U-fold,  which is  a non-geometric generalization of this bundle \cite{Hull:2004in}.
 
We have
given  the relations between the  6D fields of the duality-invariant formulation of 6D $\mathcal{N}=8$  supergravity and the 10D fields of type IIB supergravity on a four-torus explicitly.
We then  reduced this six-dimensional theory on a circle with a duality twist. For this reduction we have chosen a monodromy in the R-symmetry, depending on four independent twist parameters. This reduction yields gauged 5D $\mathcal{N}=8$ supergravity, with Minkowski vacua preserving $\mathcal{N}=6,4,2,0$ supersymmetry. The amount of supersymmetry that is preserved depends on the number of the twist parameters that are equal to zero.

This Scherk-Schwarz reduction in supergravity can be embedded in string theory as a compactification with a duality twist. For such an embedding to exist,  the monodromy must be an element of the discrete U-duality group Spin$(5,5;\mathbb{Z})$. As a consequence, the twist parameters were constrained to take certain discrete values, and could hence be thought of as being quantized. The minimum of the Scherk-Schwarz potential in such compactifications is a fixed point under the action of the monodromy. When the duality twist is a T-duality, the theory arising at the minimum of the potential is an asymmetric orbifold of the type IIB string theory and so has an exact CFT description. In this case, the stringy quantum corrections can be calculated  exactly. For more general twists in which the twist is a non-perturbative symmetry, the result is 
 a generalized orbifold of  the type IIB string theory in which it is quotiented by a U-duality symmetry.

One of our main objectives was to study black holes in this set-up with partially broken supersymmetry.
Here we considered several brane configurations -- D1-D5, F1-NS5 and D3-D3 -- that result in five-dimensional black holes after  standard  (untwisted) dimensional reduction.
In each case, we compactified on $T^4$ to a  6D solution and then
  chose the twist in such a way that all the fields that source the 6D solution remain massless 
 in 5D. This ensures that the original black hole solution remains a solution of the twisted theory with partially broken supersymmetry.
 
 Our reduction scheme yielded a rich spectrum of massive modes. In 5D, massive BPS  multiplets can be chiral.
For twists that yield chiral BPS multiplets, integrating out the chiral fields gives quantum corrections to the coefficients of the pure gauge and mixed gauge-gravitational Chern-Simons terms. This gives
an EFT with both pure gauge and mixed gauge-gravitational Chern-Simons terms and these   terms led to modifications of  the BPS black hole solutions and in particular modifies the expression for the black hole entropy.

Several interesting directions for follow-up research remain. One is to investigate the microscopic side of the macroscopic story laid out in this work. This would involve studying the effects of the duality twist on the CFT dual of the black holes in our set-up. Of these, the D1-D5 CFT has been studied the most in the literature and therefore seems to be the most practical option for this. In general, one might expect that similar supersymmetry breaking patterns arise in the dual superconformal CFT, from  (4,4) supersymmetry to e.g. (4,2), (2,2) or (4,0) supersymmetry. 
D-branes and their world-volume theories in backgrounds with a duality twist have been discussed in \cite{Hull:2019iuy}
and it will be interesting to apply the results found there to the configurations discussed here.

Another open question is the computation of the Chern-Simons coefficients in the full string theory. That is, including modes that we don't see from the supergravity point of view (such as winding modes). For twists that lie in the T-duality group, this can be worked out in detail as an asymmetric orbifold compactification of perturbative string theory. As we have seen in this work, it is possible to choose a T-duality twist that preserves the D1-D5-P black hole in reductions to the $\mathcal{N}=4$ $(0,2)$ theory. In this reduction it may be possible to combine the detailed study of the full string theory with the microscopic calculation.

Finally, it would be interesting to extend this work to the study of four-dimensional black holes in string compactifications with duality twists. For this, one could take for example the four-charge D2-D6-NS5-P black hole of type IIA string theory or a dual brane configuration. Another possibility would be to study five-dimensional black rings and their reduction to four dimensional multi-center black hole solutions \cite{Gaiotto:2005xt}.

\acknowledgments

It is our pleasure to thank C. Couzens, N. Gaddam, T. Grimm, H. het Lam, K. Mayer, R. Minasian and M. Trigiante for useful discussions. This work is supported in part by the D-ITP consortium, a program of the Netherlands Organization for Scientific Research (NWO) that is funded by the Dutch Ministry of Education, Culture and Science (OCW), and by the FOM programme ``Scanning New Horizons".

\appendix

\section{Conventions and notation}

Throughout this work, we set $c=\hbar=k_B=1$, and we work in the `mostly plus' convention for the metric, i.e. $\eta_{\mu\nu}=\diag(-,+,\ldots,+)$. The notations that we use for the coordinates and indices in various dimensions are summarized in the table below.

\renewcommand{\arraystretch}{1.2}
\vspace{4pt}
\begin{center}
	\begin{tabular}{|c|c|c|}
		\hline
		\;Space \; & Coordinate & Indices \\ \hline
		\multicolumn{3}{|c|}{} \\[-18pt] \hline
		$D = 10$ & $X^M = \left( \hat{x}^{\hat{\mu}},y^m \right)$ & $M,N,\ldots=0,1,\ldots,9$\\ \hline
		$D=6$ & $\hat{x}^{\hat{\mu}}=\left(x^{\mu},z\right)$ & $\hat{\mu},\hat{\nu},\ldots=0,1,\ldots,5$ \\ \hline
		$D=5$ & $x^{\mu}$ & $\mu,\nu,\ldots=0,1,\ldots,4$ \\ \hline
		$T^4$ & $y^m$ & $m,n,\ldots=1,\ldots,4$ \\ \hline
	\end{tabular}
\end{center}
\vspace{6pt}
\renewcommand{\arraystretch}{1}

In general, we denote form-values fields as $A^{(d)}_p$, where $p$ is the rank of the form and $d$ is the dimension in which it lives. We define the Hodge star operator on forms as
\begin{equation}
\ast A^{(d)}_p = \frac{1}{p!(d-p)!} \, \sqrt{g_{(d)}} \: \varepsilon_{\mu_1\ldots\mu_p\nu_1\ldots\nu_{d-p}} \, A^{\mu_1\ldots\mu_p} \: \diff x^{\nu_1} \wedge \ldots \wedge \diff x^{\nu_{d-p}} \,.
\end{equation}
We use the subscript or superscript $(d)$ more often to indicate the number of spacetime dimensions where necessary, e.g. $R^{(d)}$, $e_{(d)}$, etc. In all dimensions, we normalize Lagrangians such that the corresponding actions are given by
\begin{equation}
S^{(d)} = \frac{1}{2\kappa^2_{(d)}} \int \lagr^{(d)} \,,
\end{equation}
where $\kappa^2_{(d)} = 8 \pi G_N^{(d)}$ is the $d$-dimensional Newton's constant.

We use $A,B,\ldots=1,\ldots,10$ to denote $\sofivefive$ indices that transform in $\tau$-frame (as explained in Appendix \ref{sec:groupappendix}), and we use $a,b,\ldots=1,\ldots,5$ for indices transforming under the subgroup $\glfive\subset\sofivefive$. For example, in 6D we have ten tensor fields (subject to a self-duality constraint), whose field strengths we write as
\begin{equation}
G^{(6)}_{3,A}=\begin{pmatrix}
G^{(6)}_{3,a}\\[2pt]
\tilde{G}^{(6)a}_{3}
\end{pmatrix} \,.
\end{equation}
The $\glfive$ subgroup works on the index $a$ of the (dual) field strengths $G^{(6)}_{3,a}$ and $\tilde{G}^{(6)a}_{3}$. For more information on how this subgroup works, see Appendix \ref{sec:groupappendix}.

\section{Group theory}

\subsection{The group \texorpdfstring{$\textnormal{SO}(5,5)$}{SO(5,5)} and its algebra}
\label{sec:groupappendix}

In this appendix we discuss some details and our conventions concerning the group SO$(5,5)$ and its algebra $\mathfrak{so}(5,5)$. In particular, we construct two bases in which SO$(5,5)$ can be written down; we call these the $\eta$-frame and the $\tau$-frame. Furthermore, we build an explicit basis for the algebra that we use to construct a vielbein $\mathcal{V}\in\text{SO}(5,5)$ in the main text.

Canonically, an element $g \in\text{SO}(5,5)$ is represented by a $10\times10$ matrix, satisfying the conditions
\begin{equation}
\label{rel}
g^T \eta \, g \,=\, \eta \,, \qquad\quad \eta = \begin{pmatrix}
\mathbbm{1}_5 & 0 \\
0 & - \mathbbm{1}_5
\end{pmatrix} \,,
\end{equation}
and $\det(g)=1$. Henceforth, we refer to group elements satisfying these conditions as being written in the $\eta$-frame of SO$(5,5)$. 
In the $\eta$-frame, 
 a generator of the Lie algebra  $M\in\mathfrak{so}(5,5)$ can be written in  $5\times5$ blocks as
\begin{equation}
\label{generalformMeta}
M =  \begin{pmatrix}
\,a & \,b\, \\[2pt]
\,b^T & \,c\,
\end{pmatrix} \,,
\end{equation}
where $a$ and $c$ are antisymmetric and $b$ is unconstrained.

There is another (isomorphic) way of writing down the group SO$(5,5)$. We construct this other basis by conjugating the group elements as $\tilde{g}=X^{-1}gX$, where $X$ is the matrix
\begin{equation}
\label{conjugationA}
X = \frac{1}{\sqrt{2}} \begin{pmatrix}
\mathbbm{1}_5 & \mathbbm{1}_5   \\
\mathbbm{1}_5  & - \mathbbm{1}_5
\end{pmatrix} \,.
\end{equation}
Note that $X = X^{-1}=X^{T}$. We can now rewrite \eqref{rel} in terms of $\tilde{g}$, which yields the following conditions on the conjugated group elements:
\begin{equation}
\label{rel2}
\tilde{g}^T \tau \, \tilde{g} \,=\, \tau \,, \qquad\quad \tau = \begin{pmatrix}
0 & \mathbbm{1}_5   \\
\mathbbm{1}_5  &0
\end{pmatrix} \,.
\end{equation}
We see that the conjugated matrices $\tilde{g}$ preserve the matrix $\tau$ (instead of $\eta$), and therefore we refer to these matrices as being written in the $\tau$-frame of SO$(5,5)$. It is clear from the conjugation relation $\tilde{g}=X^{-1}gX$ that the two frames are isomorphic. 
The general block structure for generators of the Lie algebra $\mathfrak{so}(5,5)$ in the $\tau$-frame is of the form
\begin{equation}
\label{generalformM}
\tilde{M} = \begin{pmatrix}
\,A & B \\[2pt]
\,C & -A^{T}
\end{pmatrix} \,.
\end{equation}
Here $A$ is unconstrained and $B$ and $C$ are antisymmetric.

There is a subgroup $\glfive\subset\text{SO}(5,5)$ that is embedded diagonally in the $\tau$-frame matrices $\tilde{g}$. Generators of $\glfive$ can be represented by unconstrained $5\times5$ matrices, and these can be embedded diagonally in the block structure \eqref{generalformM} by taking $B=C=0$ and $A$ equal to the $\mathfrak{gl}(5)$ generator. By exponentiating, we find the corresponding group element to be of the form
\begin{equation}
\label{gl5embedding}
\begin{pmatrix}
\,P\, & 0   \\[2pt]
\,0\, & (P^{T})^{-1}
\end{pmatrix} \:\in\: \glfive \:\subset\: \text{SO}(5,5) \,,
\end{equation}
where $P$ is an invertible five by five matrix. The embedding in the  $\eta$-frame can be found by conjugating \eqref{gl5embedding} with the matrix $X$ given in \eqref{conjugationA}.

\paragraph{A basis for the algebra $\mathfrak{so}(5,5)$.}

Using the general form of $M$, we build a basis of generators. Since $\mathfrak{so}(5,5)$ has rank five, we have five Cartan generators, denoted by $H_n$ ($n=0,\ldots,4$). 
We choose the Cartan subalgebra to be block-diagonal in the $\tau$-frame, so that when written in
the form \eqref{generalformM}, they all have $B=C=0$. Furthermore, for convenience we choose the following form for the $A$ matrices for the $H_n$:
\begin{eqnarray}
\notag A_{H_0} &=& \frac{1}{2}\diag \, (0,0,0,-1,1) \\
\notag A_{H_1} &=& \frac{1}{\sqrt{2}} \diag \, (-1,0,0,0,0) \ , \\
\notag A_{H_2} &=& \frac{1}{\sqrt{2}}\diag \, (0,-1,0,0,0)\ , \\
\notag A_{H_3} &=& \frac{1}{\sqrt{2}}\diag \, (0,0,-1,0,0) \ ,\\
\notag A_{H_4} &=& \frac{1}{2}\diag \, (0,0,0,-1,-1) \ .
\end{eqnarray}
Apart from these Cartan generators, there are 20 root generators with $B=C=0$. We denote them by $E^{A,+}_{nm}$ and $E^{A,-}_{nm}$ ($n,m=1,\ldots,5$ and $n<m$). The $E^{A,+}_{nm}$ together fill the upper triangular part of $A$ and the $E^{A,-}_{nm}$ fill the lower-triangular part. They do so in such a way that $\left(E^{A,+}_{nm}\right)^T=E^{A,-}_{nm}$. For example, we have
\begin{eqnarray}
A_{E^{A,+}_{12}}=\begin{pmatrix}
0 & -1 & 0 & 0 & 0 \\
0 & 0 & 0 & 0 & 0 \\
0 & 0 & 0 & 0 & 0 \\
0 & 0 & 0 & 0 & 0 \\
0 & 0 & 0 & 0 & 0
\end{pmatrix}\qquad\text{and}\qquad
A_{E^{A,-}_{12}}=\begin{pmatrix}
0 & 0 & 0 & 0 & 0 \\
-1 & 0 & 0 & 0 & 0 \\
0 & 0 & 0 & 0 & 0 \\
0 & 0 & 0 & 0 & 0 \\
0 & 0 & 0 & 0 & 0
\end{pmatrix}\ .
\end{eqnarray}
Finally, there are ten root generators $E^B_{nm}$ with $A=C=0$, and ten root generators $E^C_{nm}$ with $A=B=0$. The generators $E^B_{nm}$ have
\begin{eqnarray}
B_{E^B_{12}}=\begin{pmatrix}
0 & 1 & 0 & 0 & 0 \\
-1 & 0 & 0 & 0 & 0 \\
0 & 0 & 0 & 0 & 0 \\
0 & 0 & 0 & 0 & 0 \\
0 & 0 & 0 & 0 & 0
\end{pmatrix}\ ,\quad
B_{E^B_{13}}=\begin{pmatrix}
0 & 0 & 1 & 0 & 0 \\
0 & 0 & 0 & 0 & 0 \\
-1 & 0 & 0 & 0 & 0 \\
0 & 0 & 0 & 0 & 0 \\
0 & 0 & 0 & 0 & 0
\end{pmatrix}\ ,\quad\text{etc.}
\end{eqnarray}
The generators $E^C_{nm}$ are constructed in the same way as $E^B_{nm}$, but now we have $B=0$ and $C\neq0$. The matrix $C$ that corresponds to $E^C_{nm}$ is equal to the matrix $B$ that defined $E^B_{nm}$ in the construction above. Note that this implies that $\left(E^B_{nm}\right)^T=\,-\,E^C_{nm}$.

The set of matrices defined above $\left\{H_n,\,E^{A,+}_{nm},\,E^{A,-}_{nm},\,E^B_{nm},\,E^C_{nm}\right\}$ gives a complete basis of generators of $\mathfrak{so}$(5,5). When we mention $E^A_{mn}$ below we always mean $E^{A,+}_{mn}$.

Let us now discuss the notation $T_{ij}^{F}$ used in the text. These matrices $T$ are  certain generators of the $\mathfrak{so}$(5,5) algebra described above. In particular if we let $\vec{T}^F_{ij}:=(T^F_{12},T^F_{13},\dots,T^F_{34})$, then we have the following definitions for $T$:
\begin{eqnarray}\label{Tdef}
\notag\vec{T}^{A}_{ij} &=& \left(\, E^C_{23},\, E^C_{12},\,- E^C_{13},\, -(E^A_{13})^{T},\, -(E^A_{12})^T,\, -E^A_{23} \,\right) \\
\notag\vec{T}^{B}_{ij} &=& \left(\, E^A_{14},\, E^A_{34},\, E^A_{24},\, -E^C_{24},\, E^C_{34},\, -E^C_{14} \,\right) \\
\vec{T}^{C}_{ij} &=& \left(\, E^A_{15},\, E^A_{35},\, E^A_{25},\, -E^C_{25},\, E^C_{35},\, -E^C_{15} \,\right) \\
\notag T^{a} &=& E^A_{45}  \\
\notag T^{b} &=& E^C_{45} 
\end{eqnarray}

\subsection{The isomorphism \texorpdfstring{$\mathfrak{usp}(4)\cong \mathfrak{so}(5)$}{usp(4) = so(5)}}
\label{appendix: isomorphism}

The group USp$(4)$ is the group of $4\times4$ matrices $g$ satisfying
\begin{equation}
g^{\dagger} =g^{-1} \ , \qquad \Omega \, g \,\Omega^{-1} = \left( g^{-1} \right)^{T}
\end{equation}
where $\Omega$ is the symplectic metric, given by the block matrix
\begin{equation}
\Omega^{AB} = \begin{pmatrix}
0_{2\times 2} & \mathbbm{1}_{2\times2} \\
- \mathbbm{1}_{2\times2} & 0_{2\times2}
\end{pmatrix} \ .
\end{equation} 
The  Lie algebra $\mathfrak{usp}(4)$ is represented by $4\times4$ matrices $M_A^{\;\;\,B}$ satisfying
\begin{equation}\label{conditions symplectic generator}
M^{\dagger} = - M , \qquad \Omega \, M \, \Omega^{-1} = - M^{T} \ ,
\end{equation}

The isomorphism USp$(4)$ $\cong$ Spin$(5)$ can be made explicit by introducing five $4\times4$ gamma matrices, that satisfy the Euclidean Clifford algebra
\begin{equation}
	\label{clifford algebra}
	\left\{\Gamma_a,\Gamma_b\right\}_A^{\;\;\,B}=2\,\delta_{ab}\,\delta_A^B.
\end{equation}
Here $a,b=1,\ldots,5$ are the indices corresponding to Spin$(5)$, and $A,B=1,\ldots,4$ are the indices corresponding to USp$(4)$. An explicit basis of (Hermitian and traceless) gamma matrices, that satisfies \eqref{clifford algebra}, is given by
\vspace*{4pt}
\begin{equation}
	\nonumber
	\Gamma_1=\begin{pmatrix}
	0 & i\, & \,0 & 0\\
	-i & 0\, & \,0 & 0\\
	0 & 0\, & \,0 & -i\\
	0 & 0\, & \,i & 0\\
	\end{pmatrix}, \qquad
	\Gamma_2=\begin{pmatrix}
	0 & 0 & 0 & i\,\\
	0 & 0 & -i & 0\,\\
	0 & i & 0 & 0\,\\
	-i & 0 & 0 & 0\,\\
	\end{pmatrix}, \qquad
	\Gamma_3=\begin{pmatrix}
	\,0 & 0 & \!0 & 1\,\\
	\,0 & 0 & \!-1 & 0\,\\
	\,0 & -1 & \!0 & 0\,\\
	\,1 & 0 & \!0 & 0\,\\
	\end{pmatrix},
\end{equation}
\vspace*{4pt}
\begin{equation}
	\Gamma_4=\begin{pmatrix}
		\,1 & 0 & 0 & 0\\
		\,0 & -1 & 0 & 0\\
		\,0 & 0 & 1 & 0\\
		\,0 & 0 & 0 & -1\\
	\end{pmatrix}, \qquad
	\Gamma_5=\begin{pmatrix}
		\,0\, & \,1\, & \,0\, & \,0\,\\
		1 & 0 & 0 & 0\\
		0 & 0 & 0 & 1\\
		0 & 0 & 1 & 0\\
	\end{pmatrix}.
\end{equation}\\[4pt]
It can easily be checked that the gamma matrices with upper indices, defined as $(\tilde{\Gamma}_a)^{AB}=\Omega^{AC}(\Gamma_a)_C^{\;\;\;B}$, are antisymmetric\footnote{This property is used in what follows, but it is not generally true for other choices of $\Omega$ and $\Gamma_a$.}, i.e. $(\tilde{\Gamma}_a)^T=-\tilde{\Gamma}_a$. Using this, we deduce that
\begin{equation}
	\label{transpose gamma matrix}
	(\Gamma_a)^T=(\Omega^{-1}\tilde{\Gamma}_a)^T=-\tilde{\Gamma}_a\,(\Omega^{-1})^T=\Omega\,\Gamma_a\,\Omega^{-1}.
\end{equation}
Hence, the symplectic metric $\Omega$ acts on the gamma matrices as a charge conjugation matrix. We now define $\Gamma_{ab}=\frac{1}{2}[\Gamma_a,\Gamma_b]$. From \eqref{transpose gamma matrix} and the Hermitian property of the Dirac matrices, it follows directly that $\Gamma_{ab}$ satisfies the conditions \eqref{conditions symplectic generator}. Furthermore,  using  the Clifford algebra, it is straightforward to  check that the commutator of $\Gamma_{ab}$ reads
\begin{equation}
	\big[\Gamma_{ab},\Gamma_{cd}\big]=-2\,\delta_{ac}\Gamma_{bd}+2\,\delta_{ad}\Gamma_{bc}+2\,\delta_{bc}\Gamma_{ad}-2\,\delta_{bd}\Gamma_{ac}.
\end{equation}
This is exactly the commutator of the basis elements of the $\mathfrak{so}(5)$ algebra. We conclude that the ten matrices $\Gamma_{ab}$ form a set of generators of $\uspfour\cong\spinfive$. Using these gamma matrices the explicit form of the isomorphism between the algebras can be derived \cite{Villadoro:2004ci}
\begin{equation}
	\label{isomorphism generators}
	M_{ab}=-\frac{1}{2}\,\tr\!\left[M_A^{\;\;\,B}(\Gamma_{ab})_B^{\;\;\;C}\,\right].
\end{equation}
The special orthogonal Lie algebra $\mathfrak{so}(5)$ consists of real antisymmetric matrices. We can check these properties for the found generators \eqref{isomorphism generators}. The antisymmetry follows immediately from the antisymmetry in the gamma matrices $\Gamma_{ab}=-\Gamma_{ba}$. To prove the reality condition we use that both $M_A^{\;\;\,B}$ and $(\Gamma_{ab})_A^{\;\;\,B}$ satisfy the conditions \eqref{conditions symplectic generator}. Using these constraints we find
\begin{equation}
	\begin{aligned}
		(M_{ab})^{*}&=-\frac{1}{2}\,\tr\big[M^{\,*}(\Gamma_{ab})^*\big]\\[4pt]
		&=-\frac{1}{2}\,\tr\big[\Omega\,M\,\Omega^{-1}\,\Omega\,\Gamma_{ab}\:\Omega^{-1}\big]\\[4pt]
		&=-\frac{1}{2}\,\tr\big[M\,\Gamma_{ab}\big]\;=\;M_{ab}.
	\end{aligned}
\end{equation}
Thus we find that $M_{ab}$, as given in \eqref{isomorphism generators}, is a real antisymmetric matrix, and therefore a suitable generator of SO(5). 
For completeness we also mention the inverse of the isomorphism \eqref{isomorphism generators} which maps $\mathfrak{so}(5)$ to $\mathfrak{usp}(4)$:
\begin{equation}
M^A_{\;\;\,B}=\frac{1}{4}\,M_{ab}\,(\Gamma^{ab})^A_{\;\;\,B} \ .
\end{equation}

\newpage
\section{Scalar and tensor masses after Scherk-Schwarz reduction}
\label{appendix: masses}

\subsection{The D1-D5 system}
\label{app: masses d1d5}

Here we show the masses of the fields for the D1-D5 set-up, corresponding to the mass matrices shown in \eqref{mass matrices USp(4)} and \eqref{mass matrices SO(5)}. The scalar masses are as follows:

\renewcommand{\arraystretch}{1.25}
\begin{table}[ht!]
	\scalebox{0.8}{
	\begin{tabular}{|c|c|}
		\hline
		Field $\tilde{\sigma}^i$ & Mass \\ \hline
		\multicolumn{2}{|c|}{} \\[-14.5pt] \hline
		$\frac{1}{\sqrt{2}}(\phi_4 + \Phi)$ & $0$ \\ \hline
		$\frac{1}{2}(\phi_4 - \Phi + \sqrt{2}\,\phi_3)$ & $|m_1 - m_2 - m_3 + m_4|$ \\ \hline
		$\frac{1}{2}(\phi_4 - \Phi - \sqrt{2}\,\phi_3)$ & $|m_1 - m_2 + m_3 - m_4|$ \\ \hline
		$\frac{1}{\sqrt{2}}(\phi_1+\phi_2)$ & $|m_1 + m_2 - m_3 - m_4|$ \\ \hline
		$\frac{1}{\sqrt{2}}(\phi_1-\phi_2)$ & $|m_1 + m_2 + m_3 + m_4|$ \\ \hline
		$\frac{1}{2}(A_{12} + A_{34} + C_{12} + C_{34})$ & $|m_1 + m_2 - m_3 + m_4|$ \\ \hline
		$\frac{1}{2}(A_{12} + A_{34} - C_{12} - C_{34})$ & $|m_1 + m_2 + m_3 - m_4|$ \\ \hline
		$\frac{1}{2}(A_{12} - A_{34} + C_{12} - C_{34})$ & $|m_1 - m_2 + m_3 + m_4|$ \\ \hline
		$\frac{1}{2}(A_{12} - A_{34} - C_{12} + C_{34})$ & $|m_1 - m_2 - m_3 - m_4|$ \\ \hline
		$\frac{1}{2}(A_{14} + A_{23} + C_{14} - C_{23})$ & $|m_1 - m_2 + m_3 + m_4|$ \\ \hline
		$\frac{1}{2}(A_{14} + A_{23} - C_{14} + C_{23})$ & $|m_1 - m_2 - m_3 - m_4|$ \\ \hline
		$\frac{1}{2}(A_{14} - A_{23} + C_{14} + C_{23})$ & $|m_1 + m_2 - m_3 + m_4|$ \\ \hline
		\hspace*{.3cm}$\frac{1}{2}(-A_{14} + A_{23} + C_{14} + C_{23})$\hspace*{.3cm} & \hspace*{.3cm}$|m_1 + m_2 + m_3 - m_4|$\hspace*{.3cm} \\ \hline
		$A_{13}$ & $|m_1 + m_2 - m_3 - m_4|$ \\ \hline
		$A_{24}$ & $|m_1 + m_2 + m_3 + m_4|$ \\ \hline
		$C_{13}$ & $|m_1 - m_2 + m_3 - m_4|$ \\ \hline
		$C_{24}$ & $|m_1 - m_2 - m_3 + m_4|$ \\ \hline
		$\frac{1}{\sqrt{2}}(B_{12}+B_{34})$ & $|m_1 + m_2|$ \\ \hline
		$\frac{1}{\sqrt{2}}(B_{12}-B_{34})$ & $|m_3 + m_4|$ \\ \hline
		$\frac{1}{\sqrt{2}}(B_{13}+B_{24})$ & $|m_3 - m_4|$ \\ \hline
		$\frac{1}{\sqrt{2}}(B_{13}-B_{24})$ & $|m_1 - m_2|$ \\ \hline
		$\frac{1}{\sqrt{2}}(B_{14}+B_{23})$ & $|m_1 + m_2|$ \\ \hline
		$\frac{1}{\sqrt{2}}(B_{14}-B_{23})$ & $|m_3 + m_4|$ \\ \hline
		$\frac{1}{\sqrt{2}}(a+b)$ & $|m_1 - m_2|$ \\ \hline
		$\frac{1}{\sqrt{2}}(a-b)$ & $|m_3 - m_4|$ \\ \hline
	\end{tabular}}
\end{table}
\renewcommand{\arraystretch}{1}

The tensor masses are given by:

\renewcommand{\arraystretch}{1.25}
\begin{table}[ht!]
	\scalebox{0.8}{
	\begin{tabular}{|c|c|}
		\hline
		Field $A_{2,A}^{(5)}$ & Mass \\ \hline
		\multicolumn{2}{|c|}{} \\[-16pt] \hline
		$C_2^{(5)}$ & $0$ \\ \hline
		$\tilde{C}_2^{(5)}$ & $0$ \\ \hline
		$\frac{1}{\sqrt{2}}\big(B_2^{(5)}+\tilde{B}_2^{(5)}\big)$ & $|m_1 - m_2|$ \\ \hline
		$\frac{1}{\sqrt{2}}\big(B_2^{(5)}-\tilde{B}_2^{(5)}\big)$ & $|m_3 - m_4|$ \\ \hline
		$\frac{1}{\sqrt{2}}\big(R^{(5)}_{2;\,1}+\tilde{R}^{(5)}_{2;\,1}\big)$ & $|m_1 + m_2|$ \\ \hline
		$\frac{1}{\sqrt{2}}\big(R^{(5)}_{2;\,1}-\tilde{R}^{(5)}_{2;\,1}\big)$ & $|m_3 + m_4|$ \\ \hline
		$\frac{1}{\sqrt{2}}\big(R^{(5)}_{2;\,2}+\tilde{R}^{(5)}_{2;\,2}\big)$ & $|m_1 + m_2|$ \\ \hline
		$\frac{1}{\sqrt{2}}\big(R^{(5)}_{2;\,2}-\tilde{R}^{(5)}_{2;\,2}\big)$ & $|m_3 + m_4|$ \\ \hline
		$\frac{1}{\sqrt{2}}\big(R^{(5)}_{2;\,3}+\tilde{R}^{(5)}_{2;\,3}\big)$ & $|m_1 - m_2|$ \\ \hline
		\hspace*{.5cm}$\frac{1}{\sqrt{2}}\big(R^{(5)}_{2;\,3}-\tilde{R}^{(5)}_{2;\,3}\big)$\hspace*{.5cm} & \hspace*{.5cm}$|m_3 - m_4|$\hspace*{.5cm} \\ \hline
	\end{tabular}}
\end{table}
\renewcommand{\arraystretch}{1}

\newpage
\subsection{The F1-NS5 system}
\label{appendix:NS5F1masstables}

For the reduction of the F1-NS5 system, we chose mass matrices as in \eqref{mass matrices SO(5) NS5F1} and \eqref{mass matrices USp(4) NS5F1}. The scalar masses are:

\renewcommand{\arraystretch}{1.25}
\begin{table}[ht!]
	\scalebox{0.8}{
	\begin{tabular}{|c|c|}
		\hline
		Field $\tilde{\sigma}^i$ & Mass \\ \hline
		\multicolumn{2}{|c|}{} \\[-14.5pt] \hline
$\frac{1}{\sqrt{2}}(\phi_4-\Phi)$ & $0$ \\ \hline
$\frac{1}{2} (\phi_4+\Phi +\sqrt{2}\,\phi_3)$ & $|m_1-m_2-m_3+m_4|$ \\ \hline
$\frac{1}{2} (\phi_4+\Phi -\sqrt{2}\,\phi_3)$ & $|m_1-m_2+m_3-m_4|$ \\ \hline
$\frac{1}{\sqrt{2}}(\phi_1+\phi_2)$ & $|m_1+m_2-m_3-m_4|$ \\ \hline
$\frac{1}{\sqrt{2}}(\phi_1-\phi_2)$ & $|m_1+m_2+m_3+m_4|$ \\ \hline
$\frac{1}{2} (A_{12}+A_{34}+B_{12}+B_{34})$ & $|m_1+m_2-m_3+m_4|$ \\ \hline
$\frac{1}{2} (A_{12}+A_{34}-B_{12}-B_{34})$ & $|m_1+m_2+m_3-m_4|$ \\ \hline
$\frac{1}{2} (A_{12}-A_{34}+B_{12}-B_{34})$ & $|m_1-m_2+m_3+m_4|$ \\ \hline
$\frac{1}{2} (A_{12}-A_{34}-B_{12}+B_{34})$ & $|m_1-m_2-m_3-m_4|$ \\ \hline
$\frac{1}{2} (A_{14}+A_{23}+B_{14}-B_{23})$ & $|m_1-m_2+m_3+m_4|$ \\ \hline
$\frac{1}{2} (A_{14}+A_{23}-B_{14}+B_{23})$ & $|m_1-m_2-m_3-m_4|$ \\ \hline
$\frac{1}{2} (A_{14}-A_{23}+B_{14}+B_{23})$ & $|m_1+m_2-m_3+m_4|$ \\ \hline
\hspace*{.3cm}$\frac{1}{2} (-A_{14}+A_{23}+B_{14}+B_{23})$\hspace*{.3cm} & \hspace*{.3cm}$|m_1+m_2+m_3-m_4|$\hspace*{.3cm} \\ \hline
$A_{13}$ & $|m_1+m_2-m_3-m_4|$ \\ \hline
$A_{24}$ & $|m_1+m_2+m_3+m_4|$ \\ \hline
$B_{13}$ & $|m_1-m_2+m_3-m_4|$ \\ \hline
$B_{24}$ & $|m_1-m_2-m_3+m_4|$ \\ \hline
$\frac{1}{\sqrt{2}}(C_{12}+C_{34})$ & $|m_1+m_2|$ \\ \hline
$\frac{1}{\sqrt{2}}(C_{12}-C_{34})$ & $|m_3+m_4|$ \\ \hline
$\frac{1}{\sqrt{2}}(C_{13}+C_{24})$ & $|m_3-m_4|$ \\ \hline
$\frac{1}{\sqrt{2}}(C_{13}-C_{24})$ & $|m_1-m_2|$ \\ \hline
$\frac{1}{\sqrt{2}}(C_{14}+C_{23})$ & $|m_1+m_2|$ \\ \hline
$\frac{1}{\sqrt{2}}(C_{14}-C_{23})$ & $|m_3+m_4|$ \\ \hline
$\frac{1}{\sqrt{2}}(a+b)$ & $|m_3-m_4|$ \\ \hline
$\frac{1}{\sqrt{2}}(a-b)$ & $|m_1-m_2|$ \\ \hline
 \end{tabular}}
\end{table}
\renewcommand{\arraystretch}{1}

The tensor masses are:

\renewcommand{\arraystretch}{1.25}
\begin{table}[ht!]
	\scalebox{0.8}{
	\begin{tabular}{|c|c|}
		\hline
		Field $A_{2,A}^{(5)}$ & Mass \\ \hline
		\multicolumn{2}{|c|}{} \\[-16pt] \hline
		$B_2^{(5)}$ & $0$ \\ \hline
		$\tilde{B}_2^{(5)}$ & $0$ \\ \hline
		$\frac{1}{\sqrt{2}}\big(C_2^{(5)}+\tilde{C}_2^{(5)}\big)$ & $|m_1 - m_2|$ \\ \hline
		$\frac{1}{\sqrt{2}}\big(C_2^{(5)}-\tilde{C}_2^{(5)}\big)$ & $|m_3 - m_4|$ \\ \hline
		$\frac{1}{\sqrt{2}}\big(R^{(5)}_{2;\,1}+\tilde{R}^{(5)}_{2;\,1}\big)$ & $|m_1 + m_2|$ \\ \hline
		$\frac{1}{\sqrt{2}}\big(R^{(5)}_{2;\,1}-\tilde{R}^{(5)}_{2;\,1}\big)$ & $|m_3 + m_4|$ \\ \hline
		$\frac{1}{\sqrt{2}}\big(R^{(5)}_{2;\,2}+\tilde{R}^{(5)}_{2;\,2}\big)$ & $|m_1 + m_2|$ \\ \hline
		$\frac{1}{\sqrt{2}}\big(R^{(5)}_{2;\,2}-\tilde{R}^{(5)}_{2;\,2}\big)$ & $|m_3 + m_4|$ \\ \hline
		$\frac{1}{\sqrt{2}}\big(R^{(5)}_{2;\,3}+\tilde{R}^{(5)}_{2;\,3}\big)$ & $|m_1 - m_2|$ \\ \hline
		\hspace*{.5cm}$\frac{1}{\sqrt{2}}\big(R^{(5)}_{2;\,3}-\tilde{R}^{(5)}_{2;\,3}\big)$\hspace*{.5cm} & \hspace*{.5cm}$|m_3 - m_4|$\hspace*{.5cm} \\ \hline
	\end{tabular}}
\end{table}
\renewcommand{\arraystretch}{1}

\subsection{The D3-D3 system}
\label{app: masses d3d3}

For the D3-D3 brane set-up that we consider in \autoref{sec:dualbraneconfigurations}, we use the mass matrices given in \eqref{mass matrices SO(5) D3} and \eqref{mass matrices USp(4) D3}. The scalars acquire the following masses:

\renewcommand{\arraystretch}{1.25}
\begin{table}[ht!]
	\scalebox{0.8}{
	\begin{tabular}{|c|c|}
		\hline
		Field $\tilde{\sigma}^i$ & Mass \\ \hline
		\multicolumn{2}{|c|}{} \\[-14.5pt] \hline
$\phi_1$ & $0$ \\ \hline
$\frac{1}{2} (\Phi +\sqrt{2}\,\phi_2+\phi_4)$ & $|m_1+m_2-m_3-m_4|$ \\ \hline
$\frac{1}{2} (\Phi +\sqrt{2}\,\phi_3-\phi_4)$ & $|m_1-m_2+m_3-m_4|$ \\ \hline
$\frac{1}{2} (\Phi -\sqrt{2}\,\phi_3-\phi_4)$ & $|m_1-m_2-m_3+m_4|$ \\ \hline
$\frac{1}{2} (\Phi -\sqrt{2}\,\phi_2+\phi_4)$ & $|m_1+m_2+m_3+m_4|$ \\ \hline
$\frac{1}{2} (b+a+A_{12}-A_{34})$ & $|m_1-m_2-m_3-m_4|$ \\ \hline
$\frac{1}{2} (b-a-A_{12}-A_{34})$ & $|m_1+m_2+m_3-m_4|$ \\ \hline
$\frac{1}{2} (b-a+A_{12}+A_{34})$ & $|m_1+m_2-m_3+m_4|$ \\ \hline
$\frac{1}{2} (b+a-A_{12}+A_{34})$ & $|m_1-m_2+m_3+m_4|$ \\ \hline
$\frac{1}{2} (C_{23}-C_{14}-B_{24}+B_{13})$ & $|m_1-m_2-m_3-m_4|$ \\ \hline
$\frac{1}{2} (C_{23}+C_{14}+B_{24}+B_{13})$ & $|m_1+m_2+m_3-m_4|$ \\ \hline
$\frac{1}{2} (C_{23}+C_{14}-B_{24}-B_{13})$ & $|m_1+m_2-m_3+m_4|$ \\ \hline
\hspace*{.3cm}$\frac{1}{2} (C_{23}-C_{14}+B_{24}-B_{13})$\hspace*{.3cm} & \hspace*{.3cm}$|m_1-m_2+m_3+m_4|$\hspace*{.3cm} \\ \hline
$B_{14}$ & $|m_1+m_2+m_3+m_4|$ \\ \hline
$B_{23}$ & $|m_1+m_2-m_3-m_4|$ \\ \hline
$C_{13}$ & $|m_1-m_2+m_3-m_4|$ \\ \hline
$C_{24}$ & $|m_1-m_2-m_3+m_4|$ \\ \hline
$\frac{1}{\sqrt{2}}(C_{12}+C_{34})$ & $|m_3-m_4|$ \\ \hline
$\frac{1}{\sqrt{2}}(C_{12}-C_{34})$ & $|m_1-m_2|$ \\ \hline
$\frac{1}{\sqrt{2}}(A_{13}+A_{24})$ & $|m_3+m_4|$ \\ \hline
$\frac{1}{\sqrt{2}}(A_{13}-A_{24})$ & $|m_1+m_2|$ \\ \hline
$\frac{1}{\sqrt{2}}(A_{14}+A_{23})$ & $|m_1-m_2|$ \\ \hline
$\frac{1}{\sqrt{2}}(A_{14}-A_{23})$ & $|m_3-m_4|$ \\ \hline
$\frac{1}{\sqrt{2}}(B_{12}+B_{34})$ & $|m_3+m_4|$ \\ \hline
$\frac{1}{\sqrt{2}}(B_{12}-B_{34})$ & $|m_1+m_2|$ \\ \hline
 \end{tabular}}
\end{table}
\renewcommand{\arraystretch}{1}

The tensors masses are:

\renewcommand{\arraystretch}{1.25}
\begin{table}[ht!]
	\scalebox{0.8}{
	\begin{tabular}{|c|c|}
		\hline
		Field $A_{2,A}^{(5)}$ & Mass \\ \hline
		\multicolumn{2}{|c|}{} \\[-16pt] \hline
		$R^{(5)}_{2;\,1}$ & $0$ \\ \hline
		$\tilde{R}^{(5)}_{2;\,1}$ & $0$ \\ \hline
		$\frac{1}{\sqrt{2}}\big(C_2^{(5)}+\tilde{C}_2^{(5)}\big)$ & $|m_1 + m_2|$ \\ \hline
		$\frac{1}{\sqrt{2}}\big(C_2^{(5)}-\tilde{C}_2^{(5)}\big)$ & $|m_3 + m_4|$ \\ \hline
		$\frac{1}{\sqrt{2}}\big(B^{(5)}_{2}+\tilde{B}^{(5)}_{2}\big)$ & $|m_1 - m_2|$ \\ \hline
		$\frac{1}{\sqrt{2}}\big(B^{(5)}_{2}-\tilde{B}^{(5)}_{2}\big)$ & $|m_3 - m_4|$ \\ \hline
		$\frac{1}{\sqrt{2}}\big(R^{(5)}_{2;\,2}+\tilde{R}^{(5)}_{2;\,2}\big)$ & $|m_1 + m_2|$ \\ \hline
		$\frac{1}{\sqrt{2}}\big(R^{(5)}_{2;\,2}-\tilde{R}^{(5)}_{2;\,2}\big)$ & $|m_3 + m_4|$ \\ \hline
		$\frac{1}{\sqrt{2}}\big(R^{(5)}_{2;\,3}+\tilde{R}^{(5)}_{2;\,3}\big)$ & $|m_1 - m_2|$ \\ \hline
		\hspace*{.5cm}$\frac{1}{\sqrt{2}}\big(R^{(5)}_{2;\,3}-\tilde{R}^{(5)}_{2;\,3}\big)$\hspace*{.5cm} & \hspace*{.5cm}$|m_3 - m_4|$\hspace*{.5cm} \\ \hline
	\end{tabular}}
\end{table}
\renewcommand{\arraystretch}{1}

\bibliographystyle{JHEP}
\bibliography{bibliography}

\end{document}